\documentclass[a4paper,11pt]{article}
\pdfoutput=1 % if your are submitting a pdflatex (i.e. if you have
\usepackage{jcappub} % for details on the use of the package, please
\usepackage[T1]{fontenc} % if needed

\usepackage{graphicx}
\usepackage{epstopdf, epsfig}
\usepackage{subcaption}
\usepackage{tikz}
\usepackage{framed}

\title{\boldmath Caustic Skeleton \& Cosmic Web}

\author[a]{Job Feldbrugge}
\author[b]{Rien van de Weygaert}
\author[b,c]{Johan Hidding}
\author[d]{Joost Feldbrugge}

\affiliation[a]{Perimeter Institute for Theoretical Physics, University of Waterloo,\\
Waterloo, Canada}
\affiliation[b]{Kapteyn Astronomical Institute, University of Groningen,\\ Groningen, The Netherlands}
\affiliation[c]{Netherlands eScience Center, Amsterdam, The Netherlands}
\affiliation[d]{JFA Feldbrugge Studios,\\ Lettelbert, The Netherlands}

\emailAdd{jfeldbrugge@perimeterinstitute.ca}

\newcommand{\defm}{\mathcal{M}}

\newcommand{\defmij}{M}
\newcommand{\phasespace}{\mathcal{C}}
\newcommand{\defv}{\mathcal{V}}
\newcommand{\Ham}{\mathcal{H}}

\newtheorem{thr}{Theorem:}

\abstract{

  \noindent We present a general formalism for identifying the caustic structure of a dynamically evolving mass distribution,
  in an arbitrary dimensional space. The identification of caustics in fluids with Hamiltonian dynamics, viewed in Lagrangian space,
  corresponds to the classification of singularities in Lagrangian catastrophe theory. On the basis of this formalism we develop a
  theoretical framework for the dynamics of the formation of the cosmic web, and specifically those aspects that characterize its unique nature:
  its complex topological connectivity and multiscale spinal structure of sheetlike membranes, elongated filaments and compact cluster nodes.
  Given the collisionless nature of the gravitationally dominant dark matter component in the universe, the presented formalism
  entails an accurate description of the spatial organization of matter resulting from the gravitationally driven formation of cosmic structure.  
  
  The present work represents a significant extension of the work by Arnol'd et al.\,\cite{Arnold:1982a}, who classified the caustics
  that develop in one- and two-dimensional systems that evolve according to the Zel'dovich approximation. His seminal work established
  the defining role of emerging singularities in the formation of nonlinear structures in the universe. At the transition from the linear
  to nonlinear structure evolution, the first complex features emerge at locations where different fluid elements cross to establish
  multistream regions. Involving a complex folding of the 6-D sheetlike phase-space distribution, it manifests itself in the appearance
  of infinite density caustic features. The classification and characterization of these mass element foldings can be encapsulated in
  \textit{caustic conditions} on the eigenvalue and eigenvector fields of the deformation tensor field.
 
  In this study we introduce an alternative and transparent proof for Lagrangian catastrophe theory.
  This facilitates the derivation of the caustic conditions for general Lagrangian fluids, with arbitrary dynamics. 
  Most important
  in the present context is that it allows us to follow and describe the full three-dimensional geometric and topological complexity of
  the purely gravitationally evolving nonlinear cosmic matter field. While generic and statistical results can be based on the eigenvalue
  characteristics, one of our key findings is that of the significance of the \textit{eigenvector field} of the deformation field for outlining
  the entire spatial structure of the \textit{caustic skeleton} emerging from a primordial density field.

  In this paper we explicitly consider the caustic conditions for the three-dimensional Zel'dovich approximation, extending earlier work on
  those for one- and two-dimensional fluids towards the full spatial richness of the cosmic web. In an accompanying publication, we apply
  this towards a full three-dimensional study of caustics in the formation of the cosmic web and evaluate in how far it manages to outline
  and identify the intricate skeletal features in the corresponding $N$-body simulations.
}

\begin{document}
\maketitle

\section{Introduction}
Caustics\footnote{In singularity theory, a caustic is the curve of critical values the Lagrangian mapping $q \to x_t(q)$.} that emerge in fluid flows are best
studied in a Lagrangian space. They are important features, marking the positions where fluid elements cross and multi-stream regions form.
These caustics can be associated to the regions with infinite density, corresponding to locations where shell-crossing occurs. In the present study,
we concentrate specifically on the role of caustics in the formation of the cosmic web. The gravitationally driven formation of structure in the
universe is dominated by the dark matter component. Given its collisionless nature, the formalism that we present in this study entails an accurate  
description of the spatial structure that emanates as a result of its dynamical evolution. The emerging caustics even have a direct
physical impact on the baryonic matter, given its accretion into the gravitational potential wells delineated by the evolving
dark matter distribution. Notwithstanding this cosmological focus, the caustic conditions and mathematical formalism that we have derived for
this are of a more generic nature, with a validity that extends to all systems which allow for a Lagrangian description.

The cosmic web is the complex network of interconnected filaments and walls into which galaxies and matter have aggregated
on Megaparsec scales. It contains structures from a few megaparsecs up to tens and even hundreds of megaparsecs of size.
The weblike spatial arrangement is marked by highly elongated filamentary and flattened planar structures, connecting in dense compact cluster nodes surrounding
large near-empty void regions. As borne out by a large sequence of N-body computer experiments of cosmic structure formation (e.g. \cite{springmillen2005,
  illustris2014,eagle2015}), these web-like patterns in the overall cosmic matter distribution do represent a universal but possibly transient phase in the
gravitationally driven emergence and evolution of cosmic structure (see e.g. \cite{aragon2010,cautun2014}).

According to the \emph{gravitational instability scenario} \citep{peebles1980}, cosmic structure grows from tiny primordial density and velocity perturbations.
Once the gravitational clustering process has progressed beyond the initial linear growth phase, we see the emergence of complex patterns and structures in the
density field.  The resulting web-like patterns, outlined by prominent anisotropic filamentary and planar features surrounding characteristic large underdense
void regions, are therefore a natural manifestation of the gravitational cosmic structure formation process.

The recognition of the \emph{cosmic web} as a key aspect of the emergence of structure in the Universe came with early
analytical studies and approximations concerning the emergence of structure out of a nearly featureless primordial Universe. In this respect
the Zel'dovich formalism~\cite{Zeldovich:1970} played a seminal role. The emphasis on anisotropic collapse as agent for forming and shaping structure
in the Zel'dovich "pancake'' picture \cite{Zeldovich:1970,icke1973}
was seen as the rival view to the purely hierarchical clustering view of structure formation. The successful synthesis of both elements
in the \emph{cosmic web} theory of Bond et al. \cite{bondweb1996} appears to provide a succesful description of large scale structure
formation in $\Lambda$CDM cosmology. The cosmic web theory emphasizes the intimate dynamical relationship
between the prominent filamentary patterns and the compact dense clusters that stand out as the nodes within the cosmic matter distribution
\cite{bondweb1996,colberg2005,weybond2008}. It also implies that a full understanding of the cosmic web's dynamical evolution is necessary to
clarify how its structural features are connected in the intricate network of the cosmic web. To answer this question we need to turn
to a full phase-space description of the evolving matter distribution and mass flows. 

The Zel'dovich formalism \cite{Zeldovich:1970} already underlined the importance of a full phase-space description
for understanding cosmic structure formation, however, with the exception of a few prominent studies \cite{Arnold:1982a},
the wealth of information in the full 6-D phase-space escaped attention. This changed with the publication of a number
of recent publications \cite{abel2012,falck2012,neyrinck2012,shandarin2012,ramachandra2015} (for an early
study on this observation see \cite{buchertehlers1993}) in which it was realized that the morphology of components in the evolving matter
distribution is closely related to its multistream character. This realization is based on the recognition that the emergence of
nonlinear structures occurs at locations where different streams of the corresponding flow field cross each other.

Looking at the appearance of the evolving spatial mass distribution as a 3D {\it phase space sheet} folding itself in 6D phase space,
a connection is established between the structure formation process and the morphological classification of the emerging structure. 
Caustics, which are the subject of this study, mark the regions where the cosmic web begins to form. Based on
recent advances and insights, in this study we discuss the role of caustics in the formation of the cosmic web.  By tracing the caustics during
the formation of the cosmic web we obtain a skeleton of the current three-dimensional large scale structure. 

\begin{figure}[h]
\centering
\includegraphics[width=0.9 \textwidth]{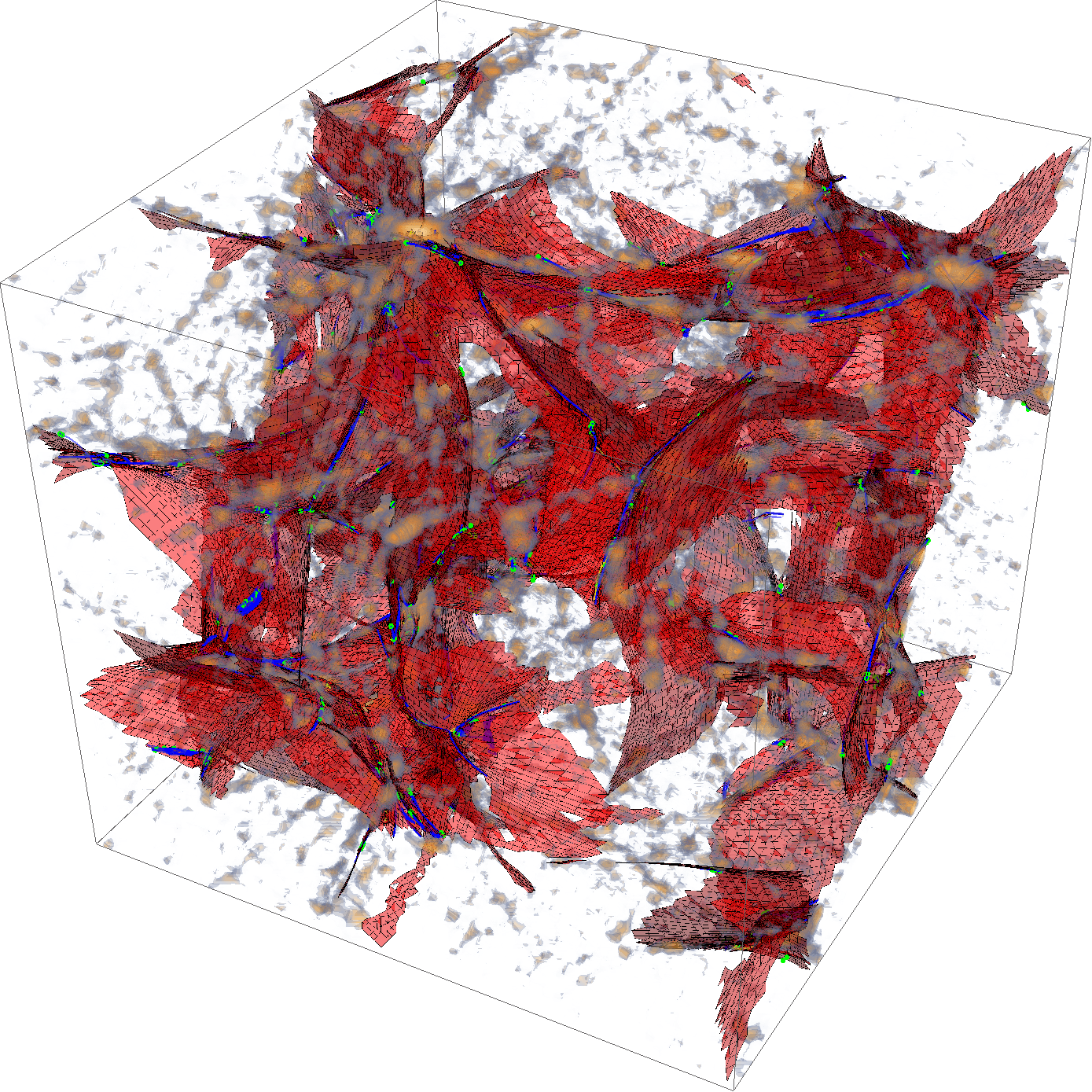}
\caption{Illustration of caustic features in the cosmic web, and its relation with the corresponding density field. The N-body simulation
  is a CDM simulation in an Einstein-de Sitter Universe. On the basis of the initial flow deformation field, the caustics in the
  matter distribution have been identified. The detailed description of these follows in section 4. 
  The red sheets represent the cusps ($A_3$) singularities which correspond to the walls or membranes of the cosmic web. The blue lines and the
  green points are the swallowtail ($A_4$) and butterfly ($A_5$) singularities corresponding to the filaments and clusters of the large scale
  structure. The dark matter distribution in the N-body simulation is represented by a log density colour scheme.}
\label{fig:exSkel}
\end{figure}

Caustics in fluids with Hamiltonian dynamics, viewed in Lagrangian space, are classified by Lagrangian catastrophe theory \citep{Arnold:1976, Arnold:1982b,
  Kravtsov:1983, Zeldovich:1983a}. Following this, these results were soon extended to fluids with generic dynamics \citep{Bruce:1985}. For the classification of
caustics emerging in the context of a one- and two-dimensional description of cosmic structure formation by the Zel'dovich approximation,
Arnol'd et al.\,\cite{Arnold:1982a} translated this into conditions on the displacement field. Following up on this seminal work,
Hidding et al.\,\cite{Hidding:2014} 
analyzed the overall morphology and connectivity of caustics that emerge in a displacement field described by the one- and two-dimensional 
Zel'dovich approximation. The visual illustration of the emerging structure, for a field of initially Gaussian random density and potential 
fluctuations, revealed how the caustics spatially outline the spine of the cosmic web. Feldbrugge et al.\, \cite{Feldbrugge:2014} elaborated this into 
an analytical evaluation of the statistical properties of caustics, assuming a random Gaussian initial density field. \\

In the current study we assess the caustics emerging in a one-parameter family of sufficiently differentiable maps $x_t:\mathbb{R}^3\mapsto \mathbb{R}^3$, mapping
the initial mass distribution to the final mass distribution at time $t$. For practical considerations we consider the evolution of a collisionless medium of matter
in 6-dimensional phase-space. The collisionless Boltzmann equation, known as the Vlasov equation, describes the development of the phase-space density $f(x,v)$ of
the medium. In a gravitational field $\Phi$, the phase-space density of mass elements with velocity $v$ at location $x$ evolves according to 
\begin{equation}
{\displaystyle \partial f \over \displaystyle \partial t}\,+\,v_k {\displaystyle \partial f \over \displaystyle \partial x_k}
-{\displaystyle \partial \Phi \over \displaystyle \partial x_k}{\displaystyle \partial f \over \displaystyle \partial v_k}\,=\,0\,.
\end{equation}
While the medium strictly speaking cannot be considered as a physical fluid, in the sense of a medium characterized by continuously varying
one-valued quantities in Eulerian space, we might use the term ``Lagrangian fluid'' or ``Vlasov fluid'' for the dark matter medium. For reasons of lucidity, in the
remainder of this study we denote a ``Vlasov fluid'' shortly as ``fluid''. 

Within this context, we give a novel proof of Lagrangian catastrophe theory and the corresponding \textit{caustic conditions} for
three-dimensional Hamiltonian fluids. These conditions are expressed in both the eigenvalue and the eigenvector fields of the mass flow deformation tensor. Moreover, our scheme allows us to extend these caustic conditions to fluids with non-Hamiltonian dynamics. Applied to the three-dimensional Zel'dovich approximation, these conditions on the initial density field lead to a \textit{caustic skeleton} of the cosmic web. In this skeleton the walls, filaments and clusters of the large scale structure are directly related to the $A_3,A_4,A_5,D_4$ and $D_5$ caustics of Lagrangian catastrophe theory. See figure \ref{fig:exSkel} for an illustration of the caustic skeleton of the Zel'dovich approximation and a dark matter $N$-body simulation. A detailed analysis of the caustic skeleton of the Zel'dovich approximation and a comparison with $N$-body simulations is the subject of a follow-up paper \citep{Feldbrugge:2017}.

It should be emphasized that the eigenvalue fields of the mass flow deformation tensor have, for a long time, been successfully used in Lagrangian studies of the cosmic web \cite{Chong:1990, Wang:2014, Leclercq:2017}. In these studies, the clusters, filaments and walls are related to the number of eigenvalues exceeding a threshold. The 
caustic skeleton here proposed complements their work in that it include the information of the eigenvector fields, which so far has been largely neglected. 

The paper begins section \ref{sec:LFD} with a concise description of Lagrangian fluid dynamics. 
The formation of caustics and derivation of the shell-crossing conditions for the occurrence of multistream regions in a flow field is studied in
section \ref{sec:ShellCrossing}.
These conditions are among the main results presented here. In section \ref{sec:CC} we apply these shell-crossing conditions to the classification of 
catastrophes, described in section \ref{sec:class}, to derive the caustic conditions. Section~\ref{sec:cosmic_web} discusses the relevance and
significance of the caustic structure in the context of the evolving cosmic mass distribution, and in particular the emergence and
morphological structure of the cosmic web. Also, it discusses the further application and development of the caustic formalism in a 
cosmological context, outlining the main elements of our project. In section \ref{sec:Dynamics} we describe 
the dynamical framework resulting from the considerations above. Finally, in section \ref{sec:Conclusion} we summarize the results and discuss possible applications.

%%%%%%%%%%%%%%%%%%%%%%%%%%%%%%%%%%%%%%%%%%%%%%%%%%%%%%%%%%%%%%%%%%%%%%%%%%%%%%%%%%%%%%%%%%%%%%%%%%%%%%%%%%%%
\section{Lagrangian fluid dynamics}
\label{sec:LFD}
\noindent There exist multiple approaches to fluid dynamics. In the Eulerian approach, the evolution of the smoothed density and velocity 
fields is analyzed. The equations of motion of Eulerian fluids are relatively concise and give a reasonably accurate description of 
the mean flow in a fluid element at a given location in the fluid. The Lagrangian view of particle flows is the appropriate tool for
following the complex dynamical evolution of fluid elements, including the evolution of multi-stream regions and the emergence of caustics, where the caustics
are the critical values of the Lagrangian map.

%One serious disadvantage of the Eulerian framework is that it does not directly relate to the motion of the particles in the fluid. 
%Because it basically restricts itself to the mean motion in a fluid element, it does not facilitate an accurate description of the 
%evolution of multi-stream regions. More suited for following 

\bigskip
In Lagrangian fluid dynamics, we assume every point in space to consist of a mass element that is moving with the fluid. 
Their motion is described by a Lagrangian map $x_t : L \to E$, mapping the initial position $q$ in the Lagrangian manifold $L$ to the position $x_t(q)$ 
of the mass element in the Eulerian manifold $E$ at time $t$.\footnote{Note that here we do not explicitly use a distinct notation for vector quantities: 
  $q$ and $x_t$ are vectors which in conventional cosmology notation are usually written as ${\vec q}$ and ${\vec x}_t$. Throughout this paper we use
  the notation familiar to the mathematics literature.} In the context of Lagrangian fluid dynamics, 
it is most convenient to describe the evolving fluid in terms of the displacement map $s_t$ defined by, 
\begin{equation}
s_t(q)=x_t(q)-q\,,
\label{eq:displac}
\end{equation}
for all $q\in L$. For the Zel'dovich approximation \cite{Zeldovich:1970} of cosmic structure formation the displacement field is given by
\begin{equation}
s_t(q) = - b_+(t) \nabla_q \Psi(q)\,,
\end{equation}
with the growing mode $b_+$ and the displacement potential $\Psi$ (appendix~\ref{sec:Zel'dovich}). The displacement potential is proportional to the linearly extrapolated gravitational potential to the current epoch $\phi_0$, \textit{i.e.} 
\begin{equation}
\Psi(q)= \frac{2}{3\Omega_0 H_0^2} \phi_0(q)\,,
\end{equation}
with $H_0$ the current Hubble parameter and $\Omega_0$ the current total energy density.
In this paper we always assume the maps $x_t$ and $s_t$ to be continuous and sufficiently differentiable. While in 
the Lagrangian description a mass element has a constant mass, it may contract, expand, deform and even rotate. This is 
described in terms of the deformation tensor $\defm$, the gradient of the displacement field with respect to the Lagrangian coordinates 
of a mass element,
\begin{equation}
\defm\,=\,\frac{\partial s_t}{\partial q}\,=\,\left(
 \begin{array}{ccc}
  \defmij_{1,1} & \defmij_{2,1} & \defmij_{3,1} \\
  \defmij_{1,2} & \defmij_{2,2} & \defmij_{3,2} \\
  \defmij_{1,3} & \defmij_{2,3} & \defmij_{3,3} 
 \end{array}\right).
\label{eq:deformtensor}
\end{equation}

\medskip
\noindent While mass elements in a Lagrangian fluid are characterized by a few fundamental quantities, which characterize 
them and remain constant throughout their evolution, most physical properties are basically derived quantities. 
A good example and illustration of a 
derived quantity is the density field. The density in a point $x'\in E$ is defined as the initial mass in 
the mass element times the ratio of the initial and final volume of the mass element. Formally, this is expressed as a change of 
coordinates involving the Jacobian of the map $x_t$,
\begin{eqnarray}
\rho(x',t)&=&\sum_{q \in A_t(x')}\rho_i(q)\left| \frac{\partial x_t(q)}{\partial q}\right|^{-1}\nonumber\\
&=&\sum_{q \in A_t(x')}\rho_i(q)\left| I+\frac{\partial s_t(q)}{\partial q}\right|^{-1}\,.
\end{eqnarray}
\noindent This can be written as
\begin{eqnarray}
\rho(x',t)&=&\sum_{q \in A_t(x')} \frac{\rho_i(q)}{|1+\mu_{t1}(q)||1+\mu_{t2}(q)||1+\mu_{t3}(q)|}\,,
\label{eq:LagrangianDensity}
\end{eqnarray}
with $A_t(x')$ the points $q$ in Lagrangian space $L$ which map to $x'$, i.e., $A_t(x')=\{q\in L| x_t(q)=x'\}$, $\rho_i$ the initial density field and $\mu_{ti}$ the eigenvalues of the deformation tensor $\defm(q)$, defined by
\begin{equation}
\mathcal{M}v_i = \mu_i v_i
\end{equation}
with eigenvector $v_i$. The equality in equation \eqref{eq:LagrangianDensity} applies to general deformation tensors\footnote{Note that here we use the general convention to represent the deformation eigenvalue field,
  with $\mu_i(q)$ the $i$-th eigenvalue of the deformation tensor, $\defm(q)$. This differs from the usual convention in cosmology to use the
  time-independent representation of the deformation field in the context of the Zel'dovich approximation. Within this formalism,
  the eigenvalues $\lambda_i(q)$ of the deformation field $\psi_{ij}=\partial^2 \Psi(q)/\partial q_i \partial q_j$, are related to the eigenvalues
  $\mu_i(q)$ via the linear relation $\mu_i(q,t) = - b_+(t) {\lambda}_i(q)$, in which $b_+(t)$ is the growing mode growth factor. See Appendix~A for
  further details.}, since the characteristic polynomial of the deformation tensor can be expressed in terms of the eigenvalues
\begin{equation}
\chi(\lambda)= \det\left[\frac{\partial s_t}{\partial q} - \lambda I\right] = (\mu_{t1}-\lambda)(\mu_{t2}-\lambda)(\mu_{t3}-\lambda)\,,
\end{equation}
by which 
\begin{equation}
\det\left[I+\frac{\partial s_t}{\partial q}\right]=\chi(-1)=(1+\mu_{t1})(1+\mu_{t2})(1+\mu_{t3})\,.
\end{equation}
By substituting derived quantities like density in the, often more familiar, Eulerian fluid equations, we may obtain a 
closed set of differential equations for the Lagrangian map $x_t$ or the displacement map $s_t$. Note that for practical
reasons in this paper we will sometimes suppress the time index of the eigenvalue fields, i.e. $\mu_{i}=\mu_{ti}$. 
 
Equation \eqref{eq:LagrangianDensity} applies to a fluid with three spatial dimensions. For simplicity, 
we will restrict explicit expressions to the $3$-dimensional case \footnote{Formally, it would be 
appropriate to describe the fluids as $(d+1)$-dimensional fluids, a combination of their embedding in a $d$-dimensional space along with their 
evolution along time dimension $t$.}. The arguments presented in this paper straightforwardly generalize to a Lagrangian fluid with an arbitrary number of spatial dimension
and it is straightforward to generalize equation~\eqref{eq:LagrangianDensity} to $d$-dimensional fluids in 
$d$-dimensional space. 

\bigskip
The appearance of singularities in equation~\eqref{eq:LagrangianDensity} is central to our discussion concerning the nature of
these singularities. They occur when a mass element reaches an infinite density. More formally stated, as we will see in
section~\ref{sec:ShellCrossing}, an infinite density occurs when for at least one of the $i=1,\ldots,d$,
\begin{equation}
1+\mu_i = 0\,.
\end{equation}
The regions, in which the mapping $x_t$ becomes degenerate and 
the density becomes infinite are known as \emph{foldings}, \emph{caustics} or \emph{shocks}. They mark important features in 
the Lagrangian fluid and are the object of study in this paper. 

\medskip While these eigenvalue conditions provide the necessary condition for a mass element to pass through a caustic, and
reach infinite density, it does not yield the full information necessary to infer the geometric structure, spatial connectivity and
identity of the
caustic. As mass elements pass through a multistream region, the spatial properties of the flow will determine the complexity of the
folding of the phase-space sheet in which they are embedded. In this study we demonstrate that the corresponding
eigenvectors are instrumental in establishing the spatial outline and identify of the corresponding caustics. This
key realization emanates from the so-called \textit{caustic conditions}. 

\medskip 
Throughout our study, we assume that the displacement map $s_t$ is continuous and sufficiently differentiable. The corresponding 
eigenvalues are the roots of the characteristic polynomial of the matrix $\defm={\partial s_t}/{\partial q}$. Since the characteristic 
equation is a non-linear equation, in principle the eigenvalues could develop singularities and become non-differentiable. However, 
it can be shown that the eigenvalues can be ordered such that they are continuous. Furthermore the eigenvalues will be assumed to be differentiable 
whenever the eigenvalues are distinct. When two eigenvalues coincide, the eigenvalue fields may become non-differentiable. 

\vskip 0.5truecm
\subsection{Hamiltonian fluid dynamics}

\label{sec:hamiltonian}
\noindent For fluids moving with no dissipation of energy, the Hamiltonian formalism may be applied. Hamiltonian fluids have a potential velocity field
\begin{equation}
v=\nabla \phi
\end{equation}
with the velocity potential $\phi$. The mass density $\rho$ and the velocity potential serve as conjugate variables for the Hamiltonian $\mathcal{H}$, with the equations of motion
\begin{eqnarray}
\frac{\partial \rho}{\partial t}&\,=\,&+ \frac{\delta \Ham}{\delta \phi}=-\nabla \cdot (\rho v)\,,\nonumber\\
\frac{\partial \phi}{\partial t}&\,=\,&- \frac{\delta \Ham}{\delta \rho}\,.
\label{eq:eqmotion}
\end{eqnarray}
A simple example of a Hamiltonian is 
\begin{equation}
\Ham\,=\,\int {\rm d}x  \left({1 \over 2} \rho (\nabla \phi)^2 + e(\rho) \right)\,,
\end{equation}
where $e(\rho)$ is the internal energy as a function of density $\rho$. The first equation of motion 
in equation~\eqref{eq:eqmotion} is equivalent to the continuity equation, while the second equation implies 
the Euler equation
\begin{equation}
\frac{\partial v}{\partial t}+v\cdot \nabla v\,=\,- \frac{1}{\rho} \nabla p\,,
\end{equation}
in which $p$ is the pressure of the fluid. For a thorough discussion of fluid mechanics we refer to the seminal volumes of \cite{Landau:1976}, and \cite{Landau:1959}.
For detailed and extensive treatments and analyses of Hamiltonian mechanics and Hamiltonian fluids, we refer to the 
reviews and textbooks by \cite{Arnold:1978}, \cite{Arnold:1992}, \cite{Goldstein:1980}, \cite{Morrison:1998}, and \cite{Salmon:1988}.

%%%%%%%%%%%%%%%%%%%%%%%%%%%%%%%%%%%%%%%%%%%%%%%%%%%%%%%%%%%%%%%%%%%%%%%%%%%%%%%%%%%%%%%%%%%%%%%%%%%%%%%%%%%%

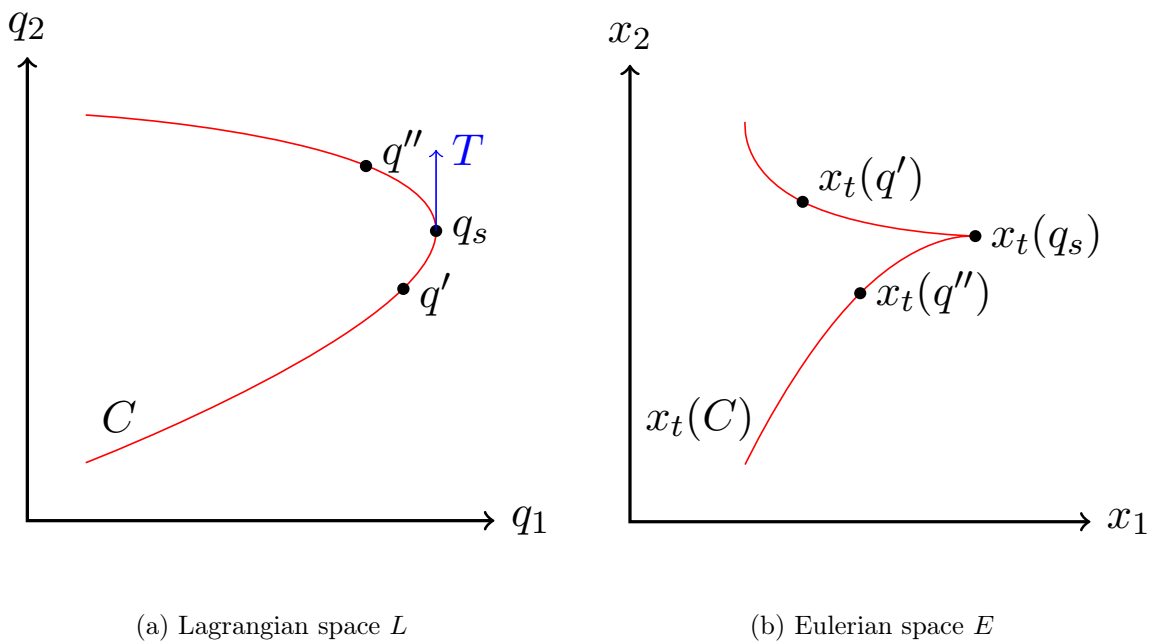
\begin{figure}
\centering
\begin{subfigure}[b]{0.5\textwidth}
\resizebox {\columnwidth}{!}{
\begin{tikzpicture}
\draw [<->,thick] (0,4) node (yaxis) [above] {$q_2$}|- (4,0) node (xaxis) [right] {$q_1$};
\draw [red, xshift=-0.5cm, yshift=-0.5cm] plot [smooth, tension=1] coordinates { (1,1) (4,3) (1,4)};
\fill[black] (3.5,2.5) circle (1.5pt);
\fill[black] (3.22,2) circle (1.5pt);
\fill[black] (2.9,3.06) circle (1.5pt);
\draw[] (3.5,2.5) node[right] {$q_s$};
\draw[] (3.22,1.95) node[right] {$q'$};
\draw[] (2.9,3.19) node[right] {$q''$};
\draw[] (0.5,0.9) node[right] {$C$};
\draw[blue,->] (3.5,2.5) -- (3.5,3.2) node[right] {$T$}; 
\draw[white] (2,0) node[below] {$x(q_s)$};
\end{tikzpicture}}
\caption{Lagrangian space $L$}
\label{fig:SC1}
\end{subfigure}~
\begin{subfigure}[b]{0.5\textwidth}
\resizebox {\columnwidth}{!}{
\begin{tikzpicture}
\draw [<->,thick] (0,4) node (yaxis) [above] {$x_2$}|- (4,0) node (xaxis) [right] {$x_1$};
\draw [red, xshift=-1cm, yshift=-0.5cm] plot [smooth, tension=1] coordinates { (2,1) (3,2.5) (4,3)};
\draw [red, xshift=-1cm, yshift=-0.5cm] plot [smooth, tension=1] coordinates { (4,3) (2.5,3.3) (2.,4)};
\fill[black] (3.,2.5) circle (1.5pt);
\fill[black] (2.,2.) circle (1.5pt);
\fill[black] (1.5,2.8) circle (1.5pt);
\draw[] (3.,2.5) node[right] {$x_t(q_s)$};
\draw[] (2.,2.) node[right] {$x_t(q'')$};
\draw[] (1.5,3.) node[right] {$x_t(q')$};
\draw[] (0.,0.9) node[right] {$x_t(C)$};
\draw[white] (2,0) node[below] {$x(q_s)$};
\end{tikzpicture}}
\caption{Eulerian space $E$}
\label{fig:SC2}
\end{subfigure}
\caption{The shell-crossing process of a curve $C$ in a Lagrangian map $x_t$. The left panel shows Lagrangian space, describing the initial positions of the fluid. The right panel shows Eulerian space, describing the positions of the fluid at time $t$. The fluid undergoes shell-crossing in point $q_s$ on the curve $C$ (red) at time $t$. The neighboring points $q'$ and $q''$ have passed through the opposing segments of $C$. The Lagrangian mapping of the curve $x_t(C)$ (red) develops a non-differentiable point in $x_t(q_x)$, which is known as a caustic. The arrow $T$ (blue) is the tangent vector of the curve $C$ in point $q_s$.}
\label{fig:SC3}
\end{figure}

\section{Shell-crossing conditions} 
\label{sec:ShellCrossing}
\noindent The caustics mentioned above result from the folding of the phase space fluid. At the initial time, $t=0$, the 
fluid has not yet evolved. The displacement map $s$ is therefore the zero map (eqn.~\eqref{eq:displac}), i.e., 
\begin{equation}
s_0(q)=0
\end{equation}
for all $q\in L$. The map $x_0(q)$ is one-to-one, i.e. each Eulerian coordinate $x$ corresponds to one Lagrangian 
position $q$. Throughout the entire volume, the fluid only contains single-stream regions. As the fluid evolves and 
nonlinearities start to emerge, we see the development of \emph{multi-stream regions} in the fluid. At the boundary of 
a multi-stream region, the volume of a mass element vanishes and its density -- following eqn.~\eqref{eq:LagrangianDensity} --  becomes 
infinite. 
At such locations in phase space the map $x_t(q)$ attains a $n$-to-one character, with $n$ an odd positive integer ($n=3,5,7,\ldots$). It means that 
at any one Eulerian location $x$, streams from $n$ different Lagrangian positions cross. 

The key question we address here is that of inferring the conditions under which a mass element with Lagrangian coordinate $q$ 
undergoes shell-crossing. Here we derive the necessary and sufficient conditions for the process of shell-crossing to occur. 
These conditions are called \emph{shell-crossing conditions}. They are the foundation on the basis of which we infer -- in 
section~\ref{sec:CC} -- the related conditions on the displacement field for the occurrence of the various classes of caustics. 
These are called the \emph{caustic conditions}. We infer the caustic conditions for generic as well as Hamiltonian fluid dynamics. 

\subsection{Shell-crossing condition: the derivation}\label{sec:shellcrossingderivation}
\noindent A typical configuration resulting from the \emph{shell-crossing process} -- the name by which it is usually indicated -- 
is illustrated in figure~\ref{fig:SC3}. It focuses on points $q=(q_1,q_2)$ that lie on a smooth curve $C$ in Lagrangian space 
$L$ (fig.~\ref{fig:SC1}). In this context, smooth refers to the assumption that the curve $C$ is $C^1$ continuous. At time $t$, the
points on the Lagrangian curve $C$ map to the variety $x_t(C)$ in Eulerian space $E$ 
(fig.~\ref{fig:SC2})\footnote{In algebraic geometry, a {\it variety} is the zero set of a function $f$, ie. the set of solutions $x \in E$ such that 
$f(x)=0$.}. The fluid in point $q_s$ undergoes shell-crossing at time $t$. The neighboring points $q'$ and $q''$ have 
passed through the opposing segments of $C$. As a result of this, the curve $C$ develops a non-differentiable point 
in $x_t(q_x)$, which is known as a \emph{caustic}. 

\begin{figure}
\centering
\begin{subfigure}[b]{0.33\textwidth}
\resizebox {\columnwidth}{!}{
\begin{tikzpicture}
\draw [<->,thick] (0,4) node (yaxis) [above] {$q$}|- (4,0) node (xaxis) [right] {$x$};
\draw [red, xshift=-0.5cm, yshift=-0.5cm] plot [smooth, tension=1] coordinates { (1,1) (2.,1.7) (3,3.3) (4,4.1)};
\fill[black,opacity = 0.0] (3.5,2.5) coordinate (c_1) circle (1.5pt);
\fill[black] (1.65,1.4) coordinate (c_2) circle (1.5pt);
\fill[black] (3.3,3.5) coordinate (c_3) circle (1.5pt);
\draw[] (3.5,2.5) node[right] {};
\draw[] (3.22,1.6) node[right] {};
\draw[] (2.5,3.) node[right] {};
\draw[] (0.4,0.9) node[right] {$\mathcal{L}$};
\draw[dashed,opacity = 0.0] (yaxis |- c_1) node[left] {$q_s$}
		 -| (xaxis -| c_1) node[below] {$x(q_s)$};
\draw[dashed] (yaxis |- c_2) node[left] {$q''$}
		 -| (xaxis -| c_2) node[below] {$x(q'')$};
\draw[dashed] (yaxis |- c_3) node[left] {$q'$}
		 -| (xaxis -| c_3) node[below] {$x(q')$};
\end{tikzpicture}}
\end{subfigure}~
\begin{subfigure}[b]{0.33\textwidth}
\resizebox {\columnwidth}{!}{
\begin{tikzpicture}
\draw [<->,thick] (0,4) node (yaxis) [above] {$q$}|- (4,0) node (xaxis) [right] {$x$};
\draw [red, xshift=-0.5cm, yshift=-0.5cm] plot [smooth, tension=1] coordinates { (1,1.0) (2.2,1.6) (2.7,3.4) (3.8,4.1)};
\fill[black] (1.97,2.1) coordinate (c_1) circle (1.5pt);
\fill[black] (1.1,0.6) coordinate (c_2) circle (1.5pt);
\fill[black] (2.9,3.5) coordinate (c_3) circle (1.5pt);
\draw[] (3.5,2.5) node[right] {};
\draw[] (3.22,1.6) node[right] {};
\draw[] (2.5,3.) node[right] {};
\draw[] (0.4,0.9) node[right] {$\mathcal{L}$};
\draw[dashed] (yaxis |- c_1) node[left] {$q_s$}
		 -| (xaxis -| c_1) node[below] {$x(q_s)$};
\draw[dashed] (yaxis |- c_2) node[left] {$q''$}
		 -| (xaxis -| c_2) node[below] {$x(q'')$};
\draw[dashed] (yaxis |- c_3) node[left] {$q'$}
		 -| (xaxis -| c_3) node[below] {$x(q')$};
\end{tikzpicture}}
\end{subfigure}~
\begin{subfigure}[b]{0.33\textwidth}
\resizebox {\columnwidth}{!}{
\begin{tikzpicture}
\draw [<->,thick] (0,4) node (yaxis) [above] {$q$}|- (4,0) node (xaxis) [right] {$x$};
\draw [red, xshift=-0.5cm, yshift=-0.5cm] plot [smooth, tension=1] coordinates { (1,1) (4,3) (1,4.2)};
\fill[black] (3.5,2.5) coordinate (c_1) circle (1.5pt);
\fill[black] (2.6,1.52) coordinate (c_2) circle (1.5pt);
\fill[black] (1.65,3.5) coordinate (c_3) circle (1.5pt);
\draw[] (3.5,2.5) node[right] {};
\draw[] (3.22,1.6) node[right] {};
\draw[] (2.5,3.) node[right] {};
\draw[] (0.4,0.9) node[right] {$\mathcal{L}$};
\draw[dashed] (yaxis |- c_1) node[left] {$q_s$}
		 -| (xaxis -| c_1) node[below] {$x(q_s)$};
\draw[dashed] (yaxis |- c_2) node[left] {$q''$}
		 -| (xaxis -| c_2) node[below] {$x(q'')$};
\draw[dashed] (yaxis |- c_3) node[left] {$q'$}
		 -| (xaxis -| c_3) node[below] {$x(q')$};
\end{tikzpicture}}
\end{subfigure}~
\caption{Folding of a one-dimensional fluid in phase space $\phasespace$. The three panels show the time evolution of the Lagrangian submanifold $\mathcal{L}$ (red) of the fluid in phase space. We track the evolution of two points $(q',x(q')),(q'',x(q''))$ forming a multi-stream region and mark the point undergoing shell-crossing by $(q_s,x(q_s))$. Left panel: the fluid -- early in its evolution -- consisting of a single-stream region. Middle panel: a fluid during the process of shell-crossing. Right panel: a fluid consisting of a multi-stream region.}
\label{fig:Plot2}
\end{figure}
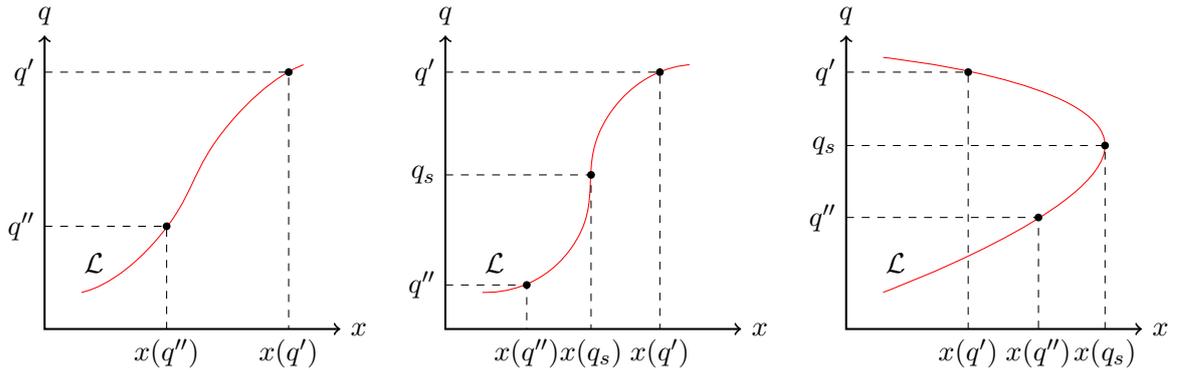

In a time sequence of three steps, figure~\ref{fig:Plot2} illustrates the dynamical process that is underlying the 
formation of the caustic at $x_t(q_s)$. The singularity at $x_t(q_s)\in x_t(C)$ forms as the result of a folding process in 
phase space. We may appreciate the emerging structure when assessing the fate of two neighboring points $q', q'' \in C$ on 
both sides of $q_s$. While the phase space sheet $x_t(C)$ is folded, the points $x_t(q')$ and $x_t(q'')$ turn around while 
passing through $x_t(q_s)$. In figure~\ref{fig:Plot2} we observe how the initially single-stream phase space sheet (lefthand panel) 
morphs into a configuration marked by shell-crossing as different mass elements $q$ pile up at the same Eulerian position $x_t(q_s)$ 
(central panel). Subsequently, around $x_t(q_s)$ we notice the formation of a multi-stream region, with the presence of mass 
elements $q'$ having passed into a region where mass elements from other Lagrangian locations $q$ are to be found. 

\medskip
To infer the shell-crossing conditions, we investigate a curve $C$ in Lagrangian space along which we have  
points $q$ that will find themselves incorporated in a singularity at Eulerian position $x_s(q_s)$. In the case 
of shell-crossing, points $q$ near the Lagrangian location $q_s$ will map onto the same Eulerian position 
$x(q_s)$. The key realization is that this occurs as points $q$ along a direction $T$ tangential to C are all 
folded on to a single Eulerian position $x_s(q_s)$. This translates the question of the shell-crossing condition 
into one on the identity of a tangential direction $T(q)$ along which shell-crossing may or will occur. In other words, 
whether on a  particular curve $C$ -- or, more general, a manifold $M$ -- there are points $q$ where along one or more 
tangential directions $T(q)$ to that curve or manifold shell-crossing may or will take place. 

Zooming in on two points $q'$ and $q''$ in the vicinity of the singularity point $q_s$, we see that as a result of the 
folding process the ratio of the distances of the two points in the Lagrangian and Eulerian manifold, must go to zero in the limit that we zoom in on points $q'$ and $q''$ along the Lagrangian curve $C$ at an infinitesimal distance from 
$q_s$, i.e.
\begin{equation}
\frac{\Delta x}{|\Delta q|}\,=\,\frac{\|x_t(q')-x_t(q'')\|}{\|q'-q''\|}\,\rightarrow 0\qquad q', q'' \rightarrow q_s\, .
\end{equation}
 The direct implication of this is, following equation~\eqref{eq:LagrangianDensity}, that the density in a caustic is 
infinite: the volume of the mass element associated to $q_s$ vanishes at time $t$. In essence it informs us that during shell crossing the 
points $q$ near Lagrangian location $q_s$, along the tangential direction $T$ to the Lagrangian curve $C$, map onto the same 
Eulerian position $x(q_s)$. This means that the norm of the directional derivative of $x_t$ along the tangential direction vanishes. 
In other words, along the non-zero tangent vector $T$ along $C$,  
\begin{equation}
\left \|\frac{\partial x_t}{\partial q} T \right \|=0\,,
\end{equation}
where $\partial x_t/\partial q$ is the Jacobian of $x_t$ evaluated in $q_s$ (see figure \ref{fig:SC1}). This is equivalent to requiring that 
\begin{equation}
\frac{\partial x_t}{\partial q} T \,=\,0\,.
\end{equation}
In terms of the displacement map $s_t$, this condition can be expressed as
\begin{equation}
\label{eq:con4}
T+\frac{\partial s_t}{\partial q} T = 0\,,
\end{equation}
with the Jacobian $\frac{\partial s_t}{\partial q}$ also evaluated in $q_s$\footnote{Unless mentioned otherwise, we will assume all Jacobians to 
be evaluated in $q_s$.}. Subsequently consider the eigenvalues $\mu_i$ and eigenvectors $v_i$ of the deformation tensor $\defm = \frac{\partial s_t}{\partial q}$, defined by
\begin{equation}
\defm v_i = \mu_i v_i\,.
\end{equation}
Under the assumption that the deformation tensor is diagonalizable\footnote{In practice, the assumption of diagonalizability is not really restrictive:
  non-diagonalizable matrices are unstable, which means that they can be turned diagonalizable by means of a small perturbation in the initial conditions.}, we can construct the diagonal
matrix $\mathcal{M}_d =$ diag$(\mu_1,\dots, \mu_d)$
and the eigenvector matrix $\mathcal{V}=(v_1,\dots,v_d)$. For an analysis of the case of non-diagonalizable deformation tensors see appendix \ref{Ap:Non-diagonalizable}. In three dimensions, with the 
eigenvalues $\mu_i$ and eigenvectors $v_i=(v_{i,1},v_{i,2},v_{i,3})$, the diagonal matrix $\defm_d$ and eigenvector matrix $\defv$ are given by 
\begin{equation}
\defm_d=\left(
 \begin{array}{ccc}
  \mu_{1} & 0 & 0 \\
  0 & \mu_2 & 0 \\
  0 & 0 & \mu_3 
 \end{array}\right),\quad \quad \quad \quad
\defv=\left(
 \begin{array}{ccc}
  v_{1,1} & v_{2,1} & v_{3,1} \\
  v_{1,2} & v_{2,2} & v_{3,2} \\
  v_{1,3} & v_{2,3} & v_{3,3} 
 \end{array}\right)\,.
\end{equation}

\noindent In terms of $\defv$ and $\defm_d$, condition \eqref{eq:con4} reduces to
\begin{equation}
0 = (I+\defm)\defv \defv^{-1} T = \defv(I+\defm_d) \defv^{-1}T
\end{equation}
since $\defv$ is always invertible\footnote{That is to say, the eigenvectors can always be chosen to be linearly independent.}, using the identity
\begin{equation}
\mathcal{M} \defv = \mathcal{M} (v_1,\dots, v_d) = (\mathcal{M}v_1,\dots, \mathcal{M} v_d) = (\mu_1 v_1,\dots \mu_d v_d) =  \defv \mathcal{M}_d\,.
\end{equation}
We thus obtain the condition
\begin{equation}
(I+\defm_d)\defv^{-1}T = 0\,, 
\label{eq:CC}
\end{equation}
which holds for general diagonalizable deformation tensors. 
Note that the rows of $\defv^{-1}$ consist of the dual vectors $\{v_i^*\}$ of the eigenvectors $\{v_i\}$, defined by $v_i\cdot v_j^* = \delta_{ij}$ 
for all $i$ and $j$. Explicitly, this means that $\defv^{-1}$ in three dimensions is given by 
\begin{equation}
\defv^{-1}=\left(
 \begin{array}{ccc}
  v^*_{1,1} & v^*_{1,2} & v^*_{1,3} \\
  v^*_{2,1} & v^*_{2,2} & v^*_{2,3} \\
  v^*_{3,1} & v^*_{3,2} & v^*_{3,3} 
 \end{array}
 \right)\,,
\end{equation}
with $v_i^*=(v_{i,1}^*,v_{i,2}^*,v_{i,3}^*)$. The product $\defv^{-1}T$ is the vector composed out of the inner product of these dual vectors with 
the tangent vector $T$, so that in three dimensions equation~\eqref{eq:CC} reduces to
\begin{equation}
\label{eq:Diagonal}
\left(
\begin{array}{ccc}
(1+\mu_1)v_1^*\cdot T\\
(1+\mu_2)v_2^*\cdot T\\
(1+\mu_3)v_3^*\cdot T
\end{array}
\right)=0\,.
\end{equation}

\medskip
This represents the proof for the shell-crossing condition for one-dimensional submanifolds. It states the condition for the 
tangential direction $T$ along which Lagrangian points get folded into an Eulerian singularity point. The obtained condition 
is a telling expression for the central role of both the deformation eigenvalues and eigenvectors in determining the occurrence of a singularity. 

\subsection{Shell-crossing condition: theorems}
\noindent Following the proof outlined in the previous subsection~\ref{sec:shellcrossingderivation}, we arrive at the following two 
theorems stipulating the conditions for the formation of singularities by curves $C$ and arbitrary manifolds 
$M$ in Lagrangian space $L$, 

\bigskip

\begin{framed}
\noindent
\begin{thr}

\medskip
\noindent A $C^1$ continuous curve $C \subset L$ forms a singularity under the mapping $x_t$ in the point $x_t(q_s) \in x_t(C) \subset E$, meaning that $x_t(C)$ is not smooth in $x_t(q_s)$, if and only if

\begin{equation} 
(1+\mu_{it}(q_s)) v_{it}^*(q_s) \cdot T =0
\end{equation}

\medskip
\noindent for all $i=1,2,\ldots,\text{dim}(L)$, with $T$ a nonzero tangent vector of $C$ in $q_s$.
\end{thr}
\end{framed}

\bigskip
Note that the derived caustic condition is independent of the dynamics of the fluid. In general, both the eigenvalue and eigenvector fields are complex-valued. For Hamiltonian fluids, the relation condition simplifies since the eigenvalue and eigenvector fields are forced to be real-valued and the eigenvectors can be chosen to coincide with their dual vectors, i.e. $v_i^*=v_i$.

\medskip
\indent A similar argument holds for higher dimensional submanifolds of $L$, e.g., sheets and volumes. These manifolds can be $n$-dimensional, with $n=1,\dots,3$ for three-dimensional fluids. Given an arbitrary manifold $M\subset L$ we can consider all curves $C \subset M$ passing through the point $q_s\in M$. The variety $x_t(M)$ contains a singularity at $x_t(q_s)$ if and only if at least one such curve $C \subset M$ gets folded under the map $x_t$. Hence for an arbitrary submanifold $M$, %not necessarily one-dimensional, 
we should consider the one-dimensional shell-crossing condition for the subset of vectors $T \in M$ \footnote{$T_{q_s}M$,  i.e. all
  tangent vectors $T$ constrained to be located in the vector space $T_{q_s}M$ \footnote{$T_{q_s}M$ is the vector space of all tangential
  vectors to the manifold $M$ in $q_s \in M$.}. In other words, the one-dimensional shell-crossing
  condition is considered for all vectors $T$ in the vector space of all tangential vectors to 
  the manifold $M$ in $q_s \in M$.}. This proves the general shell-crossing condition:

\bigskip

\begin{framed}
\noindent
\begin{thr}

\medskip
\noindent A manifold $M \subset L$ forms a singularity under the mapping $x_t$ in the point $x_t(q_s) \in x_t(M) \subset E$ at time $t$, meaning that $x_t(M)$ is not smooth in $x_t(q_s)$, if and only if there exists at least one nonzero tangent vector $T\in T_{q_s} M$ satisfying

\begin{equation} 
(1+\mu_{it}(q_s)) v_{it}^*(q_s) \cdot T =0
\label{eq:SC}
\end{equation}

\medskip
\noindent for all $i=1,2,\ldots,\text{dim}(L)$.
\end{thr}
\end{framed}

\noindent From this theorem, we immediately observe that the eigenvectors $v_i$ are of key importance in determining the nature of the
singularity, in that the shell-crossing condition is not simply that of $1+\mu_i=0$ for at least one $i$. More explicitly, the shell-crossing
condition says that

\begin{framed}
  \noindent
    
\begin{equation}
  1+\mu_{it}(q_s)\,=\,0\qquad\qquad  OR \qquad\qquad v_{it}^*(q_s) \cdot T\,=\,0\qquad\qquad \textrm{for all}\ i\,,
\end{equation}

\end{framed}

\noindent indicating that, in addition to one or more eigenvalue constraints $1+\mu_i=0$, the shell-crossing condition consists of 
complementary constraints. These single out those points $q_s$ where the eigenvectors $v_{j}^*(q_s)$ (with $j \neq i$) are orthogonal to a
vector $T$ that is restricted to be located in the plane tangent to the manifold $M$ in which the singularity emerges. It is this constraint
that is instrumental in defining the area occupied by the corresponding caustic.

Note that the shell-crossing conditions are manifestly independent of coordinate choices. While in general the eigenvalue and eigenvector fields
generally do depend on the choice of coordinates, it can be shown that they are invariant if the corresponding coordinate transformation
is orthogonal and global. These transformations include rotations and translations. See appendix~\ref{app:coordtransf} for more details. 
\bigskip

%%%%%%%%%%%%%%%%%%%%%%%%%%%%%%%%%%%%%%%%%%%%%%%%%%%%%%%%%%%%%%%%%%%%%%%%%%%%%%%%%%%%%%%%%%%%%%%%%%%%%%%%%%%%

\section{Caustic conditions}
\label{sec:CC}
\noindent In section~\ref{sec:ShellCrossing}, we inferred the general condition for shell-crossing. The condition establishes the relation 
between the eigenvalue and eigenvector fields of the deformation tensor in Lagrangian space, and the Lagrangian regions that get incorporated in features of infinite density in Eulerian space. Moreover, it allows us to establish the identity of the resulting singularity in Eulerian space.

The stable singularities that emerge can be classified by Lagrangian catastrophe theory in the $A_k$, $D_k$ and $E_k$ series (see \cite{arnold1984},
\cite{gilmore1981} and \cite{poston1978}). This is described in some detail in section~\ref{sec:class} \footnote{The classification ultimately has its
  origin in the classification of Coxeter groups}. The $A_k$ series is of co-rank $1$, in which co-rank is the number of independent directions in which the
Hessian is degenerate. The $A_k$ series 
    corresponds to the caustics for which the density diverges due to only one eigenvalue. The $D_k$ series is of co-rank $2$ and corresponds to the points for which the density diverges due to two eigenvalue fields. The $E_k$ series is of co-rank $3$ and corresponds to the points for which three eigenvalue fields. However, for three-dimensional fluids, the points for which all eigenvalues simultaneously satisfy this condition are degenerate. For this reason
we will not discuss them in the context of the present paper. 

In this section we apply the shell-crossing condition to three-dimensional Lagrangian fluids to obtain the \textit{caustics conditions} which relate the classification of caustics to the eigenvalue and eigenvector field. These conditions have not been derived in earlier work and are necessary to perform a quantitative study of caustics in large scale
structure formation. In section \ref{sec:class}, we summarize the classification of caustics in its traditional form and compare them to the caustic conditions derived here. 

\vskip 0.5truecm
\subsection{The $A$ family}
\label{sec:A-family}
\noindent The $A$ family of caustics form when
\begin{eqnarray}
1+\mu_i&\,=&\,0
\end{eqnarray}
for one $i$.
For diagonalizable deformation tensors, the eigenvector fields $\{v_i\}$ and their dual vector fields 
$\{v_i^*\}$ are linearly independent. 

For three-dimensional fluids, the $A$ family consists of $5$ classes running from the trivial $A_1$ class, corresponding to the points that 
never form caustics, the sheetlike $A_2$ fold, the curvelike $A_3$ cusp, the $A_4$ swallowtail, to the pointlike $A_5$ butterfly singularity. 

\vskip 0.25truecm
\subsubsection{The trivial $A_1$ class}
\noindent The $A_1$ class labels the points which never form caustics. According to the shell-crossing condition, $q_s$ will form a singularity at time $t$ if and only if there exists a nonzero tangent vector $T \in T_{q_s} L$ for which 
\begin{equation} 
(1+\mu_i(q_s)) v_i^*(q_s) \cdot T =0
\end{equation}
for all $i$. The point $q_s$ will not satisfy this condition if $1+\mu_i(q_s) \neq 0$ for all $i$ since the three dual vectors $\{v^*_i\}$ of the (generalized) eigenvectors span the tangent space $T_{q_s}L$ 

\bigskip
\noindent From the shell-crossing condition we therefore conclude that the three-dimensional variety $A_1$, 
\begin{equation}
A_1 = \{ q\in L| 1+\mu_{ti}(q)\neq 0 \mbox{ for all } i  \mbox{ and } t\}\,,
\end{equation}
consists of the points never forming caustics. In this respect we should note that the displacement map at the initial time is the zero map, so that 
the eigenvalues at the initial time are equal to zero, i.e. $\mu_{0i}(q)=0$ for all $q\in L$. Since the eigenvalues are a continuous function
  of time, for the cosmologically interesting case of potential flow the requirement for a point $q$ to belong to $A_1$ is equivalent to
  $\mu_{ti}(q) > -1$.

\vskip 0.25truecm
\subsubsection{The $A_2$ caustics}
\noindent Based on the discussion above, we may conclude that for a given $i$, $i=1 \ldots 3$, at time $t$ the points 
\begin{equation}
A_2^i(t)=\{ q\in L| 1+\mu_{ti}(q)=0 \}
\end{equation}
form a singularity. For three-dimensional fluids, the set $A_2(t)$ forms a two-dimensional sheet, sweeping through space as the fluid evolves. 
These singularities can be associated to the $A_2$ fold singularity class. 

\bigskip
\noindent From this, we conclude that the set of points which form a $A_2$ fold singularity at a time $t \in [0,\infty)$ is given by
\begin{equation}
A^i_2 =\{ q\in L| 1+\mu_{ti}(q)=0 \mbox{ for some } t\}\,.
\end{equation}

\vskip 0.25truecm
\subsubsection{The $A_3$ caustics} 
\label{sec:A3}
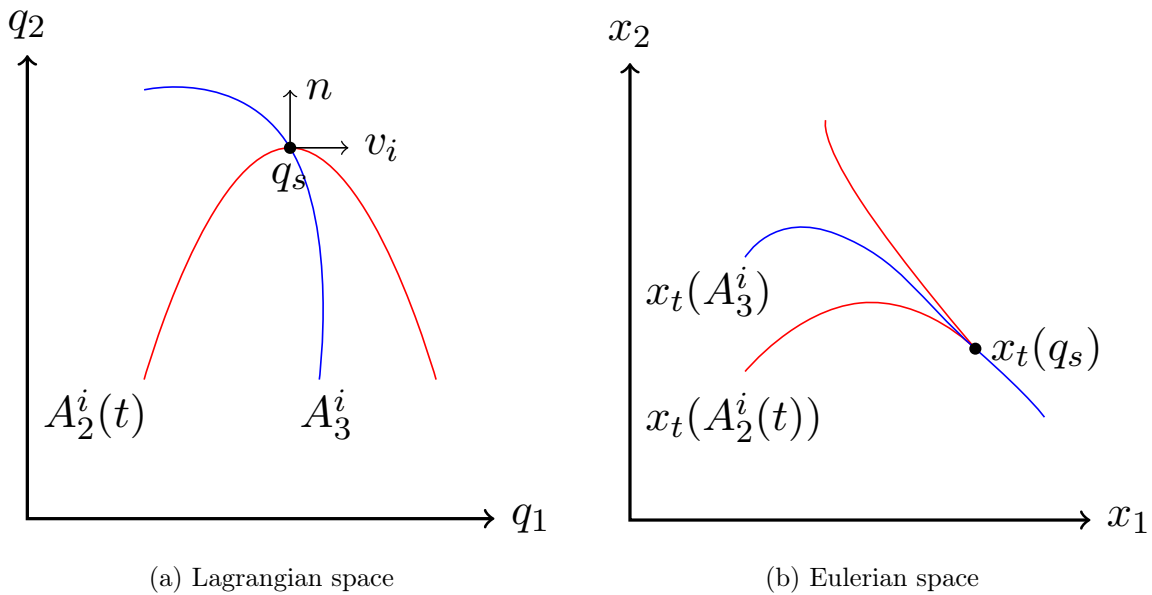
\begin{figure}
\centering
\begin{subfigure}[b]{0.5\textwidth}
\resizebox {\columnwidth}{!}{
\begin{tikzpicture}
\draw [<->,thick] (0,4) node (yaxis) [above] {$q_2$}|- (4,0) node (xaxis) [right] {$q_1$};
\draw [red, xshift=0.cm, yshift=0.2cm] plot [smooth, tension=1] coordinates { (1,1) (2.25,3) (3.5,1)};
\draw [blue, xshift=0.cm, yshift=0.2cm] plot [smooth, tension=1] coordinates { (1,3.5) (2.25,3) (2.5,1)};
\fill[black] (2.25,3.2) coordinate (c_1) circle (1.5pt);
\draw[] (c_1) node[below] {$q_s$};
\draw[] (0.0,0.9) node[right] {$A_2^i(t)$};
\draw[] (2.2,0.9) node[right] {$A_3^i$};
\draw[black,->] (c_1) -- (2.75,3.2) node[right] {$v_i$}; 
\draw[black,->] (c_1) -- (2.25,3.7) node[right] {$n$}; 
\end{tikzpicture}}
\caption{Lagrangian space}
\end{subfigure}~
\begin{subfigure}[b]{0.5\textwidth}
\resizebox {\columnwidth}{!}{
\begin{tikzpicture}
\draw [<->,thick] (0,4) node (yaxis) [above] {$x_2$}|- (4,0) node (xaxis) [right] {$x_1$};
\draw [red, xshift=-0.5cm, yshift=0.0cm] plot [smooth, tension=1] coordinates { (1.5,1.3) (2.5,1.9) (3.5,1.5)};
\draw [red, xshift=-0.5cm, yshift=0.0cm] plot [smooth, tension=1] coordinates { (2.2,3.5) (2.5,2.8) (3.5,1.5)};
\draw [blue, xshift=-0.5cm, yshift=0.0cm] plot [smooth, tension=1] coordinates {(1.5,2.3) (2.3,2.5) (3.5,1.5) (4.1,0.9)};
\fill[black] (3.,1.5) coordinate (c_1) circle (1.5pt);
\draw[] (c_1) node[right] {$x_t(q_s)$};
\draw[] (0.,0.9) node[right] {$x_t(A_2^i(t))$};
\draw[] (0.,2) node[right] {$x_t(A_3^i)$};
\end{tikzpicture}}
\caption{Eulerian space}
\end{subfigure}
\caption{The formation of a cusp ($A_3$) singularity in a Lagrangian map $x_t$. The left panel shows Lagrangian space, describing the initial positions of the fluid. The right panel shows Eulerian space, describing the positions of the fluid at time $t$. The fluid undergoes shell-crossing along the fold $A_2^i(t)$ (red) at time $t$. The fold gets mapped under the Lagrangian map to $x_t(A_2)$ (red), which is folded into a cusp in the point $x_t(q_s)$ corresponding to $q_s$. The cusp forms if and only if the normal $n$ of $A_2^i(t)$ is orthogonal to the eigenvector field $v_i$ in $q_s$. Over time, the cusp traces out the curve $A_i$ (blue) which is mapped to $x_t(A_3^i)$ (blue).}
\label{fig:A3}
\end{figure}

\noindent 
Following up on the folding of the fluid to the $A^i_2$ singularity, the $A^i_2$ manifold itself may be 
folded into a more complex configuration. The result is a so-called $A_3$ singularity. To guide understanding in 
the emergence of cusps we may refer to the eigenvalue contour map of figure~\ref{fig:A3}. 

To infer the identity of the $A^i_3$ caustic, we restrict the criterion for shell-crossing to points on the $A^i_2$ manifold. 
In other words, we look for points $q_s$ on the surface of the sheetlike variety $A_2^i(t)$ that fulfill the criterion for 
shell-crossing. 

A point $q_s \in A_2^i(t)$ forms a singularity if there exists a nonzero tangent vector T, $T\in T_{q_s}A_2^i(t)$,  
orthogonal to the $\mbox{Span}_{\mathbb{C}} \{v^*_j|j\neq i\}$. Writing the tangent vector $T$ as a linear
combination of the eigenvectors $v_i$,
\begin{equation}
  T\,=\,\alpha_1 v_1+\alpha_2 v_2+\alpha_3 v_3\,, 
  \end{equation}
with $\alpha_i \in \mathbb{C}$. The caustic conditions tell us that
\begin{equation}
\alpha_j\,=\,v_j^*(q_s) \cdot T = 0 \quad \textrm{for } j \neq i \,.
  \end{equation}
Given that we know that the $i$th eigenvalue is real, $\mu_i \in \mathbb{R}$, the eigenvector $v_i$ is
also real. This means that this condition is satisfied if and only if the tangent vector $T$ is 
parallel to $v_i$. This is equivalent to the condition that $v_i$ is orthogonal to the normal $n=\nabla \mu_{ti}$ of the 
manifold $A_2^i(t)$ in the point $q_s$. Explicitly, this means that the inner product of $n$ with $v_i$ is equal to 0,
\begin{equation}
\mu_{ti,i} \equiv v_i \cdot \nabla \mu_{ti} = 0\,.
\end{equation}
Note that this is the condition that Arnol'd \citep{Arnold:1982b} found for the $A_3$ line for the 2-dimensional 
Zel'dovich approximation. As we see from the derivation above, the condition is valid in any dimensional space and for
general flow configurations. 

\begin{figure}[t]
\vskip -0.5truecm
\centering
\includegraphics[width=0.90\textwidth]{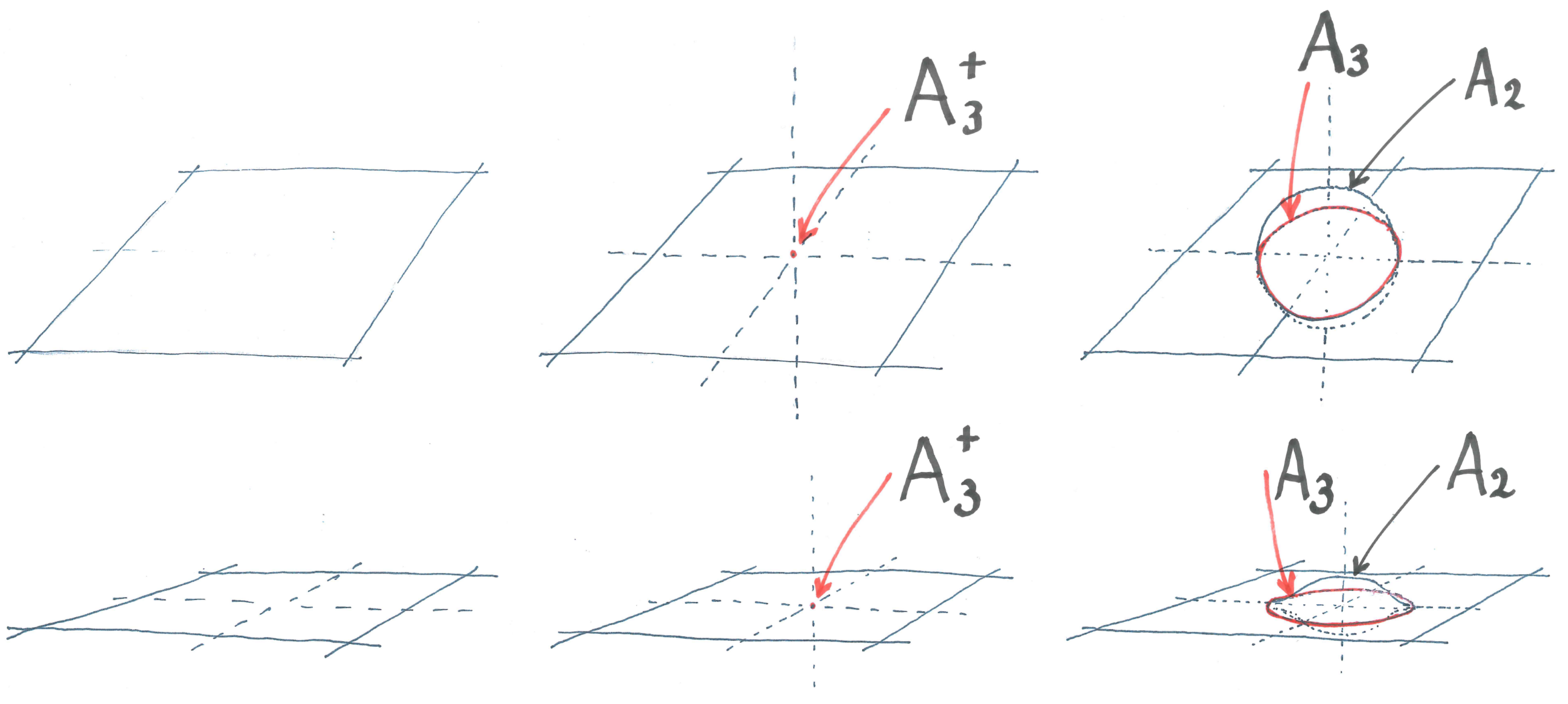}
\caption{The creation/annihilation of a fold ($A_2$) sheet in a $A_3^+$ point. The upper three panels show the unfolding of a $A_3^+$ singularity in Lagrangian space. The lower three panels show the corresponding unfolding in Eulerian space. The two panels on the left show the cusp ($A_3$) plane on which the cusps form. The middle panels show the appearance of a $A_3^+$ singularity in which a fold sheet is formed/removed. The right panels show the resulting fold ($A_2$) sheet. The fold sheet gets folded into a cusp ($A_3$) curve (red). This configuration is known as the {\it Zel'dovich pancake} (Zel'dovich 1970).}
\label{fig:A3p}
\centering
\includegraphics[width=0.90\textwidth]{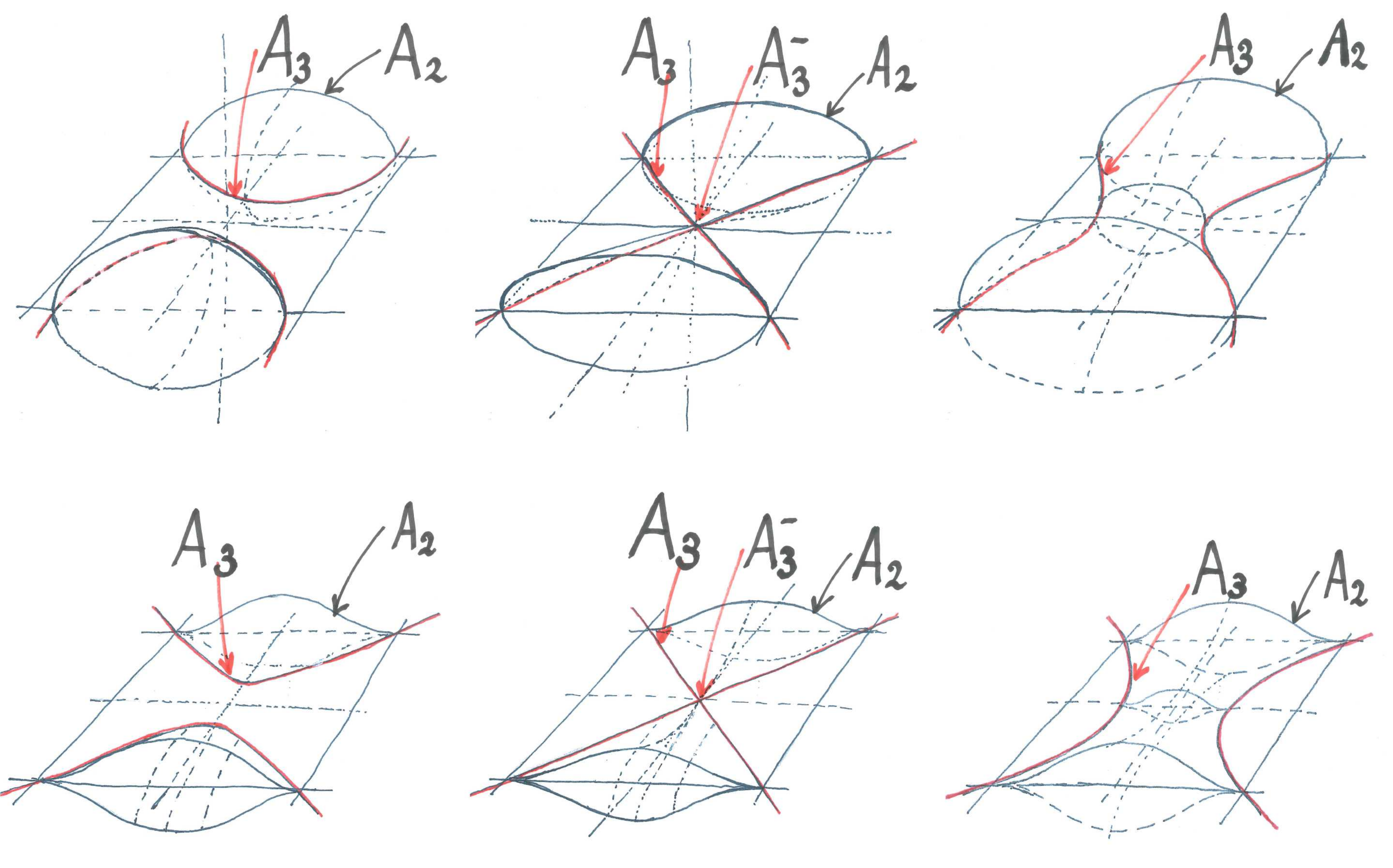}
\caption{The merger/splitting of a fold ($A_2$) sheet in a $A_3^-$ point. The upper three panels show the unfolding of a $A_3^-$ singularity in Lagrangian space. The lower three panels show the corresponding unfolding in Eulerian space. The two panels on the left show two fold ($A_2$) sheets, two cusp ($A_3$) curves (red) and the cusp ($A_3$) plane on which the cusps form. The middle panels show the merger/splitting of the two fold ($A_2$) sheets in a $A_3^-$ singularity. The right panels show the resulting merged fold ($A_2$) sheet. This configuration is known as the {\it Kissing Lips}.}
\label{fig:A3m}
\end{figure}

\begin{figure}

\centering
\includegraphics[width=\textwidth]{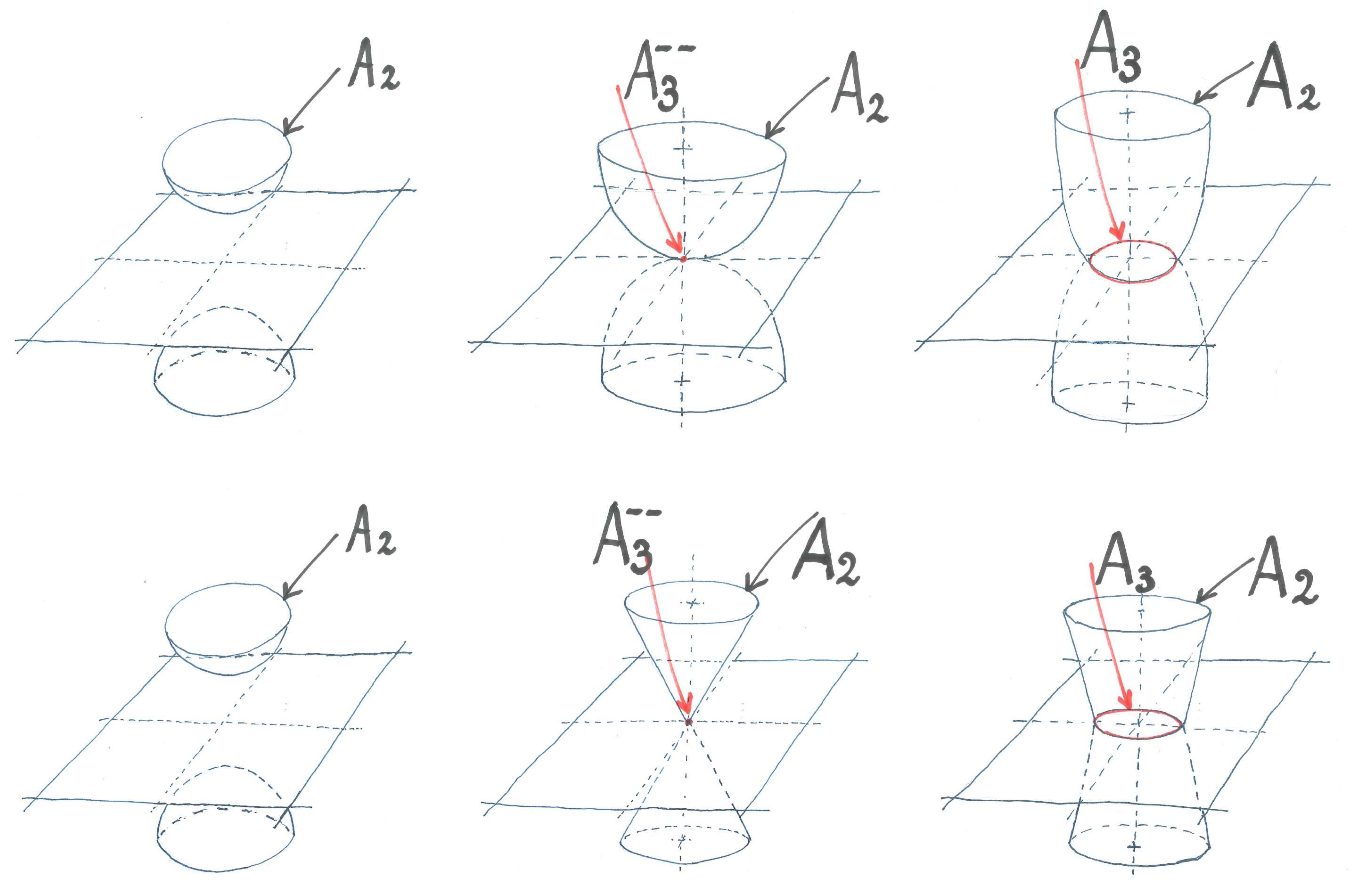}
\caption{The merger/splitting of a fold ($A_2$) sheet in a $A_3^{--}$ point. The upper three panels show the unfolding of a $A_3^{--}$ singularity in Lagrangian space. The lower three panels show the corresponding unfolding in Eulerian space. The two panels on the left show two fold ($A_2$) sheets, and the cusp ($A_3$) plane on which the cusps form. The middle panels show the merger/splitting of the two fold ($A_2$) sheets in a $A_3^{--}$ singularity. The right panels show the resulting merged fold ($A_2$) sheet with the corresponding cusp ($A_3$) curve.}
\label{fig:A3mm}
\end{figure}

\bigskip 
\noindent The points $q$ forming a cusp at time $t$ corresponding to eigenvalue field $\mu_i$ is given by the one-dimensional variety 
\begin{equation}
A_3^i(t)=\{q\in L|q \in A_2^i(t)\ \wedge \ \mu_{ti,i}(q)=0  \}\, .
\end{equation}

\begin{figure}
\centering
\begin{subfigure}[b]{0.5\textwidth}
\resizebox {\columnwidth}{!}{
\begin{tikzpicture}
\draw [<->,thick] (0,4) node (yaxis) [above] {$q_2$}|- (4,0) node (xaxis) [right] {$q_1$};
\draw [red, xshift=0.cm, yshift=0.2cm] plot [smooth, tension=1] coordinates { (1,1) (2.25,3) (3.5,1)};
\draw [blue, xshift=0.cm, yshift=0.2cm] plot [smooth, tension=1] coordinates { (0.5,3.5) (2.25,3) (3.,3.5)  };
\draw [green, xshift=0.cm, yshift=0.2cm] plot [smooth, tension=1] coordinates { (1.3,3.5) (2.25,3) (2.5,1)};
\fill[black] (2.25,3.2) coordinate (c_1) circle (1.5pt);
\draw[] (c_1) node[below] {$q_s$};
\draw[] (0.0,0.9) node[right] {$A_2^i(t)$};
\draw[] (0.0,3.4) node[right] {$A_3^i$};
\draw[] (2.2,0.9) node[right] {$A_4^i$};
\draw[black,->] (c_1) -- (2.75,3.2) node[right] {$v_i$}; 
\draw[black,->] (c_1) -- (2.25,3.7) node[right] {$n$}; 
\end{tikzpicture}
}
\caption{Lagrangian space}
\end{subfigure}~
\begin{subfigure}[b]{0.5\textwidth}
\resizebox {\columnwidth}{!}{
\begin{tikzpicture}
\draw [<->,thick] (0,4) node (yaxis) [above] {$x_2$}|- (4,0) node (xaxis) [right] {$x_1$};
\draw [red, xshift=-0.5cm, yshift=0.0cm] plot [smooth, tension=1] coordinates { (1.5,1.3) (2.5,1.9) (3.5,1.5)};
\draw [red, xshift=-0.5cm, yshift=0.0cm] plot [smooth, tension=1] coordinates { (2.2,3.5) (2.5,2.8) (3.5,1.5)};
\draw [green, xshift=-0.5cm, yshift=0.0cm] plot [smooth, tension=1] coordinates {(1.5,2.3) (2.3,2.5) (3.5,1.5) (4.1,0.9)};
\draw [blue, xshift=-0.5cm, yshift=0.0cm] plot [smooth, tension=1] coordinates {(2.5,0.75) (2.75,1.25) (3.5,1.5) };
\draw [blue, xshift=-0.5cm, yshift=0.0cm] plot [smooth, tension=1] coordinates {(2.75,3.5) (3,2.5) (3.5,1.5) };
\fill[black] (3.,1.5) coordinate (c_1) circle (1.5pt);
\draw[] (c_1) node[right] {$x_t(q_s)$};
\draw[] (0.,0.9) node[right] {$x_t(A_2^i(t))$};
\draw[] (2.,0.9) node[right] {$x_t(A_3^i)$};
\draw[] (0.,2) node[right] {$x_t(A_4^i)$};
\end{tikzpicture}}
\caption{Eulerian space}
\end{subfigure}
\caption{The formation of a swallowtail ($A_4$) singularity in a Lagrangian map $x_t$. The left panel shows the Lagrangian space describing the initial positions of the fluid. The right panel shows the Eulerian space describing the positions of the fluid at time $t$. The fluid undergoes shell-crossing along $A_2^i(t)$ (red) at time $t$. The fold gets mapped in Eulerian space, under the Lagrangian map, to $x_t(A_2)$ (red), which is folded into a cusp in the point $x_t(q_s)$ corresponding to $q_s$. The cusp forms if and only if the normal $n$ of $A_2^i(t)$ is orthogonal to the eigenvector field $v_i$ in $q_s$. Over time, in Lagrangian space the cusp traces out the curve $A_i$ (blue) which in Eulerian space is mapped to $x_t(A_3^i)$ (blue). Since the cusp ($A_3^i$) curve is tangential to the fold ($A_2$) curve in $q_s$, the cusp curve $x_t(A_3^i)$ forms a swallowtail ($A_4$) singularity. Over time, the swallowtail traces out $A_4^i$ (green), which in Eulerian space 
is mapped into $x_t(A_4^i)$ (green).}
\label{fig:A4}
\end{figure}
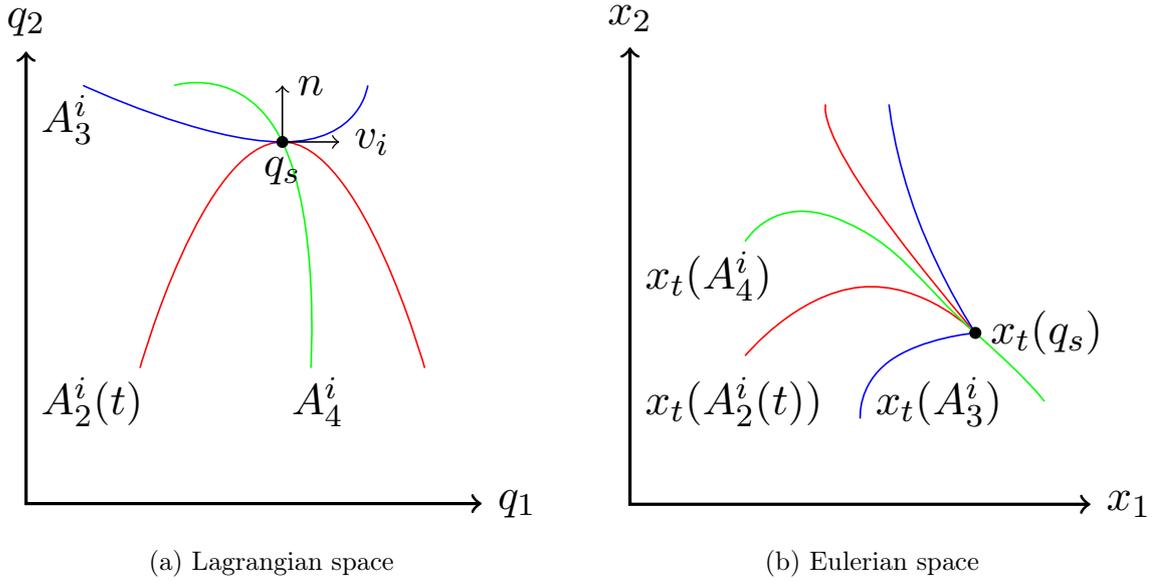

\begin{figure}
\vskip -1.5truecm
\centering
\includegraphics[width=0.90\textwidth]{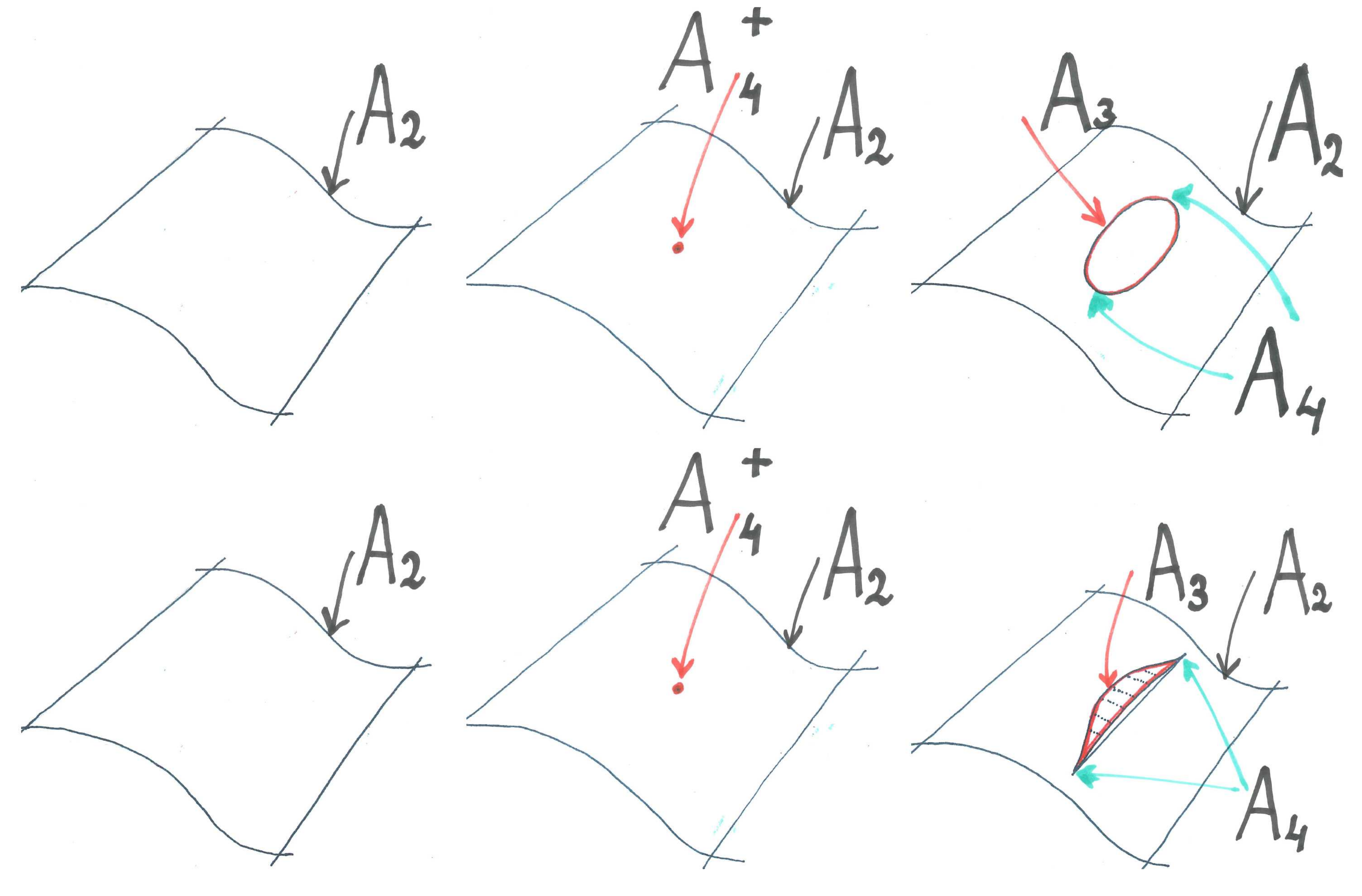}
\caption{The creation/annihilation of a swallowtail ($A_4$) singularity in a $A_4^{+}$ point. The upper three panels show the unfolding of a $A_4^+$ singularity in Lagrangian space. The lower three panels show the corresponding unfolding in Eulerian space. The two panels on the left show a fold ($A_2$) sheet. The middle panels show a $A_4^+$ point on the fold ($A_2$) sheet. The $A_4^+$ point leads to the creation/annihilation of two swallowtail ($A_4$) singularities. The right panels show the resulting cusp ($A_3$) curves and swallowtail ($A_4$) singularities.}
\label{fig:A4p}
\centering
\includegraphics[width=0.92\textwidth]{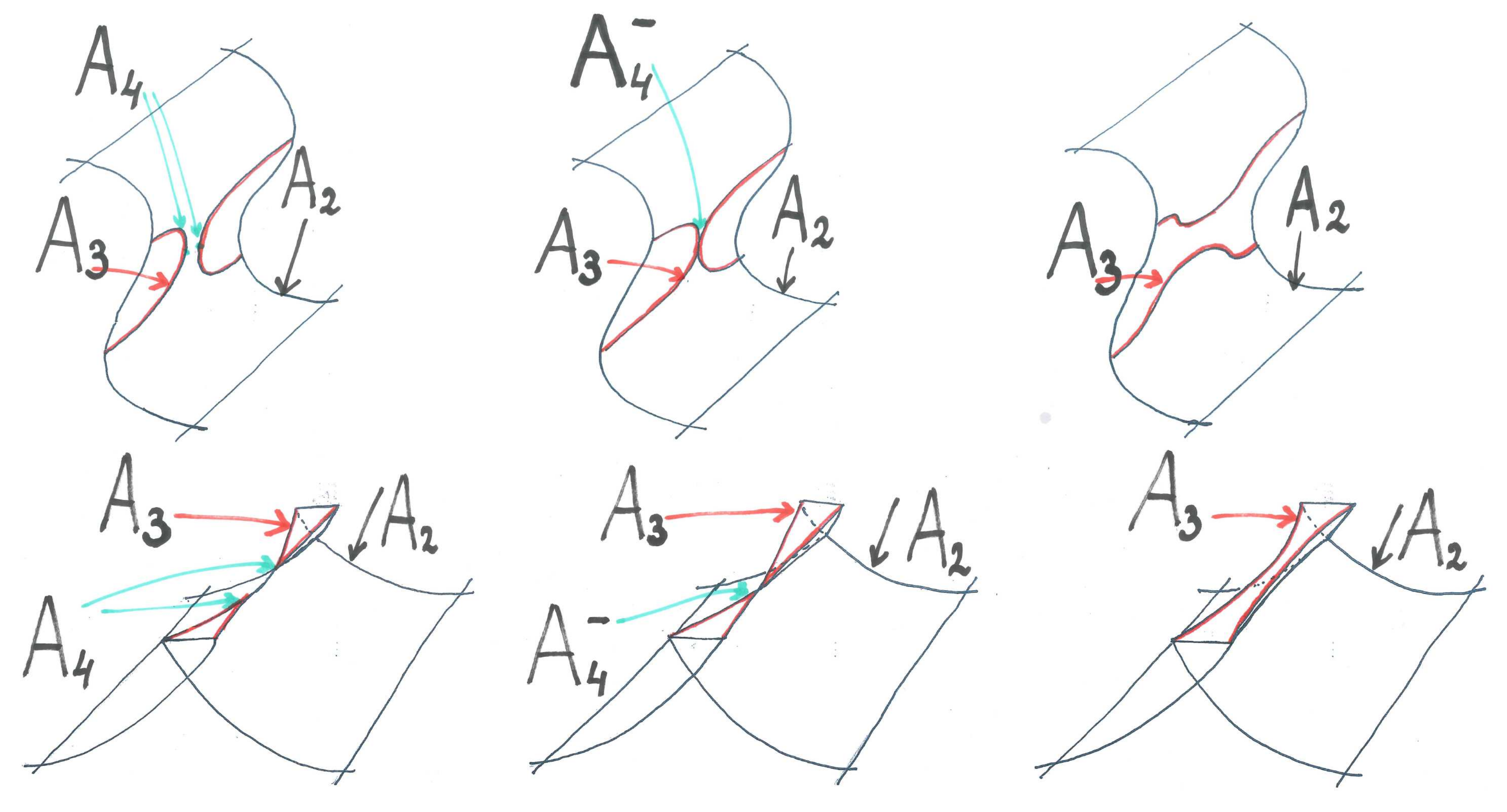}
\caption{The merger/splitting of a cusp ($A_3$) curve in a $A_4^-$ point. The upper three panels show the unfolding of a $A_4^-$ singularity in Lagrangian space. The lower three panels show the corresponding unfolding in Eulerian space. The two panels on the left show a fold ($A_2$) sheet, cusp ($A_3$) curves and swallowtail ($A_4$) singularities. The middle panels show the merger/splitting of the cusp ($A_3$) curves in a $A_4^-$ point. The right panels show the resulting fold ($A_2$) sheet and cusp ($A_3$) curves singularities.}
\label{fig:A4m}
\end{figure}

\medskip
\noindent Extrapolating this to the set of all points $q$ that at some time $t \in [0,\infty)$ have belonged to 
or will be incorporated in a cusp singularity defines a two-dimensional variety 
\begin{equation}
A_3^i=\{q\in L|q \in A_2^i(t)\ \wedge\ \mu_{ti,i}(q)=0 \mbox{ for some } t \} \,,
\end{equation}
which is the assembly of all $A_3^i(t)$ over the time interval $t \in [0,\infty)$.

\vskip 0.25truecm
\subsubsection{The $A_3^\pm$ points}
\noindent The topology of the sheetlike $A_2^i(t)$ variety changes as a function of time. These topological changes occur at critical points of 
the corresponding eigenvalue field $\mu_{ti}$. It is at these points where in Eulerian space we see the emergence of new features, the disappearance of 
features and/or the merging of features. The critical points are classified as cusp singularities. 

At minima of the $\mu_i$ field, a feature gets created. At maxima, a feature gets annihilated. Particularly interesting points are the saddle points. 
In three-dimensional space, there are two classes of saddles in the eigenvalue field $\mu_{ti}$. The index~1 saddles have a Hessian signature $(--+)$, 
with 1 positive eigenvalue, while the index 2 saddles have a signature $(-++)$. 

Based on their impact on caustic structure, Arnol'd used a slightly different classificiation scheme, in which the distinguished between 
$A_3^{++}$, $A_3^{+-}$ and $A_3^{--}$ points \citep{Arnold:1982b}. The $A_3^{++}$ point are identified with the minima\footnote{Note that in Arnol'd's 
notation, related to the Zel'dovich formalism (see appendix~A), these are the maxima of the eigenvalue field}, while 
the $A_3^{+-}$ points are the saddle points for which the $A_3$ sheet intersects the two disjoint $A_2$ sheets. This is illustrated in the upper left 
panel in figure \ref{fig:A3m}. The additional $A_3^{--}$ points correspond to saddle points for which the $A_3$ sheet does not intersect the disjoint 
$A_2$ sheets. Because this concerns a non-generic situation, we do not treat it here. Also note that higher dimensional fluids will have 
additional $A_3$ points.

In the context of this paper we therefore use a slightly shorter notation for the maxima, minima and saddles, classifying them as the cusp singularities 
$A_3^+$ an $A_3^-$, 
\begin{eqnarray}
A_3^{i+}&=&\{ q \in L| q \in A_2^i(t) \wedge \mu_{ti}(q) \mbox{ max-/minimum of } \mu_{ti} \mbox{ at some time } t\}\,,\nonumber\\
A_3^{i-}&=&\{ q \in L| q \in A_2^i(t) \wedge q \mbox{ saddle point of } \mu_{ti} \mbox{ at some time } t\}.
\end{eqnarray}
Note that in this scheme, the saddle points with index $1$ and $2$ belong to the same singularity class $A_3^{i-}$. For an illustration of the $A_3^+, A_3^-$ and $A_3^{--}$ singularities, we refer to figures \ref{fig:A3p}, \ref{fig:A3m} and \ref{fig:A3mm}. From the caustics conditions we may directly infer 
that the $A_3^{i\pm}$ points are located on the $A_3^{i}$ variety. 

\vskip 0.25truecm
\subsubsection{The $A_4$ caustics}
\noindent In Eulerian space the $A_3^i(t)$ variety gets folded in points associated with $A_4$ swallowtail singularities. The identity of the points defining the variety $A_4^i(t)$ can be inferred by the application of the general shell-crossing 
condition (eqn. \eqref{eq:SC}) to the $A_3^i(t)$ variety (see figure \ref{fig:A4}). As a consequence, the $A_4^i$ variety is defined as 
\begin{equation}
A_4^i(t)=\{q\in L|q \in A_3^i(t) \wedge \mu_{ti,ii}(q)=0  \},
\label{eq:a4var}
\end{equation}
with $\mu_{ti,ii}(q)$ the inner product of the normal $n=\nabla \mu_{ti,i}$ with the eigenvector $v_i$, 
\begin{equation}
\mu_{ti,ii} \equiv v_i \cdot \nabla \mu_{ti,i}\,.
\end{equation}

\noindent Integrated over time, the points on the varieties $A_4^i(t)$ trace out the 1-dimensional variety $A_4^i$, 
i.e. the 1-dimensional line $A_4^i$ is the set of all points $A_4^i(t)$ over the time interval $t \in [0,\infty)$, 
\begin{equation}
A_4^i=\{q\in L|q \in A_3^i(t) \wedge \mu_{ti,ii}(q)=0 \mbox{ for some } t \}\,.
\end{equation}

%%%%%%%%%%%%%%%%%%%%%%%%%%%%%%%%%%%%%%%%%%%%%%%
\vskip 0.25truecm
\subsubsection{The $A_4^\pm$ points}
\noindent The topology of the variety $A_3^i(t)$ changes as a function of time. To this end, we identify the 
critical points of the real field $\mu_{i,i}$, 
\begin{equation}
\mu_{ti,i} \equiv v_i \cdot \nabla \mu_{ti}\,.
\end{equation}
Constraining the location of these singularities to the one-dimensional curvelike variety $A_3^i(t)$, and thus implicitly also to the 
two-dimensional membrane of the variety $A_2^i(t)$, these $A_4^\pm$ points mark the locations at which topological changes occur. 
They represent the sites at which we see the birth of new singularities in Eulerian space, or the annihilation of and/or merging of 
such features. These singularities are classified as swallowtail singularities. 

The birth or death of features on $A_3^i(t)$ takes place at maxima and minima of $\mu_{ti,i}$, and is identified with 
$A_4^{i+}$ singularities. The merging or splitting of features happens at the saddle points of the same 
field $\mu_{ti,i}$. The latter mark the $A_4^{i-}$ singularities,  
\begin{eqnarray}
A_4^{i+}&=&\{ q \in L| q \in A_3^i(t), \mu_{ti,i}(q) \ \mbox{max-/minimum of } \mu_{ti,i}|_{A_2^i(t)} \mbox{ for some } t\},\nonumber\\
A_4^{i-}&=&\{ q \in L| q \in A_3^i(t)\ \mbox{saddle point of } \mu_{ti,i}|_{A_2^i(t)} \mbox{ for some } t\}.
\end{eqnarray}

The $A_4^\pm$ critical points are constrained to lie on the curvelike variety $A_2^i(t)$. Their identity is 
therefore determined by the interplay between the geometric properties of two entities. One of these is the 
geometry of the field $\mu_{ti,i}$, the other that of the geometry of the curvelike variety $A_3^i(t)$. 
For illustrations of the $A_4^+$ and $A_4^-$ singularities we refer to figure~\ref{fig:A4p} and \ref{fig:A4m}. 

From the caustic conditions -- as expressed in eqn.~\eqref{eq:a4var} -- we may also immediately observe that the 
$A_4^{i\pm}$ points belong to the $A_4^i$ variety. In fact, this also represents a condition on the topology of 
the field $\mu_{ti,i}$ and that of the $A_2^i(t)$ variety.

%%%%%%%%%%%%%%%%%%%%%%%%%%%%%%%%%%%%%%%%%%%%%%%%5

\begin{figure}
\vskip 0.25truecm
\centering
\includegraphics[width=0.99\textwidth]{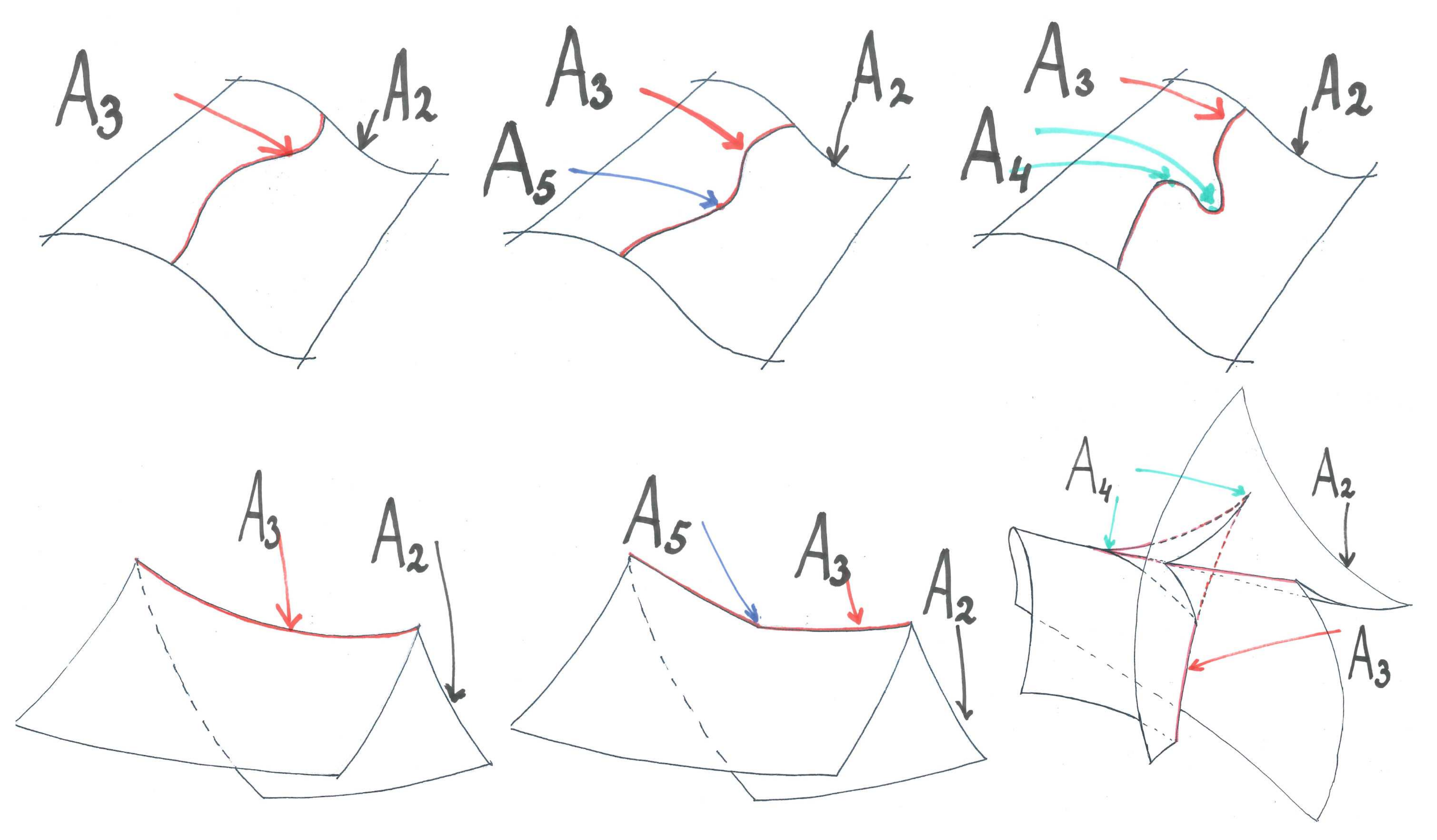}
\caption{The creation/annihilation of swallowtail singularities in a butterfly ($A_5$) singularity. The upper three panels show the unfolding of a $A_5$ singularity in Lagrangian space. The lower three panels show the corresponding unfolding in Eulerian space. The two panels on the left show a fold ($A_2$) sheet, and cusp ($A_3$) curve. The middle panels show the creation/annihilation of the butterfly ($A_5$) singularity on the cusp ($A_3$) curve. The right panels show the resulting fold ($A_2$) sheet, cusp ($A_3$) curve and swallowtail ($A_4$) singularities.}
\label{fig:A5}
\end{figure}

\vskip 0.25truecm
\subsubsection{The $A_5$ caustics}
\noindent Finally, also the swallowtail curves $A_4^i$ curve get folded in Eulerian space. It leads to the emergence of so-called butterfly 
singularities, or $A_5$ singularities. Following the same reasoning as for the $A_3^i$ and $A_4^i$ varieties, we may infer from 
the general shell-crossing condition that the $A_4^i$ curve gets folded in the points $A_5^{i}$. In general this happens when there exists a tangent vector of $A_4$ parallel to $v_i$, i.e.
\begin{equation}
A_5^{i} =\{ q \in L| q \in A_4^{i} \text{ and } v_i \in T_q A_4^{i}  \}
\end{equation}
In the case three-dimensional case, when the displacement field $s_t(q)$
is separable into temporal and spatial parts, time evolution can be seen as a progression through a series of surfaces. The folding points
can then be found from the relation, 
\begin{equation}
A_5^{i}=\{ q \in L| q \in A_4^i(t) \mbox{ and } \mu_{ti,iii}=0 \mbox{ for some time } t\}.
\end{equation}
Figure \ref{fig:A5} shows an illustration of a $A_5$ singularity.

The butterfly singularity is the highest dimensional singularity that may surface in three-dimensional Lagrangian fluids. 
It is important to realize that the butterfly singularity only exists at one point in space-time. 

\vskip 0.5truecm
\subsection{The $D$ family  }
\noindent The $D$ family of caustics correspond to manifolds for which the condition
\begin{equation}
1+\mu_i = 0\,,
\label{eq:A2D}
\end{equation}
holds for two eigenvalue fields simultaneously. From this, we may immediately infer that these caustics form at the intersection of two $A_2(t)$ fold sheets, the 
$A_2^i(t)$ and $A_2^j(t)$ varieties. In all, for three-dimensional fluids three classes of $D$ caustics can be identified, the $D_4^-$ elliptic, the $D_4^+$ hyperbolic
and the $D_5$ parabolic umbilic caustic. 

\vskip 0.25truecm
\subsubsection{The $D_4$ caustics}\label{sec:D4}
\noindent The $D_4$ caustics are defined by the points $q$ in Lagrangian space, at which two of the 
eigenvalues have the same value. For instance, the $D_4^{ij}(t)$ caustic, with $i\neq j$, is outlined by the points 
$q$ for which 
at the time $t$ the eigenvalues $\mu_i$(t) and $\mu_j(t)$ are equal, $\mu_{ti}=\mu_{tj}$. While the eigenvalue 
$\mu_{ti}$ defines the fold sheet $A_2^i$, and the eigenvalue $\mu_{ti}$ the fold sheet $A_2^j$, the umbilic 
$D_4^{ij}$ caustic consist of the set of points $q$ for which
\begin{equation}
D_4^{ ij}(t)=\{q\in L| q \in A_2^i(t) \cap A_2^j(t) \}\,.
\end{equation}
\noindent In three-dimensional space, one would expect that the intersection of the two sheets $A_2^i(t)$ and 
$A_2^j(t)$ to consist of one-dimensional curves. This would certainly be true for two sheets that would be 
entirely independent of each other. However, the situation at hand concerns a highly constrained situation, in 
which the two eigenvalues $\mu_i$ and $\mu_j$ are strongly correlated.

Because of the latter, the intersection between the folds $A_2^i$ and $A_2^j$ is considerably more complex. Instead of a 
continuous curve, the intersection consists of isolated, singular points. A telling illustration -- and discussion -- of this, 
for the two-dimensional situation, can be found in \cite{Hidding:2014}. 

\bigskip 
\noindent {\it $D_4$ singularities and $A_3$ varieties}\\ 
\noindent To investigate the geometry and structure of the set $D_4^{ ij}(t)$ we focus on the particular situation of the 
set $D_4^{12}(t)$, in which the two first eigenvalues $\mu_1$ and $\mu_2$ have the same value, $\mu_{t1}=\mu_{t2}=-1$. Without loss 
of generality, we transform the coordinate system such that the third eigenvector $v_3$ defines the $q_3$ axis. This 
transformation makes the $q_1q_2$-plane the one in which we see the folding and collapse of the phase space sheets to 
the $A_2^1$ and $A_2^2$ caustics.

Assuming that the deformation tensor $\defm$ is diagonalizable, in this coordinate system
it has the form, 
\begin{equation}
\defm\,=\, 
\left(
\begin{array}{ccc}
\defmij_{11} & \defmij_{12} & 0 \\ \defmij_{12} & \defmij_{22} & 0 \\ 0 & 0 & \mu_{3}
\end{array}
\right)\,,
\label{eq:defm12}
\end{equation} 
\noindent in which $\mu_3$ is the third eigenvalue of $\defm$. Because the eigenvalues are equal, we get 
the following 2 conditions for the $D_4^{12}$ caustic. 
\begin{eqnarray}
\defmij_{11}(q)&\,=\,&\defmij_{22}(q)\,,\nonumber\\
\defmij_{12}(q)&\,=\,&0\,.
\label{eq:Dconst}
\end{eqnarray}
Hence, the deformation tensor is 
\begin{equation}
\defm =  \left(\begin{array}{ccc} \mu & 0 & 0 \\ 0 & \mu & 0 \\ 0 & 0 & \mu_3 \end{array}\right)\,. 
\end{equation}

\medskip 
%\bigskip
\noindent
%An additional important
As a consequence of the inferred constraints~\eqref{eq:Dconst} for the $D_4$ singularities is that $D_4^{ij}$ points 
will always be located on the two corresponding $A_3$ varieties, $A_3^i$ and $A_3^j$. We may infer this from the following observation. In the 
coordinate system introduced above (cf. eq.~\eqref{eq:defm12}), the eigenvector for the third eigenvalue $\mu_3$ is given by $v_3=(0,0,1)$. 
The eigenvectors $v_1$ and $v_2$ both lie in the $q_1q_2$-plane, and since the matrix upper $2\times 2$ matrix is degenerate we have the freedom to 
take them to be orthogonal to the gradient of the corresponding eigenvalue fields which will also lay in the $q_1q_2$-plane. This means that 
\begin{eqnarray}
v_1 \cdot \nabla \mu_1&\,=\,&\mu_{1,1}=0\,,\nonumber\\ 
v_2 \cdot \nabla \mu_2&\,=\,&\mu_{2,2}=0.
\end{eqnarray}
This proves the unfolding $D_4^{ij} \to A_3^i$ and $D_4^{ij} \to A_3^j$. For the relations between the singularity classes see section \ref{sec:unfolding}. For a formal proof see \cite{Hidding:2014}. For the case of a non-diagonalizable deformation tensor we note that a small perturbation in the initial condition generically makes the deformation tensor diagonalizable.
 
\begin{figure}
\vskip 0.5truecm
\centering
\includegraphics[width=0.99\textwidth]{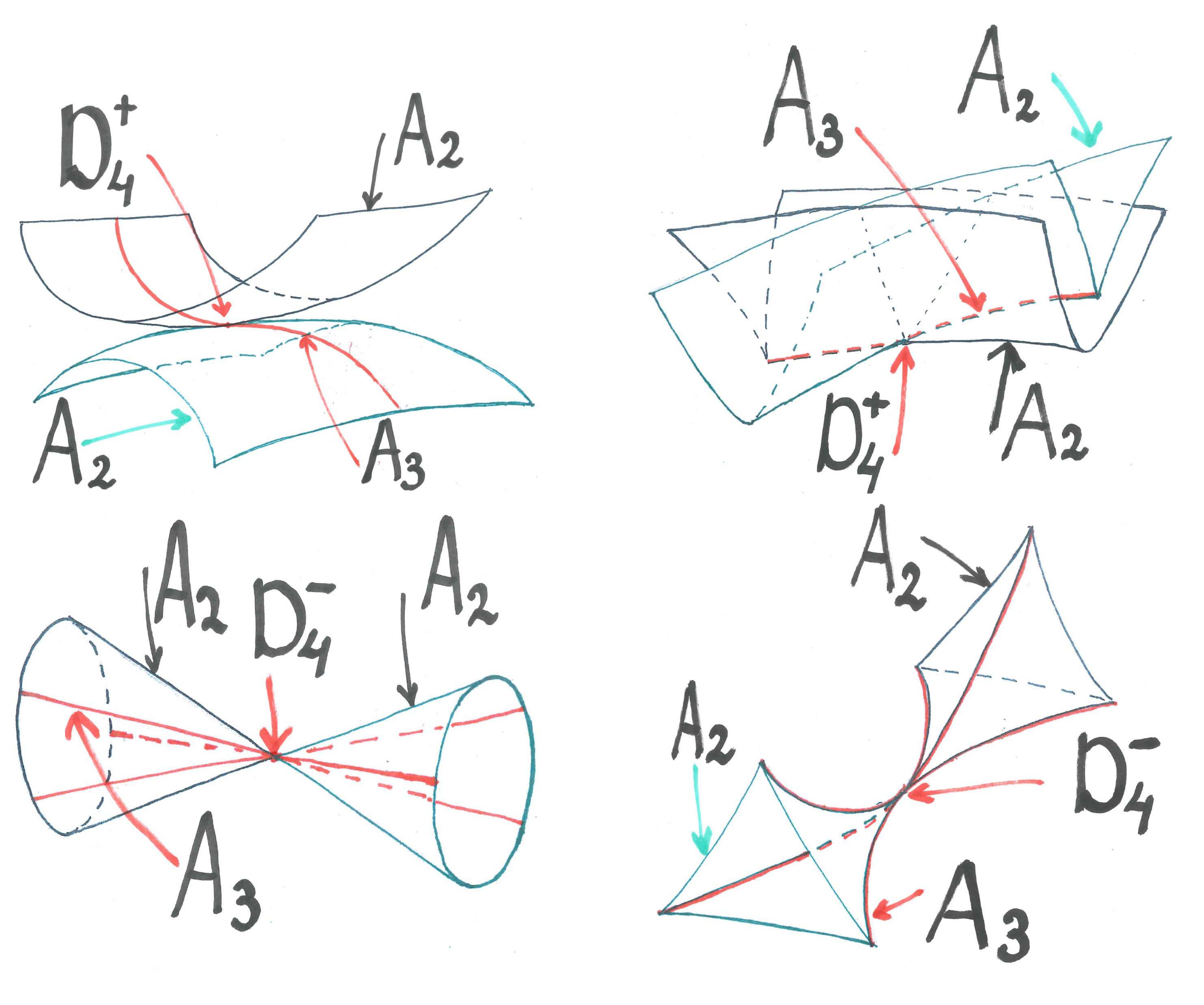}
\caption{The hyperbolic/elliptic umbilic ($D_4^\pm$) singularities. The upper two panels show the elliptic umbilic $(D_4^+$) singularity. The lower panels show the hyperbolic umbilic ($D_4^-$) singularity. The two panels on the left are their representations in Lagrangian space and the two panels on the right their representation in Eulerian space. The black sheets are fold ($A_2$) sheets corresponding to one eigenvalue field. The green sheets are fold ($A_2$) sheets corresponding to a second eigenvalue field. The red lines are cusp ($A_3$) curves. The point in the center depict the hyperbolic/elliptic umbilic ($D_4^\pm$) singularities. The hyperbolic umbilic ($D_4^+$) and elliptic umbilic ($D_4^-$) singularity are also known as the purse and pyramid singularity.}
\label{fig:D4}
\end{figure}

\bigskip 
  \noindent {\it The $D_4^+$ and $D_4^-$ caustics}\\
  
\noindent Shell-crossing for $A$ caustics is a one-dimensional process. A direct implication of this is that the related critical points are equivalent up to 
diffeomorphisms. For the $D$ family this is no longer true. Shell-crossing for the $D$-family is two dimensional. As a consequence, the $D_4$ class consist of hyperbolic ($D_4^+$) and elliptic ($D_4^-$) umbilic points, i.e.
\begin{equation}
D_4^{ij}(t)=D_4^{+ij}(t) \cup D_4^{-ij}(t)\,.
\end{equation}
\noindent In order to infer the corresponding caustic conditions we consider the two constraint quantities $Q_1(q)$ and $Q_2(q)$ (see eq.~\eqref{eq:Dconst}),
\begin{eqnarray}
Q_1(q)&\,=\,&\frac{M_{11}(q)-M_{22}(q)}{2}\,,\nonumber\\
Q_2(q)&\,=\,&M_{12}(q)\,,
\end{eqnarray}
\noindent which at the $D_4$ singularity location vanish, i.e. $Q_1(q_s)=0$ and $Q_2(q_s)=0$. By
a Taylor expansion of $Q_1(q)$ and $Q_2(q)$ 
in a neighbourhood around the $D_4$ singularity, we find that for points located in the $q_1q_2$-plane,
\begin{eqnarray}
Q_1(q) &\,=\,& a\, q_1 + b\, q_2\,,\nonumber\\
Q_2(q) &\,=\,& c\, q_1 + d\, q_2\,.
\end{eqnarray}
In this expansion, we have taken the $D_4$ singularity to define the origin of the coordinate system. The parameters $a$, $b$, $c$ and $d$ are the
derivatives of $Q_1(q)$ and $Q_2(q)$ at the $D_4$ location, 
\begin{equation}
a= \frac{1}{2} \frac{\partial(\defmij_{11}-\defmij_{22})}{\partial q_1},\ b= \frac{1}{2} \frac{\partial(\defmij_{11}-\defmij_{22})}{\partial q_2},\ c = \frac{\partial \defmij_{12}}{\partial q_1},\ d = \frac{\partial \defmij_{12}}{\partial q_2}\,.
\end{equation}
As proposed by \cite{Delmarcelle:1995}, the determinant $S_\defm$ of the corresponding $Q_1 Q_2$ map,
\begin{equation}
S_\defm = bc -ad = \frac{1}{2} \left[ (\defmij_{112}-\defmij_{222})\defmij_{112}- (\defmij_{111}-\defmij_{122})\defmij_{122}\right]\,,
\end{equation}
is invariant under rotations in the $q_1q_2$-plane \footnote{In fact, it can be shown that this determinant is a third-order invariant 
under rotations \citep{Delmarcelle:1995}.}. In the expression above, we have used the notation
\begin{equation}
M_{iik}\,=\,\frac{\partial \defmij_{ii}}{\partial q_k}\,,\quad M_{ikk}\,=\,\frac{\partial \defmij_{ik}}{\partial q_k}\,.
\end{equation}
Using the relations between the matrix elements $\defmij_{11}$, $\defmij_{22}$ and $\defmij_{12}$ and the eigenvalues 
$\mu_1$ and $\mu_2$, we may recast the determinant $S_\defm$ in an explicit expression incorporating these eigenvalues,
\begin{equation}
S_\defm \,=\, \frac{1}{2} \left[ (\mu_1-\mu_2)_{,2} \mu_{1,2}-(\mu_1-\mu_2)_{,1} \mu_{2,1}\right]\,.
\end{equation}
As \cite{Delmarcelle:1995} pointed out, the transformation can be shown to consist of two branches. Their 
identification surfaces via a rescaling of the determinant via the multiplication by a {\it positive} number. We 
then find that the two branches correspond to two separate singularity classes of the $D_4$ family, 
\begin{equation}
D_4^{\pm ij}(t)=\{q\in L|q \in A_2^i(t) \cap A_2^j(t)\wedge \mbox{sign}(S_\defm)= \pm1\}\,,
\end{equation}
\noindent where the points $q \in A_2^i(t) \cap A_2^j(t)$ are the points for whom at time $t$ the caustic conditions are simultaneously 
valid for two eigenvalues, i.e. $1+\mu_i=1+\mu_j=0$. Integrated over time, these $D_4^{\pm ij}(t)$ points trace out the curves $D_4^{\pm ij}$,
\begin{equation}
D_4^{\pm ij}\,=\,\{q\in L| q \in A_2^i(t) \cap A_2^j(t)\wedge \mbox{sign}(S_\defm)=\pm 1, \mbox{ for some time } t\}.
\label{eq:D4ijcurve}
\end{equation}
For an illustration of the hyperbolic/elliptic  umbilic ($D_4^\pm$) caustic see figure \ref{fig:D4}.

\vskip 0.25truecm
\subsubsection{The $D_4^\pm$ points}
\noindent The topology of the $D_4^{\pm ij}(t)$ variety changes at $D_4^\pm$ and $D_5$ points. The $D_4^\pm$ points are analogous to the $A_4^\pm$ points of the $A$-family. The $D_4^\pm$ points occur when 
$i$th and $j$th eigenvalue field, $\mu_i$ and $\mu_j$, restricted to the points $q$ in the $D_4^{\pm ij}$ variety 
reaches a minimum or maximum, i.e.
\smallskip
\begin{eqnarray}
D_4^{ij +}&\,=\,&\{q\in L| q \in D_4^{+ij}(t)\ \wedge \ \mu_{tk}(q) \mbox{ max-/min. of } \mu_{tk}|_{D_4^{+ ij}} (k=i \mbox{ or } k=j) \mbox{ for some } t \}\,\nonumber\\
D_4^{ij -}&\,=\,&\{q\in L| q \in D_4^{-ij}(t)\ \wedge \ \mu_{tk}(q) \mbox{ max-/min. of } \mu_{tk}|_{D_4^{- ij}} (k=i \mbox{ or } k=j) \mbox{ for some } t \}\,\nonumber \\
\end{eqnarray}

\medskip
\noindent Particularly interesting is the fact that the $D_4^\pm$ points are always created as a pair. Two $D_4^+$ points are created simultaneously, 
as are $D_4^-$ points. By implication, also the $D_4^\pm$ curves (eq.~\eqref{eq:D4ijcurve}) are always created in pairs. This is in contrast to the $D_5$ points, which go along with the creation of a pair consisting of a $D_4^+$ and a $D_4^-$ point. 

\vskip 0.25truecm
\subsubsection{The $D_5$ caustics}
\noindent The shell-crossing condition applied to the $D_4^{ij}$ variety yields the caustic conditions for the $D_5$ parabolic umbilic singularity.  The manifold $D_4^{ij}$ forms a singularity in the point $q_s \in  D_4^{ij}(t)$ if and only if the tangent vector $T \in T_{q_s} D_4^{ij}$ is normal 
to $v_k^*$, with $k\neq i,j$. Hence, the tangent vector $T \in \text{span}_{\mathbb{C}}\{v_i,v_k\}$, i.e.
\begin{equation}
D_5^{ij} =\{ q \in L| q \in D_4^{ij} \text{ and } \text{span}\{v_i,v_j\} \cap T_q D_4^{ij} \neq \emptyset \}
\end{equation}
For three dimensional fluids in which the deformation tensor is separable in a time factor and a spatial factor, the normal $n=\nabla (\mu_{ti}-\mu_{tj})$, is orthogonal to both $v_i$ and $v_j$, 
\begin{eqnarray}
(\mu_i-\mu_j)_{,i} \equiv v_i \cdot \nabla (\mu_{ti}-\mu_{tj})&\,=\,&0\,,\nonumber\\
(\mu_i-\mu_j)_{,j} \equiv v_j \cdot \nabla (\mu_{ti}-\mu_{tj})&\,=\,&0\,.
\end{eqnarray}
\noindent The collection of all such points form the variety
\begin{equation}
D_5^{ij} = \{q \in L| q \in D_4^{ij}(t) \ \ \wedge \ \ (\mu_i-\mu_j)_{,i} = (\mu_i-\mu_j)_{,j}=0 \ \ \mbox{ for some time } t\}\,.
\end{equation}

\medskip
\noindent The $D_5^{ij}$ lays on the $A_4^i$ and $A_4^j$ variety. The elliptic and hyperbolic umbilic ($D_4^\pm$) points merge in parabolic umbilic ($D_5$) points, since $D_5^{ij}(t)\subset D_4^{ij}(t)$ and
\begin{equation}
S_\defm =\frac{1}{2} \left\{ (\mu_i-\mu_j)_{,j} \mu_{i,j}-(\mu_i-\mu_j)_{,i} \mu_{j,i}\right\}\,=\,0.
\end{equation}
\noindent The $D_5$ points are stable singularities in the classification of Lagrangian singularities. For general dynamics they are unstable and not included in the classification scheme.

\vskip 0.5truecm
\subsection{Caustic conditions: physical significance}
\noindent For a visual appreciation of the process leading to the formation of the various classes of caustics identified in the subsections above, it is helpful
to consider the phase-space manifold on which all mass elements are located in 6-D phase-space $L \times E$. This is called the \textit{phase-space sheet}
\citep[see e.g.][]{abel2012,shandarin2012}. The dynamical evolution of a system leads to the folding of this phase-space sheet. In a sense, we can recognize
a hierarchical process in which the phase-space sheet is wrapped into an increasingly complex pattern. In this process we see the emergence of
a hierarchy of complex spatial folds.

\begin{figure}
%\vskip 0.5truecm
\centering
\includegraphics[width=0.83\textwidth]{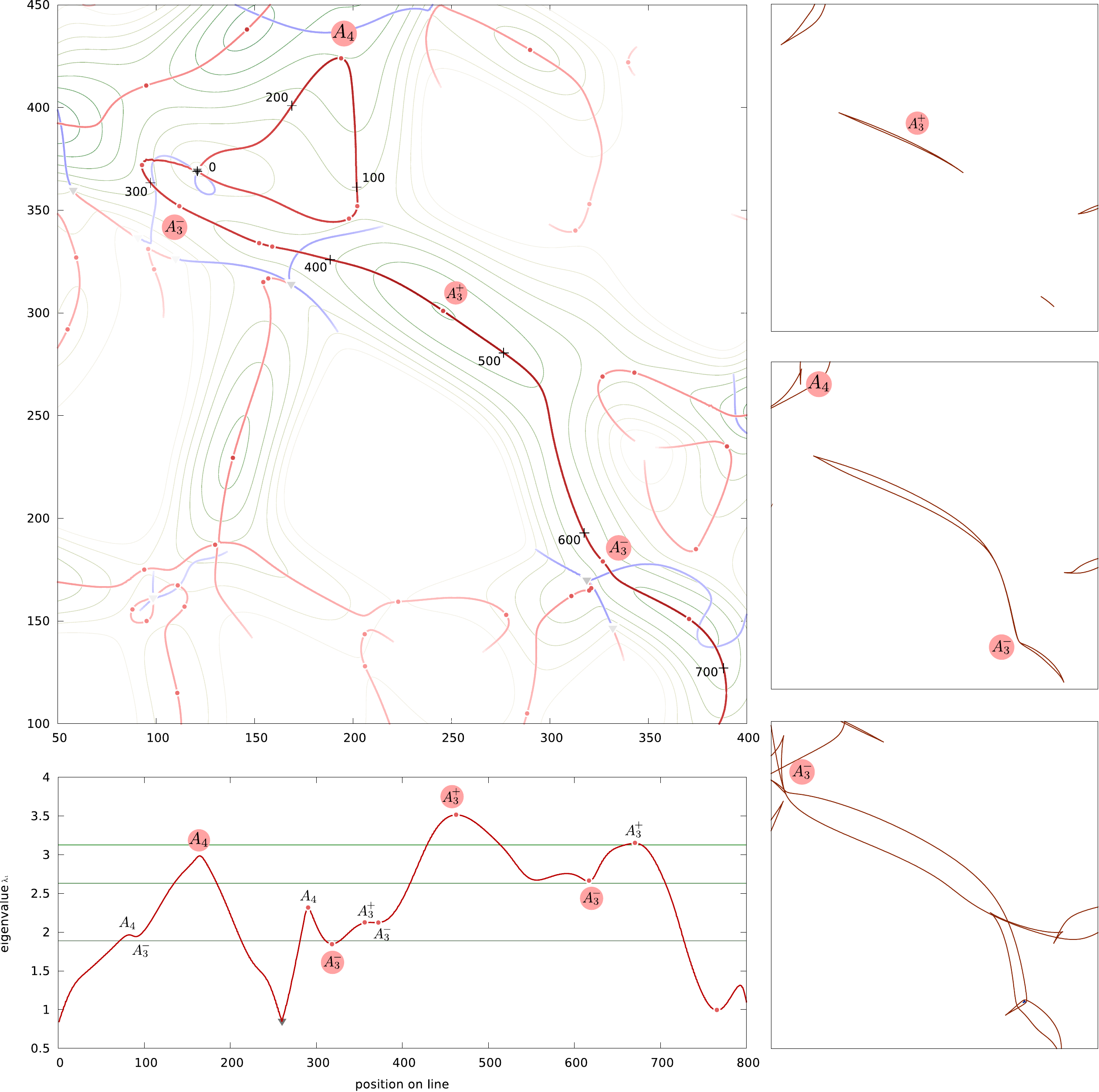}
\vskip 0.5truecm
\caption{Eigenvalue field and singularity points. In the case of the two-dimensional Zel'dovich approximation (see appendix \ref{sec:Zel'dovich}). The Zel'dovich approximation
  concerns the specific situation of potential flow, for which the eigenvalues and
    eigenvectors are real, and shows the field of the lowest eigenvalue. Top lefthand frame: the contour map illustrates the typical structure of
  the eigenvalue field $\lambda_1$ corresponding to 
  a 2-D Gaussian random density field. Indicated are the positions of different $A$-family singularity points and varieties. The run of the $A_3$ line is
  particular noteworthy. One may appreciate how the identity 
of the various singularities is determined by the specific geometric character of the eigenvalue field $\lambda_1(q)$, as expressed in its derivatives. Bottom lefthand frame: the panel depicts the run of the eigenvalue field along the $A_3$ curve (in the contour map of 
top lefthand frame). Note the location of the $A_3^\pm$ points and $A_4$ points on the extrema of the curve. The green curve represents the level $b_+(t)^{-1}$ indicating
which parts of the $A_3$ line has formed at three instances depicted in the righthand panels. Righthand panels: the three panels show the evolution,
in Eulerian space, of the $A_3$ line. Note the appearance of the corresponding caustics and the relation between the geometry of the $A_3$ line in Eulerian space and the
$A_3^\pm$ and $A_4$ points corresponding to the three green lines in the lower lefthand panel.
This is a more extensive version of figure 8 in Hidding et al. 2014.}
\label{fig:eigvalfield}
\end{figure}

The phase-space sheet folding process generates higher order singularities within the $A_2$ caustic itself. These can only be
identified with the help of the complementary eigenvector conditions. Restricting the manifold $M$ to the points $q_s$ located in
the $A_2$ caustic, one may identify the subset of points for whom a nonzero vector $T$ exists that (a) is tangent to the $A_2$ manifold
and (b) is orthogonal to the span of dual eigenvectors $\mbox{Span} \{v_j^*|j\neq i\}$. This subset fulfils the shell-crossing
conditions and maps into a higher order singularity. Proceeding along the sequence of caustic conditions leads to the identification
of the entire hierarchy of caustics.

\bigskip
The classification of A family caustic involves one eigenvalue for which $1+\mu_i=0$. It is straightforward to see that a similar procedure follows
for configurations involving more than one eigenvalue for which $\mu_k=-1$. For example, if both $1+\mu_1=0$ and $1+\mu_2=0$,
then $T$ will be a vector orthogonal to the dual eigenvector $v_3^*$. The eigenvalue conditions therefore trace
a line through three-dimensional Lagrangian space. The points $q$ along this line are singularity points. Along this line we
subsequently seek to identify higher-order singularities, by identifying points $q_s$ along the line for which a tangent
vector $T$ exists fulfilling the shell-crossing conditions. 

Conversely, note that if $1+\mu_i \neq 0$ for all $i$, then there does not exist any $T$ satisfying the general shell-crossing condition. 

\subsection{Spatial Connectivity: Singularities and Eigenvalue Fields}
With the purpose to provide a guide that evokes  a visual intuition for the connection between the structure and geometry of the eigenvalue 
fields and the formation of the various singularities, in particular those of the $A$-family, we include 
figure~\ref{fig:eigvalfield}. It shows a contour map representing the typical structure of the eigenvalue field $\mu_i$. This field  
corresponds to a two-dimensional Gaussian random density field. For reasons of convenience, we have assumed higher eigenvalues 
to correspond to earlier collapse, and negative ones to no collapse (in other words, we have mirrored $\mu_i$). The geometry and topology 
of the eigenvalue landscape is decisive for the occurrence of singularities. This may already be inferred from the positions of different $A$-family 
singularity points and varieties, whose positions are indicated on the contour map. 

The landscape defined by the eigenvalue contours is varied, characterized by several peaks, connected by ridges with lower $\mu_i$ values. These, 
in turn, are connected to valleys in which $\mu_i$ attains negative values that will prevent collapse -- along the direction of the eigenvector $v_i$ -- 
of the corresponding mass elements at any time. From the density relation (eqn.~\eqref{eq:LagrangianDensity}), we know that the region of space 
that has undergone collapse before the current epoch (i.e. attained an infinite density) is the superlevel set of the eigenvalue field defined by 
the current value $\mu_{ti}$. For each time $t$, the positive value contours correspond to the $A_2(t)$ fold sheets. Collapse occurs first at the 
maxima in the field. These mark the birth of new features, and are designated by the label of $A_3^+$ points. Evidently, the steepness 
of the hill around these maxima, i.e. the gradient $\nabla \mu_i (q)$, will determine how and which mass elements around the hill 
will follow in outlining the emerging feature around the $A_3^+$ points. 

The run of the $A_3$ line is particularly noteworthy. The key significance of the $A_3$ curve is evident from the observation that all 
$A$-family singularities are aligned along the ridge. In two-dimensional space, the $A_3$ curves delineate the points where the eigenvalues 
$\mu_i$ are maximal along the direction of the corresponding local eigenvector. At these points, along the eigenvector direction, the gradient 
of the eigenvalues is zero, i.e. they are the points where the eigenvector $v_i$ is perpendicular to the local gradient of $\nabla \mu_i$ of the 
eigenvalue field. Below, in section~\ref{sec:A3}, we will see that this follows directly from the shell-crossing conditions that were derived in 
the previous section. Because of this there is a line-up and accumulation of neighbouring mass elements that simultaneously pass through the 
singularity. When mapped to Eulerian space, this evokes the formation of an $A_3$ cusp. 

To illustrate the connection between $A_3$ curves and the various singularities even more strongly, the bottom lefthand panel depicts 
the run of the eigenvalue field along the $A_3$ curve. In particular noteworthy is the location of the $A_3^\pm$ points and $A_4$ points on the 
extrema of the curve. A prominent aspect of this is the presence of the $A_3^-$ points at saddle junctions in the 
eigenvalue field. These are topologically the most interesting locations, as they evoke the merging of separate fold sheets into 
a single structure. In other words, they are the points where the topological structure of the field undergoes a transition and 
where the connectivity of the emerging structural features is established. To establish this even more strongly, the three righthand 
panels of figure~\ref{fig:eigvalfield} represent a time sequence of the evolving structure along the $A_3$ line as it is mapped 
to its appearance in Eulerian space. The evolution follows the linear Lagrangian Zel'dovich approximation (see \cite{Zeldovich:1970} and 
appendix~\ref{sec:Zel'dovich}). We may note the appearance and merging of the corresponding caustics.

\section{Classification of singularities}
\label{sec:class}
\noindent The form and morphology in which the various singularities that were inventorized in the previous section 
will appear in the reality of a physical system depends on several aspects. The principal influence concern 
the dynamics of the system, as well as its dimensionality. The dynamics determines the way the fluid evolves, to 
a large extent via its dominant influence on the accompanying flow of the fluid. This affects the morphology of the 
fluid, and in particular the occurrence of singularities. Evidently, also the dimensionality of the fluid process 
will bear strongly on the occurrence and appearance of singularities. Higher spatial dimensions may enlarge the number 
of ways in which a singularity may form. It also influences the ways in which singularities can dynamically transform 
into one another. 

In this section, we provide an impression of the variety in appearance of singularities. To this 
end, we will first discuss the generic singularity classification scheme that we follow. It is not the 
intention of this study to provide an extensive listing of all possible classes of fluids. Instead, to make clear in 
how different physical situations may affect the appearance of singularities, we restrict our presentation of 
classification schemes to two different classes of fluids. We also restrict our inventory to fluids in a three-
dimensional context. It is the most representative situation, and at the same time offers a good illustration of 
other configurations.

\subsection{Classes of Lagrangian fluids}
\noindent To appreciate the role of the dynamics in constraining the evolution and appearance of a fluid, and that of the formation 
and fate of the singularities in the fluid, it is important to understand and describe its evolution in terms of 
six-dimensional phase space. 

\medskip
One way of defining phase space $\phasespace$ is in terms of the Cartesian product of Lagrangian and Eulerian manifolds $L$ and $E$, 
i.e. $\phasespace=L\times E$. In this context, the phase space coordinates of a mass element are $(q,x)$. Every point in 
phase space $(q,x)\in \phasespace$ represents the initial and final position $q$ and $x$ of a mass element at some time $t$. 
Evidently, one may also opt for the more conventional definition consisting of space coordinates $x$ and canonical momenta $p$, 
in which case the phase space coordinate of a mass element are given by $(x,p)$. However, for the description of Lagrangian 
fluid dynamics it is more convenient to follow the first convention. We should note that for this description of phase space 
Liouville's theorem does not apply, specifically not for the Euclidean notion of volumes. 

At the initial time $t=0$, the Lagrangian map is the identity map, i.e. for all $q\in L$ $x_0(q)=q$. In phase space $\phasespace$, the 
fluid then occupies the submanifold $\mathcal{L}_0=\{(q,x_0(q))\in \phasespace|q\in L\}$. If we equip $\phasespace$ with a symplectic 
structure $\omega$, we can prove this to be a so-called Lagrangian submanifold (for a precise definition of Lagrangian submanifolds 
see appendix~\ref{Ap:LE}).

\medskip
Differences in the dynamics of a fluid reveal themselves in particular through major differences in the phase space structure and 
topology of the manifolds delineated by the mass elements. To provide an impression of the differences in 
morphology and classification of singularities emerging in fluids of a different nature, 
specifically that of fluids with a different dynamical behaviour, we concentrate the discussion on two different classes 
of Lagrangian fluids:
\bigskip
\begin{enumerate}
\item {\it Generic Lagrangian fluids}. \\ Lagrangian fluids for which the map $x_t:L \to E$ is a generic continuous and differentiable mapping from $L$ to $E$ for every time $t$. The dynamics does not restrict the map $x$ to any extent. We describe the classification up to local diffeomorphisms, i.e. two singularities are considered equivalent if and only if there exist local coordinate transformations, which map them into each other. 
\medskip
\item {\it Lagrangian fluids with Hamiltonian dynamics}. \\ The evolution of the fluid is governed by a Hamiltonian. This assumption restricts the possible evolution of the fluid. Formally, the map $x$ corresponds uniquely to a so-called Lagrangian map. The singularities of Lagrangian maps, known as Lagrangian singularities, are classified up to Lagrange equivalence.
\end{enumerate}
\bigskip
\noindent Lagrangian fluids with Hamiltonian dynamics form an important class of fluids: fundamental theories of particle physics generally allow 
for a Hamiltonian description. Nonetheless, in a range of practical circumstances we may encounter fluids that are either more or less constrained. 
An example are fluids with effective dynamics. They contain friction terms which are not described by Hamiltonian systems. Such fluid systems are less 
restrictive than those that are specifically Hamiltonian. On the other hand, there are also Hamiltonian fluids that are characterized by additional 
constraints. 

\subsection{Singularity classification: generic fluids}
\label{sec:condnH}
\noindent For the classification of singularities of generic one-family maps $x:L \to E$, with $L$ and $E$ three-dimensional, we 
follow the classification by Bruce \cite{Bruce:1985}. Bruce showed that the singularities that 
emerge in generic mappings are equivalent to those emerging in the simple linear maps 
\begin{equation}
x_t(q)= q+t\,u(q)\,,
\end{equation}
in which $u$ is a vector field on $L$. In general, the vector field $u(q)$ consists of both a longitudinal and a transversal part, 
\begin{equation}
u(q)\,=\,u_l(q) + u_t(q)\,.
\end{equation}
The longitudinal component corresponds to potential motion and has curl zero, $\nabla \times u_l = 0$, while the transversal component 
has divergence zero, $\nabla \cdot u_t=0$.

\bigskip
The classification of singularities in general Lagrangian fluid dynamics is expressed by theorem~3. We restrict ourselves to listing 
the classification scheme, in terms of the generic expressions for the maps $x_t(q)$ of each of the classified 
singularities. In appendix \ref{ap:normalform} we show that these normal forms indeed satisfy the corresponding caustic conditions. Note that
the classification was derived using the classification of jet-spaces. It successfully cauterized the properties of caustics appearing in
Lagrangian maps but did not provide a practical way to detect them in realizations.

\begin{thr}
A stable singularity occurring in a Lagrangian fluid with generic dynamics is, up to local diffeomorphisms, equivalent to one 
of the following classes:

\medskip
\begin{center}
\begin{tabular}{lll}
    \hline
    \mbox{Singularity} & & \mbox{Singularity}\\
    \mbox{class} & \mbox{Map} $x_t(q)$ & \mbox{name}\\ \hline 	
$A_1$& 	$x_t(q) =q$ & \mbox{trivial case}\\ 
$A_2$&	$x_t(q) =q + t \left(0,0,q_3^2 - q_3\right)$ & \mbox{fold}\\ 
$A_3$& 	$x_t(q) =q + t \left(0,0,q_1 q_3 + q_3^3 - q_3\right)$ & \mbox{cusp}\\ 
$A_4$& $x_t(q)=q +  t \left(0,0,q_1 q_3  + q_3^4-q_3\right)$ & \mbox{swallowtail}\\ 
$A_5$&	$x_t(q)=q +  t \left(0,0, q_1q_3 + q_2 q_3^2 + q_3^5 - q_3\right)$  &\mbox{butterfly}\\ 
$D_4^{\pm}$& $x_t(q)=q +  t \left(0,q_2q_3-q_2, q_2^2 \pm q_3^2 + q_1 q_2 - q_3\right)$ & \mbox{hyperbolic/elliptic}\\
\hline
$A_3^\pm$& $x_t(q)=q +  t \left(0,0,(q1^2 \pm q_2^2) q_3 +q_3^3 - q_3\right)$ & \\ 
$A_4^\pm$& $x_t(q)=q +  t \left(0,0,q_1 q_3 \pm q_2^2 q_3^2 + q_3^4-q_3\right)$ &\\ 
\hline
\end{tabular}
\end{center}

\medskip 
Note: The normal forms $x_t(q)$ form the singularity at the origin $q=0$, at $t=1$. The first five singularity classes are the $A$-family. The subsequent class is the $D$-family. The last two are the normal forms of the $A_3$ and $A_4$ points.
The $A_k$ class has co-rank $1$ and co-dimension $k-2$. The $D_4^\pm$ singularities have co-rank $2$ and are one-dimensional.
\medskip
\end{thr}

\subsection{Singularity classification: Hamiltonian fluids}
\label{sec:LagrangianSingularities}
\noindent The evolution of Lagrangian fluids with Hamiltonian dynamics is more constrained than that of generic Lagrangian fluids. 
As the fluid develops complex multistream regions, the phase space submanifold $\mathcal{L}_t = \{(q,x_t(q))|q\in L\}$ for fluids 
with Hamiltonian dynamics remains a Lagrangian submanifold. 

A key step in evaluating the emerging singularities is that of connecting the displacement map $s_t(q)$ to the Lagrangian map. In 
appendix~\ref{app:displlagr}, we describe in some detail how a given Lagrangian map can be constructed from a Lagrangian 
submanifold $\mathcal{L}$. A Lagrangian map can develop regions in which multiple points in the Lagrangian manifold are mapped to 
the same point in the base space. 

Lagrangian singularities are those points at which the number of pre-images of the Lagrangian 
map undergoes a change. Lagrangian catastrophe theory \cite{Arnold:1972,Arnold:2012a} classifies the stable singularities. 
This refers to the stability of singularities with respect to small deformations of the Lagrangian manifold 
of $\mathcal{L}$. This is true up to Lagrangian equivalence, a concept that is a generalization of equivalence 
up to coordinate transformation. For a more formal and precise definition of Lagrangian equivalence see 
appendix \ref{Ap:LE}. 

\bigskip 
It can be demonstrated \citep[see][]{Arnold:2012a} that every Lagrangian map $l:\mathcal{L} \to \phasespace \to E$ is 
locally Lagrangian equivalent to a so-called gradient map, i.e. the map $x_t$ is 
locally equivalent to
\begin{equation}
x_t(q)= \nabla_q S_t\,,
\end{equation}
for some $S_t: L \to \mathbb{R}$. By recasting $S_t$ in terms of a function $\Psi_t:L \to \mathbb{R}$, 
\begin{equation}
S_t\,=\,\frac{1}{2} q^2+\Psi_t(q)\,,
\end{equation}
we find that locally the map $x$ can be written in the form
\begin{equation}
x_t(q)= q +  \nabla_q \Psi_t(q)\,.
\end{equation}
Evidently, this implies that the displacement map is longitudinal, and that the corresponding Jacobian ${\partial s_t}/{\partial q}$ is symmetric. 

\bigskip
The classification of singularities of a Lagrangian fluid with Hamiltonian dynamics is expressed by theorem~4. 
In appendix \ref{ap:normalform} it is shown that these normal forms indeed satisfy the corresponding caustic conditions.  For proofs we refer 
to Arnol'd \cite{Arnold:1972}. Note that the classification was derived using the classification of critical points of scalar functions and the theory of generating functions. It successfully characterized the properties of caustics appearing in Lagrangian maps but did not provide a practical way to detect them in realizations.

\begin{thr}
A stable Lagrangian singularity of a Lagrangian fluid with Hamiltonian dynamics, is locally Lagrange equivalent to 
one of the following classes:

\begin{center}
  \begin{tabular}{lll }
    \hline
    \mbox{Singularity} & & \mbox{Singularity}\\
    \mbox{class} & \mbox{Map} $x_t(q)$ & \mbox{name}\\ \hline 	
$A_1$& 		$x_t(q)=q$  &\mbox{trivial case}\\ 
$A_2$&		$x_t(q)=q + t \left(0,0,q_3^2 - q_3\right)$  &\mbox{fold}\\ 
$A_3$& 		$x_t(q)=q + t \left(\frac{1}{2} q_3^2,0,q_3(q_1-1)\right)$  &\mbox{cusp}\\ 
$A_4$&     		$x_t(q)=q + t   \left(\frac{1}{2} q_3^2 ,0 ,q_1 q_3  + q_3^4-q_3\right)$ & \mbox{swallowtail}\\ 
$A_5$&		$x_t(q)=q + t \left(\frac{1}{2} q_3^2,\frac{1}{3} q_3^3 , q_1q_3 + q_2 q_3^2 + q_3^5 - q_3\right)$ &\mbox{butterfly}\\
$D_4^{\pm}$&	$x_t(q)=q + t  \big(\pm q_1 q_2-q_1, $ &\mbox{hyperbolic/elliptic}\\ 
& \hskip 2.2truecm $\pm\left(\frac{1}{2}q_1^2+\frac{3}{2}q_2^2\right)+2 q_2 q_3 + 2 q_2^3-q_2,q_2^2\big)$ &\\
%& \hskip 2.2truecm $ q_2^2\big)$ &\\
$D_5$&		$x_t(q)=q + t   \left(0, q_2^3 - q_2, q_3^3 - q_3\right)$ &\mbox{parabolic}\\
\hline
$A_3^\pm$&	$x_t(q)=q + t  \left(q_1 q_3^2,\pm q_2 q_3^2,(q1^2 \pm q_2^2) q_3 +q_3^3 - q_3\right)$  &\\ 
$A_4^\pm$&	$x_t(q)=q + t   \left(\frac{1}{2} q_3^2,\pm \frac{2}{3} q_2 q_3^3 ,q_1 q_3 \pm q_2^2 q_3^2 + q_3^4-q_3\right)$ &\\ 
\hline
\end{tabular}
\end{center}

\medskip 
Note: The normal forms $x_t(q)$ form the singularity at the origin $q=0$, at $t=1$. The first five singularity classes are the $A$-family. The subsequent two are the $D$-family. The last two are the normal forms of the $A_3$ and $A_4$ points.
The $A_k$ class has co-rank $1$ and co-dimension $k-2$. The $D_k$ singularities have co-rank $2$ and co-dimension $k-2$ \cite{Arnold:1972}.
\medskip
\end{thr}

\medskip
\indent Comparing the classification schemes for generic Lagrangian singularities and those for Lagrangian fluids with Hamiltonian dynamics, 
we may note the similarities. Both classifications have an $A$ and a $D$ family. It can be demonstrated that the 
$A$ singularity classes of the scheme for Lagrangian fluids with Hamiltonian dynamics are contained in those corresponding to the 
generic Lagrangian fluid. Concretely, this means that a displacement field corresponding to the Hamiltonian $A_k$ class is also an element 
of the generic $A_k$ class. 

The $D$ families are some what different. The Hamiltonian $D_4$ class is contained in the generic $D_4$ class. However, the Hamiltonian 
$D_5$ class has no analogue in the generic classification scheme. This is a result of the $D_5$ singularity not being stable under 
coordinate transformations. 

\medskip
A final remark concerns the singularity classification schemes for higher dimensional fluids. For these a more elaborate classification 
scheme applies. This classification scheme is described in appendix \ref{Ap:LE}. 

\subsection{Unfoldings}
\noindent Singularities generally change their class upon small, but finite, deformations of the displacement map $s_t$. The corresponding evolution of a singularity follows the universal unfolding process of singularities. The general behavior is described in the following unfolding diagram, in which the 
arrows indicate the singularity into which specific singularities can transform. 

\begin{center}
\begin{tikzpicture}
\node (A1) {$A_1$};
\node [right of=A1] (A2) {$A_2$};
\node [right of=A2] (A3) {$A_3$};
\node [right of=A3] (A4) {$A_4$};
\node [right of=A4] (A5) {$A_5$};
\node [below of=A4] (D4) {$D_4$};
\node [below of=A5] (D5) {$D_5$};
  \draw[<-] (A1) to  (A2);
  \draw[<-] (A2) to  (A3);
  \draw[<-] (A3) to  (A4);
  \draw[<-] (A4) to  (A5);
  \draw[<-] (D4) to  (D5);
  \draw[<-] (A3) to  (D4);
  \draw[<-] (A4) to  (D5);
\end{tikzpicture}
\end{center}
For $i\geq 2$, the $A_i$ singularities decay into $A_{i-1}$ singularities. For $i \geq 5$, the $D_i$ singularities decay into either $A_{i-1}$ or $D_{i-1}$ singularities. In section \ref{sec:Dynamics} we will describe how the decay of singularities is connected to the evolution of the large-scale 
structure in the Universe and in outlining the spine of the cosmic web. 

%%%%%%%%%%%%%%%%%%%%%%%%%%%%%%%%%%%%%%%%%%%%%%%%%%%%%%%%%%%%%%%%%%%%%%%%%%%%%%%%%%%%%%%%%%%%%%%%%%%%%%%%%%%%

\vskip 0.5truecm
\section{The caustic skeleton \& the cosmic web}
\label{sec:cosmic_web}
\noindent The process of formation and evolution of structure in the Universe is driven by the gravitational growth of tiny primordial 
density and velocity perturbations. When it reaches a stage at which the matter distribution starts to develop nonlinearities, we see 
the the emergence of complex structural patterns. In the current universe we see this happening at Megaparsec scales. On these scales, 
cosmic structure displays a marked intricate weblike pattern. Prominent elongated filamentary features define a pervasive 
network. Forming the dense boundaries around large tenuous sheetlike membranes, the filaments connect up at massive, compact clusters 
located at the nodes of the network and surround vast, underdense and near-empty voids.  

The gravitational structure formation process is marked by vast migration streams, known as cosmic flows. Inhomogeneities in 
the gravitational force field lead to the displacement of mass out of the lower density areas towards higher density regions. 
Complex structures arise at the locations where different mass streams meet up. Gravitational collapse sets in as this happens. 
In terms of six-dimensional phase space, it corresponds to the local folding of the phase space sheet along which matter -- in particular
the gravitationally dominant dark matter component -- has distributed itself. 

\begin{figure}[h]
  \vskip -0.25truecm
\centering
\includegraphics[width=0.9\textwidth]{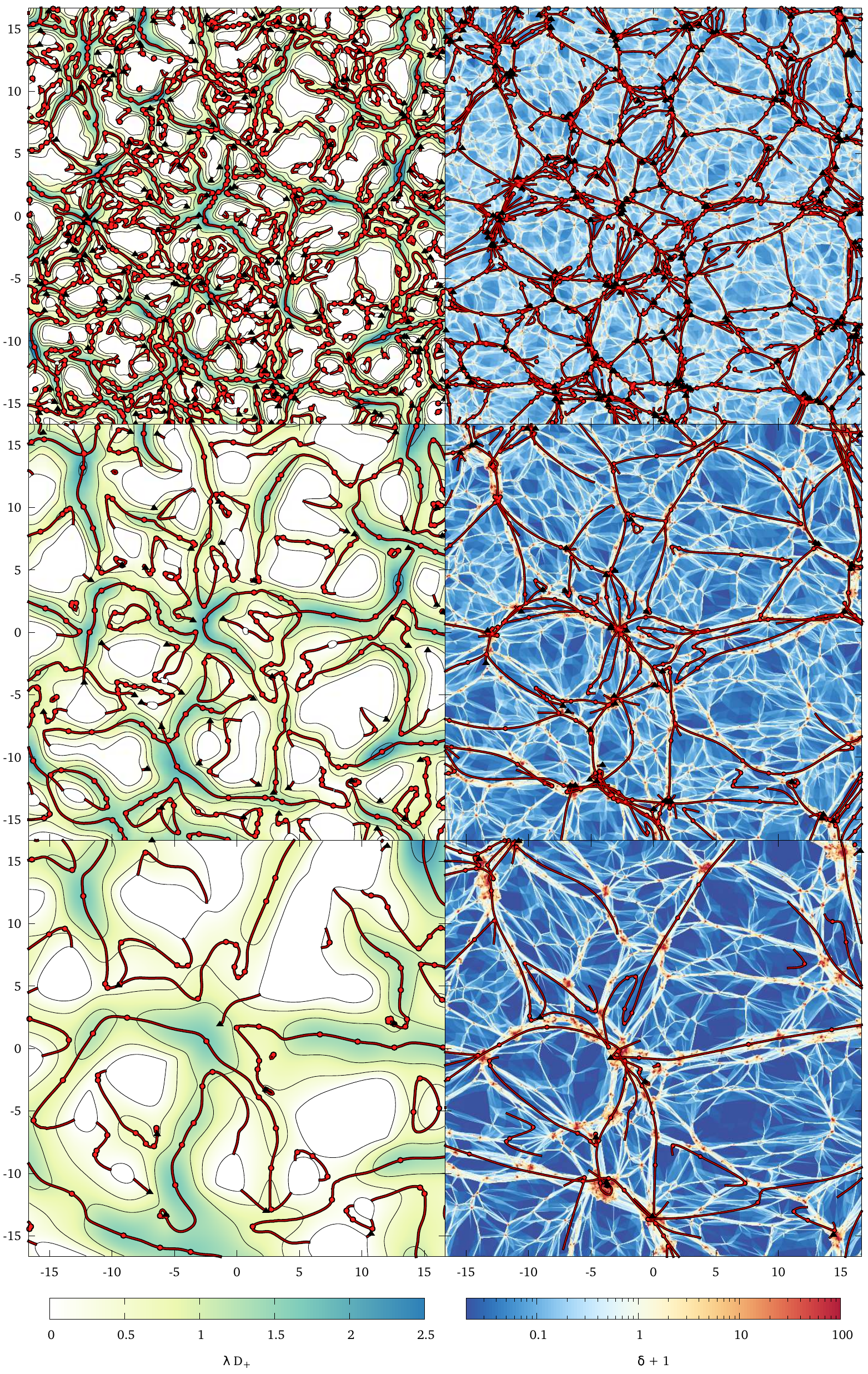}
\end{figure}
\begin{figure}[t]
\caption{Spatial distribution of singularities in the Lagrangian and Eulerian cosmic web. The figure compares the spine of the cosmic web with the mass 
distribution in a 2-D $N$-body simulation. Left panel: initial field of density fluctuations and the skeleton of identified singularities/catastrophes. Right panel: 
density field of an evolved 2D cosmological $N$-body simulation, in which the Lagrangian skeleton of singularities is mapped by means of the Zel'dovich 
approximation. From Feldbrugge et al.~ \cite{Feldbrugge:2014b}.}
\label{fig:Nbody}
\end{figure}

\subsection{the Caustic Skeleton}
\noindent The positions where streams of the dark matter fluid cross are the sites where gravitational collapse occurs. The various types of 
caustics described and classified in our study mark the different configurations in which this process may take place. Their locations  
trace out a Lagrangian skeleton of the emerging cosmic web, marking key structural elements and establishing their connectivity 
(also see the discussion in \cite{Hidding:2014}). In other words, the $A_3,A_4,A_5,D_4,D_5$ varieties, in combination with the 
corresponding $A_3^\pm, A_4^\pm,$ and $D_4^\pm$ points, are the dynamical elements whose connectivity defines the weaving of the 
the cosmic web \cite{Zeldovich:1970,bondweb1996,weybond2008,aragon2010,cautun2014}. On the basis of this observation, we may obtain the skeleton of the cosmic web
by mapping the caustic varieties defined above to Eulerian space with the Lagrangian map $x_t$. Following the identification of the various 
caustic varieties and caustic points in Lagrangian space, the application of the map $x_t$ will 
produce the corresponding weblike structure in Eulerian space.

Of central significance in our analysis and description of the cosmic web is the essential role of the deformation tensor
\textit{eigenvector} fields in outlining the caustic skeleton and in establishing the spatial connections between the
various structural features. So far, Lagrangian studies of the cosmic web have usually been based on the role of the \textit{eigenvalues}
of the deformation tensor (for recent work see \cite{Chong:1990, Wang:2014, Leclercq:2017}). Nearly without exception, they ignore the
information content of the eigenvectors of the deformation tensor. In this work we actually emphasize that the eigenvectors are of
key importance in tracing the spatial locations of the different types of emerging caustic features and, in particular, in establishing their
mutual spatial connectivity. This important fact finds its expression in terms of the \textit{caustic conditions} that we have derived
in this study.

The study by Hidding et al. \cite{Hidding:2014} illustrated the important role of the deformation field eigenvectors in outlining the
skeleton of the cosmic web, for the specific situation of $A_3$ cusp lines in the 2-D matter distribution evolving out of a Gaussian initial
density field. The present study describes the full generalization for the evolving matter distribution
(a) for each class of emerging caustics in (b) in spaces of arbitrary dimension $D$. 

\subsection{2-D Caustic Skeleton and Cosmic Web}
\noindent A telling and informative illustration of the intimate relationship between the caustic skeleton defined by the
derived caustic conditions and the evolving matter distribution is that offered by the typical patterns emerging
in the two-dimensional situation. Figure~\ref{fig:Nbody} provides a direct and quantitative comparison between the
caustic skeleton of the cosmic web and the fully nonlinear mass distribution in an N-body simulation. 
The three panels in the lefthand column show the Lagrangian skeleton for a two-dimensional fluid. The fluid is taken
to evolve according to the Zel'dovich approximation \cite{Zeldovich:1970} (see appendix~\ref{sec:Zel'dovich}), which represents a surprisingly accurate 
first-order Lagrangian approximation of a gravitationally evolving matter distribution 
\citep[see e.g.][]{shandarinzeld:1989}. The initial density field of the displayed models is that of 
a Gaussian random density field \cite{adler:1981,bbks:1986}, which according to the 
latest observations and to current theoretical understanding is an accurate description 
of the observed primordial matter distribution \cite{Planck:2016,WMAP:2003,Creminelli:2006}.

To enable our understanding of the hierarchical process of structure formation and the resulting multiscale 
structure of the cosmic web, we assess the caustic structure of the Lagrangian matter field at three 
different resolutions. In figure~\ref{fig:Nbody} the field resolution decreases from the top panels to the 
bottom panels, as the initial density field was smoothed by an increasingly large Gaussian filter. 
The contour maps that form the background in these panels represent the resulting initial 
density fields. The red lines trace the $A_3$ variety, i.e. the $A_3$ lines, 
for the largest eigenvalue $\mu_1$ field (also see fig.~\ref{fig:eigvalfield} to appreciate how they 
are related). Also the $A_3^\pm$ points and $D_4^\pm$ points are shown, 
the first as red dots, the latter as black triangles. 

\begin{figure}[h]
  \centering
  \vskip -0.85cm
\includegraphics[width=0.88\textwidth]{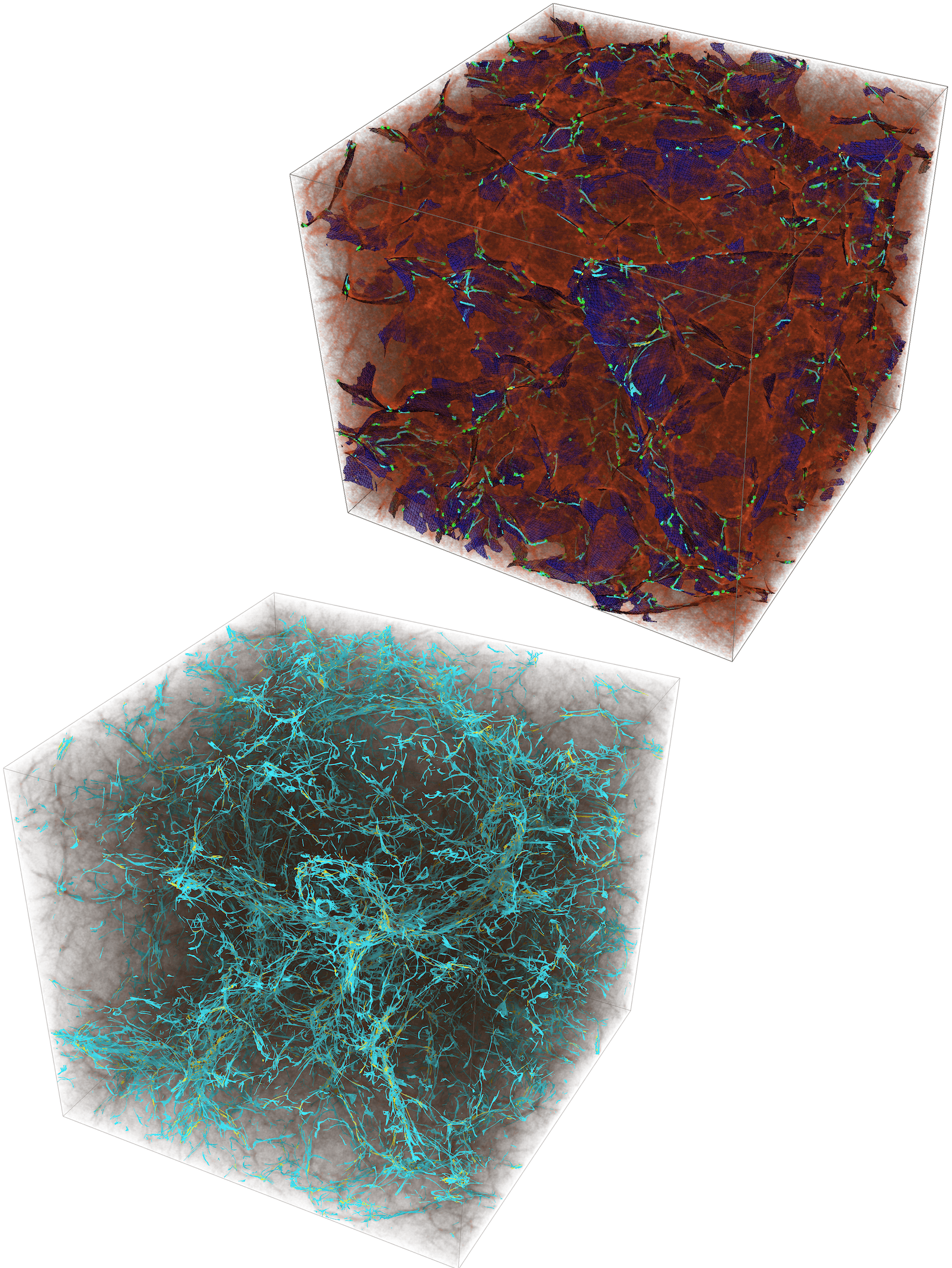}
\vskip -0.25cm
\caption{The log density field of a dark matter $N$-body simulation with $\Lambda$CDM cosmology in a box of $200h^{-1}$ Mpc with $512^3$ particles and elements of the caustic skeleton of the Zel'dovich approximation \cite{Libeskind:2017}. Top right panel: the cusp ($A_3$) sheets (dark blue), the swallowtail ($A_4$) lines (light blue) and the elliptic/hyperbolic umbilic lines (yellow) corresponding to the lowest eigenvalue field of the caustic skeleton. Note that
    the Zel'dovich approximation concerns a potential flow, which means that the eigenvalue fields can be ordered. The initial density field was smoothed on the
  scale $6.3h^{-1}$ Mpc. Bottom left panel: the swallowtail ($A_4$) lines (light blue) and the elliptic/hyperbolic umbilic lines (yellow) corresponding to
  the lowest eigenvalue field of the caustic skeleton. The initial density field is smoothed at $3.1h^{-1}$ Mpc. }\label{fig:3D}
\end{figure}

The resulting weblike structure in Eulerian space is depicted in the corresponding righthand panels. The 
$A_3$ lines, $A_3^\pm$ points and $D_4^\pm$ points are mapped to their Eulerian location by means of the 
Zel'dovich approximation. The red lines, red dots and black triangles represent the Eulerian skeleton corresponding 
to the Zel'dovich approximation. These are superimposed on the density field of the corresponding N-body 
simulations. The comparison between the latter and the Eulerian skeleton reveal that the caustic skeleton -- 
the assembly of $A_3$ lines, $A_3^\pm$ points and $D_4^\pm$ points -- trace the principal elements and connections
of the cosmic web seen in the N-body simulations remarkably well  (see table \ref{tab:table1} for the identification of the lines and points to
the cosmic web). Moreover, by assessing the caustic structure at different resolutions of the density field, one obtains considerable insight
into the multiscale structure and topology of the cosmic web.

\subsection{3-D Caustic Skeleton and Cosmic Web}
\noindent One of the unique features facilitated by the caustic conditions that we have derived in the previous sections is the
ability to go beyond the two-dimensional case and construct and explore the full caustic skeleton of the three-dimensional
mass distribution. In the case of the skeleton of the cosmic web defined by the three-dimensional mass distribution,
the cusp ($A_3$) sheets correspond to the walls or membranes of the large scale structure \cite{bondweb1996,weybond2008,
  cautun2014,Libeskind:2017}. The swallowtail ($A_4$) and elliptic/hyperbolic umbilic ($D_4^\pm$) lines correspond to the filaments of
the cosmic web and the butterfly ($A_5$) and parabolic umbilic ($D_5$) points correspond to the cluster nodes of the network
\cite{bondweb1996,weybond2008,aragon2010b,cautun2014,Libeskind:2017}. The identification of the caustics in the three dimensional cosmic web
is summarized in table~\ref{tab:table1}. 

To appreciate the impressive level at which the caustic skeleton is outlining the three-dimensional weblike
mass distribution, figure~\ref{fig:3D} provides an instructive illustration. The figure depicts elements of the caustic
skeleton of the Zel'dovich approximation in a $200h^{-1}$ Mpc box. The resulting skeleton is superposed on the log density field
of a dark matter $N$-body simulation in a $\Lambda$CDM cosmology with $512^3$ particles \cite{Libeskind:2017}. We should emphasize
that the Zel'dovich approximation is linear and that the corresponding skeleton is completely local in the initial conditions. 
While a full and detailed analysis of these three-dimensional weblike patterns is the subject of an upcoming accompanying paper
\cite{Feldbrugge:2017}, the illustrations of figure~\ref{fig:3D} already give a nice impression of the ability of the caustic
conditions to outline the spine of the cosmic web. 

The top righthand panel contains the cusp ($A_3$) sheet (dark blue colour) and the swallowtail ($A_4$) and elliptic/hyperbolic
umbilic ($D_4^\pm$) lines (light blue colour) corresponding to the lowest eigenvalue field, superimposed on the density field
of the $N$-body simulation (red shaded log density field values). The pattern concerns the caustics obtained for
a displacement field that is filtered at a length scale of $6.3h^{-1}$ Mpc. Close inspection reveals the close correspondence
between the cusp sheets of the caustic skeleton and the flattened - two-dimensional - features in the mass distribution of the
cosmic web. Notwithstandig this, one may also observe that the two-dimensional skeleton does not capture all the structures present
in the $N$-body simulation. This is predominantly an issue of scale, as the corresponding displacement field cannot resolve and trace
features whose size is more refined than the $6.3h^{-1}$ Mpc filter scale.

\medskip
\begin{table}
\center
  \begin{tabular}{llll }
    \hline
    \mbox{Singularity} & \mbox{Singularity} & \mbox{Feature in the } & \mbox{Feature in the}\\
  \mbox{class} & \mbox{name} & \mbox{2D cosmic web}  & \mbox{3D cosmic web} \\ \hline 	
$A_2$		&\mbox{fold} & \mbox{collapsed region} & \mbox{collapsed region} \\ 
$A_3$          &\mbox{cusp} & \mbox{filament}  &\mbox{wall or membrane} \\ 
$A_4$          & \mbox{swallowtail}& \mbox{cluster or knot} & \mbox{filament}  \\ 
$A_5$          &\mbox{butterfly}& \mbox{not stable} & \mbox{cluster or knot} \\ 
 &  \ \\
$D_4$          &\mbox{hyperbolic/elliptic}& \mbox{cluster or knot} & \mbox{filament} \\ 
$D_5$	        &\mbox{parabolic}& \mbox{not stable} & \mbox{cluster or knot} \\ 
\hline
\end{tabular}
\caption{The identification of the different caustics in the $2$- and $3$-dimensional cosmic web}\label{table1}
\label{tab:table1}
\end{table}
\medskip

An impression of the more refined structure can be obtained from the bottom left panel of figure~\ref{fig:3D}, which
follows the line-like elements of the caustic skeleton at a length scale of $3.1 h^{-1}$ Mpc. More specifically, it shows the
swallowtail ($A_4$) and elliptic/hyperbolic umbilic ($D_4^\pm$) lines of the caustic skeleton. The correspondence of these
with the prominent and intricate filamentary pattern in the cosmic mass distribution is even more outstanding than that of the
$A_3$ sheets with the membranes in the density field. It is important to realize, and emphasize, that blue curves were generated using only the eigenvalue field
corresponding to the first collapse. This already creates a filament in the network of caustics, without the need to involve the
second eigenvalue.  In other words, collapse along the second eigenvector is not necessary to create a filament-like structure
(also see \cite{Hidding:2014}). This leads to a radical new insight on structure formation, in that it suggests the different possible late-time morphologies for
filaments \cite{hidding2016}. We may even relate this to the prominence of the corresponding filamentary features: as they
concern features that have experienced collapse along two directions, the umbilic $D_4^\pm$ filaments will have a higher density
and contrast than the filigree of more tenuous $A_4^\pm$ filaments. An additional observation of considerable interest is that the
line-like $A_4$ and $D_4^\pm$ features trace the connectivity of the cosmic web in meticulous detail. 

\subsection{Caustic Density profiles}
\noindent Also of decisive interest in their embedding in the cosmic web, is the expected mass distribution in and around
the various classes of caustics.

Vesilev \cite{Vasilev:1978} inferred the density profiles of the various classes of singularities, in case they emerge 
as a result of potential motion in a collision-less self-gravitating medium. For each of the mass concentrations in and 
around these singularities, he found scale free power-law profiles. The radially average profiles display 
the following decrease of density $\rho(r)$ as a function of radius $r$.

\medskip
\begin{center}
  \begin{tabular}{lll }
    \hline
    \mbox{Singularity} & \mbox{Singularity} & Profile $\rho(r)$\\
    \mbox{class} & \mbox{name} & \\ \hline 	
$A_2$		&\mbox{fold} & $\rho(r) \propto r^{-1/2}$ \\ 
$A_3$          &\mbox{cusp} & $\rho(r) \propto  r^{-2/3}$ \\ 
$A_4$          & \mbox{swallowtail}& $\rho(r) \propto  r^{-3/4}$  \\ 
$A_5$          &\mbox{butterfly}& $\rho(r) \propto  r^{-4/5}$ \\ 
& &  \ \\
$D_4$          &\mbox{hyperbolic/elliptic}& $\rho(r) \propto  r^{-1}$ \\ 
$D_5$	        &\mbox{parabolic}& $\rho(r) \propto  r^{-1} \log{(1/r)}$ \\ 
\hline
\end{tabular}
\end{center}

\medskip
\noindent With respect to these radially averaged profiles, we should realize that the mass 
distribution in and around the singularities is highly anisotropic. This is true for any 
dimension in which we consider the structure around the singularities. 

Notwithstanding this, we do observe that the steepest density profiles are those around the point singularities 
$A_5$ and $D_5$. However, they are mere transient features that will only exist for a single moment in time. The point 
singularities $A_4$ and $D_4$ display a less pronounced behaviour. However, they move over time. Also, we see that 
the cusp singularity $A_3$ possesses a steeper mass distribution that that in and around the sheet singularity $A_2$.

\subsection{Higher order Lagrangian perturbations}
Evidently, the details of the dynamical evolution will bear a considerable influence on the developing caustic structure. 
This not only concerns the dynamics of the system itself, but also its description. The examples that we presented in
the previous sections showed the caustic features developing as the dynamics is predicated on the first-order Lagrangian
approximation of the Zel'dovich formalism \cite{Zeldovich:1970}. The visual comparison with the outcome of the
corresponding $N$-body simulations demonstrated the substantial level of agreement. Nonetheless, given the nature
of singularities, the process of caustic formation might be very sensitive to minor deviations of the mass element
deformations and hence the modelling of the dynamics. This may even strongly affect the predicted population
of caustics and their spatial organization in the skeleton of the cosmic web. Some indications on the level to which the
spatial mass distribution is influenced may be obtained from an early series of papers by Buchert and collaborators
\cite{buchert1992,buchert1993,buchertehlers1993,buchert1994a,buchert1994b}, who were the first to explore the
formation of structure in higher-order Lagrangian perturbation schemes and investigate in how far they would
effect the occurrenc and location of multistream regions. An important finding from their work is that 2nd
order effects are substantial, while 3rd order ones are minimal. Elaborated and augmented by additional
work \cite{bouchet1995,scoccimarro2000}, 2nd order Lagrangian perturbations -- usually designated by the
name 2LPT -- have been established as key ingredients of any accurate analytical modeling of cosmic structure
growth. In a follow-up to the present study, we investigate in detail the repercussions of different analytical
prescriptions for the dynamical evolution of the cosmic mass distribution for the full caustic skeleton of the
cosmic web.

In addition to 2LPT, we will systematically investigate the caustic skeleton in the context of the
\textit{adhesion approximation} \cite{gurbatov1989,shandarinzeld:1989,vergassola1994,gurbatov2012,hidding2012,hidding2018}. Representing 
a fully nonlinear extension of the Zel'dovich formalism, it includes an analytically tractable
gravitational source term for the later nonlinear stages. It accomplishes this via an artificial
viscosity term that emulates the effects of gravity, resulting in the analytically solvable Burger's equation.
With the effective addition of a gravitational interaction term for the emerging structures, unlike the
Zel'dovich approximation the adhesion model is capable of following the hierarchical buildup of structure and the
cosmic web \cite{hidding2012,hidding2016,hidding2018}. At early epochs, the resulting matter streams coincide with
the ballistic motion of the Zel'dovich approximation. At the later stages, as the mass flows approach multistream
regions a solid structure is created at the shell-crossing location. Matter inside these structures is confined to
stay inside, while outside collapsed structures the results from the Zel'dovich approximation and adhesion are identical. The caustics from
the Zel'dovich approximation are compressed to infinitesimally thin structures, hence unifying the Zel'dovich' idea of 
collapsed structures in terms of shell crossing with a hierarchical formation model. While offering a complete model
for the formation and hierarchical evolution of the cosmic web, it does accomplish this by seriously altering the
flow pattern involved in the buildup of cosmic structure. This, in turn, is expected to affect at least to some
extent the properties and evolution of the caustic population and its connectivity. 

\subsection{Gaussian statistics of the caustic skeleton}
In addition to characterizing the geometric and topological outline of the cosmic web in terms of the
caustic skeleton, our study points to another important and related application of the formalism described. The fact
that the linear Zel'dovich approximation provides such an accurate outline of the skeleton of the cosmic web establishes
an important relation between the primordial density and flow field and the resulting cosmic web. 
Via the Zel'dovich approximation, we may relate the caustic skeleton directly to the statistical nature and characteristics
of the primordial density field. In other words, we may directly relate the structure of the cosmic web to the nature of the
Gaussian initial density field. This, in turn, establishes a direct link between the geometric and topological properties
of the cosmic web and the underlying cosmology. Hence a probabilistic analysis of the caustic skeleton may define a path towards
a solidly defined foundation and procedure for using the structure of the observed cosmic web towards constraining global cosmological
parameters and the cosmic structure formation process.

The fact that we may invoke Gaussian statistics facilitates the calculation of a wide range of geometric and topological
characteristics of the cosmic web, as they are directly related to the primordial Gaussian deformation field, its eigenvalues and
eigenvectors. For an example of such a statistical treatment of $2$-dimensional fluids, we refer to \cite{Feldbrugge:2014}. It
describes how one may not only analytically compute the distribution of maxima, or minima, but also the population of singularities and
the length of caustic lines. In an accompaying study, we present an extensive numerical analysis of the statistics of $2$- and $3$-dimensional
caustic skeleton will follow in \cite{Feldbrugge:2017}. This will establish the reference point for the subsequent solid analytical study of
interesting geometric properties of the cosmic web (for the initial steps towards this program see \cite{Feldbrugge:2014b}). 

This will represent a major extension of statistical descriptions that were solely based on the eigenvalue fields. The latter
would make it possible to study the number density of clusters and void basins, make predictions on the statistical properties of
angular momentum, and even several aspects of the cosmic skeleton (e.g. \cite{Doroshkevich:1970,Pogosyan:2009}). As we have argued extensively
in previous sections, it is only by invoking the information contained in the corresponding
eigenvector fields that we may expect to obtain a more complete census of intricate spatial properties of the cosmic web. 

%%%%%%%%%%%%%%%%%%%%%%%%%%%%%%%%%%%%%%%%%%%%%%%%%%%%%%%%%%%%%%%%%%%%%%%%%%%%%%%%%%%%%%%%%%%%%%%%%%%%%%%%%%%%

\section{Dynamics and evolution of caustics}
\label{sec:Dynamics}
\noindent The caustic conditions presented in this study reveal the profound relationship between the various classes of singularities
that may surface in Lagrangian fluids. Besides the aspect of the identification and classification of singularities, we need to have insight
in the transformation and evolution of caustics and caustic networks that accompanies the dynamical evolution of a fluid. The evolution of
the fluid, dictated by the dynamics of the system, generally involves the development of ever more distinctive structures and the proliferation
of complex structural patterns.

Tracing the evolution of a fluid starts at an initial time $t=0$. At that time, the displacement map $s_t$ is the 
zero map. Amongst others, this implies the fluid does not (yet) contain singularities. Starting from these near uniform initial
conditions, the structure in the evolving fluid becomes increasingly pronounced. The phase space sheet that it occupies in
six-dimensional space gets increasingly folded. Its projection on Euclidian space follows this process, and it is
as a result of the folding process that we see the fluid developing singularities. While the dynamical evolution
proceeds to more advanced stages, we not only see the appearance of more singularities, but also the transformation of
one class of singularities into another one. A complementary process that may underlie the changes of local geometry that of
the merging of singularities into a new singularity, itself a manifestation of the hierarchical buildup of structural
complexity. 

The eigenvalue landscape in figure~\ref{fig:eigvalfield} offers an instructive tool for facilitating and
guiding our understanding and visual intuition for the iterative folding of singularities in phase space and
the accompanying caustic transformations. 

\subsection{Caustic mutations and transformations: evolutionary sequence}
\label{sec:unfolding}
\noindent The dynamical evolution of a fluid goes along with a rich palet of local processes. These involve fundamental mutations in the 
local singularity structure that lead to significant topological changes of the spatial pattern forming in the fluid. In some systems 
and situations this will be a key element in the hierarchical buildup of structure.
 
The fundamental notion in these structural mutations in the evolving fluid is that of the ruling dynamics of the system evoking 
changes in the deformation field. Small deformations will lead to the decay of singularities into different ones belonging 
to other singularity classes. Conversely, they may get folded according to a rigid order.

The sequence of singularity mutations is not random and arbitrary. Due to the strict geometric conditions and constraints corresponding to
the various singularities, expressed in the caustic conditions discussed extensively in this study, a given singularity is only allowed to
transform into a restricted set of other singularities. Conversely, a given singularity may only have emanated from a restricted set of
other singularities.

In most situations a particular singularity can have decayed from only one distinctive class of singularities. Some may have 
descended from two other singularity classes. Likewise, most singularities can decay only into one distinctive other class 
of singularity. This is true for all $A$-family singularities. $D$-family singularities have a richer diversity of options, 
with the $D_5$ points being able to decay into 3 different ones, while the $D_4$ points may decay into 2 distinct $A_3$ points. 
The entire singularity transformation and unfolding sequence may be transparently summarized in the unfolding diagram below.

\medskip 

\begin{center}
\begin{tikzpicture}
\node (O) {};
\node [below of=O] (A1) {$A_1$};
\node [right of=O] (A2) {$A_2^i$};
\node [right of=A2] (A3) {$A_3^i$};
\node [right of=A3] (A4) {$A_4^i$};
\node [right of=A4] (A5) {$A_5^i$};
\node [below of=A4] (D4) {$D_4^{ij}$};
\node [below of=A5] (D5) {$D_5^{ij}$};
\node [below of=D5] (A5p) {$A_5^j$};
\node [below of=D4] (A4p) {$A_4^j$};
\node [left of=A4p] (A3p) {$A_3^j$};
\node [left of=A3p] (A2p) {$A_2^j$};
  \draw[<-] (A1) to  (A2);
  \draw[<-] (A2) to  (A3);
  \draw[<-] (A3) to  (A4);
  \draw[<-] (A4) to  (A5);
  \draw[<-] (D4) to  (D5);
  \draw[<-] (A3) to  (D4);
  \draw[<-] (A4) to  (D5);
  \draw[<-] (A3p) to  (D4);
  \draw[<-] (A4p) to  (D5);
  \draw[<-] (A4p) to (A5p);
  \draw[<-] (A3p) to (A4p);
  \draw[<-] (A2p) to (A3p);
  \draw[<-] (A1) to (A2p);
\end{tikzpicture}
\end{center}
\medskip

\noindent The unfolding diagram follows directly from Lagrangian catastrophe theory, although it can also be derived from the caustic conditions. The 
unfoldings of an $A_k^i$ singularities into an $A_{k-1}^i$ singularities, with $k\geq 2$, follow trivially from the caustic conditions. The 
same holds for the unfolding of the $D_5^{ij}$ singularities into the $D_4^{ij}$ singularities. The decay from the $D_4$ to the 
$A_3$ singularities are proven in section \ref{sec:D4}. The mutations $D_5^{ij}\to A_4^i$ and $D_5^{ij}\to A_4^j$ follow directly since the 
shell-crossing of the $D_5^{ij}$ caustic is analogous to the shell-crossing condition on the $A_4^i$ and $A_4^j$ caustics.

\subsection{Singularity transformations}
\label{sec:caustic_transform}
\noindent The principal family of singularities -- principal in terms of rate of occurrence and spatial dominance -- 
is the $A$-family. They are induced by singularities in the geometric structure of one of the eigenvalue fields. 
In physical terms, they involve one-dimensional collapse on to the emerging singularity. Of a more challenging nature within
the evolutionary unfolding of the patterns emerging in fluid flow is the formation of the $D$-family of singularities. They occur
when two fold sheets corresponding to different eigenvalue fields intersect. Amongst others, this means that the $D$ singularities
connect $A$ singularities corresponding to two eigenvalue fields. 

\subsubsection{Evolving $A$-family caustics}
The most prominent and abundant singularities are those of the two-dimensional fold sheets $A^i_2(t)$. In Eulerian space, they 
mark the regions where mass elements are turned inside out as the density attains infinity. This happens while 
they represent the locations where separate matter streams are crossing each other. As time proceeds, the fold sheets 
$A^i_2(t)$ sweep over an increasingly larger Lagrangian region. Ultimately, integrating over time, they mark an entire 
Lagrangian volume, which is labelled as $A_2^i$. The $A_2^i$ set forms a three-dimensional variety. 

When we wish to identify where a particular individual fold sheet is born, we turn to the cusp points $A_3^{i+}$. They are 
the points on the fold sheets where the corresponding eigenvalue field attains an extremum. Because of this, they mark the sites 
of birth of the fold singularities. As the $A_2^i(t)$ sheets unfold, at the edges their surface gets wrapped in 
a higher order singularity, the cusp curves $A_3^i(t)$. In time, these curves move through space and trace out cusp sheets 
$A_3^i$. In the context of the Megaparsec scale matter distribution in the Universe, the cusp sheets are to be associated 
with the walls or membranes in the cosmic web \cite{bondweb1996,weybond2008,aragon2010b,cautun2014,Libeskind:2017}. 

\bigskip
A dynamically interesting process occurs at the cusp points $A_3^{i-}$, which are the saddle points of the corresponding 
eigenvalue field $\mu_{ti}$ that at a given time are encapsulated by the fold sheet $A_2^i$. At the $A_3^{i-}$ points, we see 
the merging or annihilation of fold sheets $A_2^i$ into a larger structure (cf. figure~\ref{fig:eigvalfield}). Mathematically,
they mark the key locations where the topology of the eigenvalue field changes abruptly. Physically, they are associated with
the merging of separate structural components, a manifestation of the hierarchical buildup of structural complexity
\cite{weybond2008,cautun2014}. 

\bigskip
Also the cusp curves $A_3^i(t)$ can get folded. In Eulerian space, the folding of the cusp curves manifests itself as 
$A_4^i(t)$ swallowtail points. As time proceeds, these points move through space and define the swallowtail curve $A_4^i$. 
It is of interest to note that the swallowtail curve is embedded in the cusp sheet, i.e. $A_4^i \subset A_3^i$. 
In the context of the cosmic structure formation process, the swallowtail curves outline and trace perhaps the most outstanding feature 
of the cosmic web, the pronounced elongated filaments that form the of spine the weblike network
\cite{weybond2008,aragon2010b,cautun2014}. 

Also these features build up in a hierarchical process of small filaments merging into ever larger and more prominent arteries. 
In the context of the evolving singularity structure that we study, this process is represented by the $A_4^{i+}$ points and $A_4^{i-}$ 
points. They define the decisive junctions where significant changes in topology occur. For the $A_4^{i\pm}$ points this concerns their 
identity in the gradient of the eigenvalue field, in which the $A_4^{i+}$ are maxima and minima and $A_4^{i-}$ points are the 
saddle points. The implication of this is that cusp curves get created or annihilated at $A_4^{i+}$ points, while they 
merge or separate at $A_4^{i-}$ points.

\bigskip
The final morphological constituent in this structural hierarchy of singularities is that of the butterfly points $A_5^i$. 
They conclude the $A$-family of singularities, i.e. the family of singularities that correspond to the spatial characteristics 
of the field of one eigenvalue $\mu_i$. The swallowtail curves $A_4^i$ get folded at $A_5^i$ butterfly points. In the 
three-dimensional structural pattern that formed in the fluid, these will represent nodes. In the cosmic web, they define 
the nodal junctions, connecting to the various filamentary extensions that outline its spine \cite{bondweb1996,
  Colberg:2005,weybond2008,aragon2010b,cautun2014}. In principle, for a given initial field and dynamical evolution, one might use
these identifications to e.g. evaluate how many filaments are connected to the network nodes \cite{aragon2010,Pogosyan:2009}. 

\subsubsection{Evolving $D$-family caustics}
The $A_2^i(t)$ and $A_2^j(t)$ sheets, with $i\neq j$, intersect in the elliptic and hyperbolic umbilic points $D_4^{\pm ij}(t)$. 
In contrast to the $A$ family of singularities, the collapse into $D$ singularities is two-dimensional. It leads to the 
birth of the socalled {\it umbilic} points. Over time, they trace out the umbilic curve $D_4^{\pm ij}$. The collapse 
process may occur in two distinctive ways, indicated by the labels $+$ and $-$. 

The topology of the variety $D_4^{\pm ij}(t)$ changes at $D_4^{ij\pm}$ and $D_5$ points. An interesting characteristic 
of umbilic curves is that they are always created or annihilated in pairs. The $D_4^{ij\pm}$ points correspond to the creation 
or annihilation of two $D_4^{\pm ij}$ curves of the same signature. By contrast, the $D_5^{ij}$ points correspond to the 
creation or annihilation of a pair with one $D_4^{+ij}$ and one $D_4^{-ij}$ point.

%%%%%%%%%%%%%%%%%%%%%%%%%%%%%%%%%%%%%%%%%%%%%%%%%%%%%%%%%%%%%%%%%%%%%%%%%%%%%%%%%%%%%%%%%%%%%%%%%%%%%%%%%%%%

\section{Discussion \& Conclusions}\label{sec:Conclusion}
\noindent In this study we have developed a general formalism for identifying the caustic structure of a dynamically
evolving mass distribution, in an arbitrary dimensional space. Through a new and direct derivation of the caustic conditions
for the classification and characterization of singularities that will form in an evolving matter field, our study enables the
practical implementation of a toolset for identifying the spatial location and outline of each relevant class of emerging
singularities. By enabling the development of such instruments, and the application of these to any cosmological primordial
density and velocity field, our study opens the path towards further insight into the dynamics of the formation and evolution
of the morphological features populating the cosmic web. In particular significant is that it will enable us to obtain
a fundamental understanding of the spatial organization of the cosmic web, i.e. of the way in which these structural components are
arranged and connected. 

\subsection{Phase-Space structure of the Cosmic Web}
\noindent Caustics are prominent features emerging in advanced stages of dynamically evolving fluids. They mark the positions where
fluid elements cross and multi-stream regions form. They are associated with regions of infinite density, and
often go along with the formation of shocks. In the context of the gravitationally evolving mass distribution in
the universe, caustics emerge in regions in which nonlinear gravitational collapse starts to take place. As such,
they are a typical manifestation of the structure formation process at the stage where it transits from the initial
linear evolution to that of more advanced nonlinear configurations involving gravitational contraction and collapse. 
The overall spatial organization of matter at the corresponding scale is that of the cosmic web, which assembles flattened
walls, elongated filaments and tendrils and dense, compact cluster nodes in an intricate multiscale weblike network that
pervades the Universe.  

Over the past decades our understanding of the formation and evolution of the cosmic web has advanced considerably.
The availability of large computer simulations have been instrumental in this, as they enabled us to follow the cosmic
structure formation process in detail (see e.g. \cite{springmillen2005,illustris2014,eagle2015}). In combination with new theoretical
insights \cite{bondweb1996,weybond2008}, this has led to the development of a general picture of the emergence of
the weblike matter and galaxy distribution. The full phase-space dynamics of the process and its manifestation in the
emerging matter distribution is an instrumental aspect of this that only recently received more prominent attention. While the
study by Zel'dovich \cite{Zeldovich:1970} already underlined the importance of a full phase-space description
for understanding cosmic structure formation (see also \cite{shandarinzeld:1989,shandarinsuny:2009}), with the exception of a
few prominent studies \cite{Arnold:1982a} the wealthy information content of full 6-D phase-space escaped attention.

A series of recent publications initiated a resurgence of interest in the phase-space aspects of the cosmic structure

formation process. They realized that the morphology of components in the evolving matter distribution is closely related
to its multistream character \cite{abel2012,falck2012,neyrinck2012,shandarin2012,ramachandra2015} (for an early
study on this observation see \cite{buchertehlers1993}). This realization is based on the recognition that the emergence of
nonlinear structures occurs at locations where different streams of the corresponding flow field
cross each other. Looking at the appearance of the evolving spatial mass distribution as a 3D {\it phase space sheet} folding itself
in 6D phase space, this establishes a connection between the structure formation process and the morphological classification
of the emerging structure. Moreover, to further our understanding of the dynamical evolution and buildup of the cosmic matter
distribution, we also need to answer the question in how far the various emerging structural features connect up in the overall
weblike network of the cosmic web.

\subsection{Singularities and Caustics}
\noindent To be able to answer the questions, we study the emergence of singularities and caustics in a dynamically evolving mass
distribution. Our analysis is built on the seminal work by Arnol'd, specifically his classification of singularities in Lagrangian
catastrophe theory. In a three-dimensional setting we can recognize two series of singularities, the $A_k$ and $D_k$ series. The 4 classes of
$A_k$ singularities -- $A_2$, $A_3$, $A_4$ and $A_5$ -- are the singularities for which the caustic condition holds for one eigenvalue.
The $D$-family of umbilic singularities -- including the $D_4^+$, $D_4^-$ and $D_5$ -- are caustics for which the caustic conditions
are satisfied by two eigenvalue simultaneously. In three-dimensional fluids, the case in which all three eigenvalues simultaneously
satisify the caustic conditions, the $E$-family caustics, is non-degenerate.

In order to detect these caustics in practice, we derived the \textit{caustic conditions}, which classify them in terms of both \textit{eigenvalue} and the \textit{eigenvector} fields of the deformation tensor. The derivation differs from the classical derivation of catastrophe theory, in terms of generating functions and the classification of its degenerate critical points, in that we work with the geometry of the system. Moreover, the caustic conditions are not restricted to Hamiltonian dynamics and apply to all systems which allow for a description with a sufficiently differentiable Lagrangian map.

\subsection{Caustic Skeleton and Cosmic Web}
\noindent On the basis of the derived formalism, we show how the caustics of a Lagrangian fluid form an intricate skeleton of  the
nonlinear evolution of the fluid. The family of newly derived caustic conditions allow a significant extension and elaboration
of the work described in Arnold et al. (1982) \cite{Arnold:1982a}. Arnol'd et al.\, classified the caustics that develop in one- and two-dimensional systems that
evolve according to the Zel'dovich approximation. While \cite{Arnold:1982b} did offer a qualitative description of caustics
in the three-dimensional situation, this did not materialize in a practical application to the full three-dimensional cosmological setting.
The expressions derived in our study, and the specific identification of the important role of the deformation tensor eigenvectors,
have enabled us to breach this hiatus. To identify the full spatial distribution and arrangement of caustics in the evolving
three-dimensional cosmic matter distribution, we follow the philosophy exposed in the two-dimensional study by Hidding et al. 2014
\cite{Hidding:2014,Feldbrugge:2014}. By relating the singularity distribution to the spatial properties of the initial Gaussian deformation
field, \cite{Hidding:2014} managed to identify and show the spatial connectivity of singularities and establish how in a hierarchical
evolutionary sequence they evolve and may ultimately merge with surrounding structures. 

\bigskip
When applied to the Zel'dovich approximation for cosmic structure formation, the caustic conditions form a skeleton of the caustic
web.  In the context of the cosmic web, we may identify these singularities with different components. This observation
by itself leads to some radically new insights into the origin of the structural features in the cosmic web. The $A_3$ cusp
singularities are related to the \textit{walls} of the skeleton of the comsic web. The $A_4$ swallowtail singularities trace the
filamentary ridges and tendrils in the cosmic web. Also the $D_4^\pm$ hyperbolic and elliptic umbilic singularities are related to
the filamentary spine of the spine, as they define the dense filamentary extensions of the cluster nodes. The butterfly ($A_5$) and
parabolic umbilic ($D_5$) singularities are both connected with the nodes of the weblike pattern. One immediate observation of
considerable interest is that the line-like $A_4$ and $D_4^\pm$ features trace the connectivity of the cosmic web in meticulous detail.
Perhaps equally or even more interesting, and of key importance for our understanding of the dynamical evolution of the cosmic web,
is the observation that both filaments and tendrils, as well as nodes, may have formed due to the folding by the phase-space sheet induced by
only one deformation eigenvalue: the filamentary $A_4$ caustics and nodal $A_5$ caustic belong to the one eigenvalue $A$ family of caustics.
In other words, collapse along the second eigenvector is not necessary to create a filament-like structure, and not even collapse along both
second and third eigenvector is needed for the appearance of nodes (see \cite{Hidding:2014,hidding2016}). This is a new insight as it suggests
the existence of different possible late-time morphologies for filaments and nodes \cite{hidding2016}. 

A realization of key importance emanating from our work is that it is not sufficient to limit a structural analysis to the eigenvalues
of the deformation tensor field. Usually neglected, we argue -- and show by a few examples -- that it is necessary to include the information
contained in the (local) deformation tensor eigenvectors, our study has demonstrated and emphasized that for the identification of the full
spatial outline of the cosmic web's skeleton.  In an accompanying numerical study of the caustic skeleton in cosmological $N$-body simulations,
we illustrate how essential it is to invoke the deformation eigenvectors in the analysis \cite{Feldbrugge:2017}. This study will present a numerical
and statistical comparison between the matter distribution in the simulation and the caustic skeleton of the three-dimensional cosmic web.\\

\subsection{Extensions and Applications}
\noindent Amongst the potentially most important applications of the current project is the fact that the caustic skeleton inferred
from the Zel'dovich approximation adheres closely to the spine of the full nonlinear matter distribution. The direct implication is
that we may directly link the outline of the cosmic web to the initial Gaussian density and velocity field. On the basis of the
corresponding deformation field, one may then attempt to calculate a range of properties analytically.
The fact that we may invoke Gaussian statistics facilitates the calculation of a wide range of geometric and topological
characteristics of the cosmic web, as they are directly related to the primordial Gaussian deformation field, its eigenvalues and
eigenvectors. The first step towards this program were taken by \cite{Feldbrugge:2014b}. A few examples of results of such a statistical
treatment for $2$-dimensional fluids are described in \cite{Feldbrugge:2014}. It describes how one may not only analytically compute
the distribution of maxima, or minima, but also the population of singularities and the length of caustic lines. 
This will represent a major extension of statistical descriptions that were solely based on the eigenvalue fields
(see e.g. \cite{Doroshkevich:1970,Pogosyan:2009}). Moreover, the ability to infer solid analytical results for a
range of parameters quantifying the cosmic web will be a key towards identifying properties of the cosmic web that
are sensitive to the underlying cosmology. This, in turn, would enable the use of these properties to infer
cosmological parameters, investigate the nature of dark matter and dark energy, trace the effects of deviations from
standard gravity, and other issues of general cosmological interest.

Notwithstanding the observation that the caustic skeleton inferred from the Zel'dovich approximation appears to
closely adhere to the full nonlinear structure seen in $N$-body simulations, an aspect that still needs to
be addressed in detail is the influence of the dynamical evolution on the the developing caustic structure. 
This concerns in particular the description of the dynamics of the system. Given the nature
of singularities, the process of caustic formation might be very sensitive to minor deviations of the mass element
deformations and hence the modelling of the dynamics. This may even strongly affect the predicted population
of caustics and their spatial organization in the skeleton of the cosmic web. The Zel'dovich formalism \cite{Zeldovich:1970}
is a first-order Lagrangian approximation. A range of studies have shown that second order
Lagrangian descriptions, often named 2LPT, provide a considerably more accurate approximation of in
particular the mildly nonlinear phases that are critical for understanding the cosmic web
\cite{buchert1992,buchertehlers1993,buchert1994a,bouchet1995,scoccimarro2000}. In addition to a follow-up
study in which we explore the caustic structure according to 2LPT and possible systematic differences with
that predicated by the Zel'dovich approximation, we will also systematically investigate the caustic skeleton
in the context of the adhesion formalism \cite{gurbatov1989,gurbatov2012,hidding2012,hidding2018}. Representing 
a fully nonlinear extension of the Zel'dovich formalism through the inclusion of an effective gravitational interaction
term for the emerging structures, it is capable of following the hierarchical buildup of structure. While it
provides a highly insightful model for the hierarchically evolving cosmic web, it also affects the flow patterns
and hence the multistream structure in the cosmic mass distribution. In how far this will affect the caustic
skeleton remains a major question for our work. 

Finally, of immediate practical interest to our project will be identification of the various classes of
singularities that are populating the Local Universe. On the basis of advanced Bayesian reconstruction
techniques, various groups have been able to infer constrained realizations of the implied Gaussian
primordial density and velocity field in a given cosmic volume \cite{jasche2010,kitaura2013,leclercq2015,Leclercq:2017}.
From these constrained initial density and deformation fields, we may subsequently determine the
caustic structure in the Local Universe (see e.g. \cite{hidding2016}). The resulting caustic
skeleton of the local cosmic web may then be confronted with the structures -- clusters, groups and galaxies -- that surveys have
observed. Ultimately, this will enable us to reconstruct the cosmic history of objects and structures
in the local Universe. 

\subsection{Summary}
\noindent In summary, the ability to relate the formation and hierarchical evolution of structure in the Universe to the tale of the
emergence and fate of singularities in the cosmic density field provides the basis for a dynamical
theory for the development of the cosmic web, including that of its substructure. This will be the principal
question and subject of the sequel to the work that we have presented here. 

%%%%%%%%%%%%%%%%%%%%%%%%%%%%%%%%%%%%%%%%%%%%%%%%%%%%%%%%%%%%%%%%%%%%%%%%%%%%%%%%%%%%%%%%%%%%%%%%%%%%%%%%%%%%

\section*{Acknowledgements}
We thank Sergei Shandarin for having raised our interest in caustics as a 
key towards the dynamical understanding of the cosmic web. We are very grateful to
Bernard Jones for a careful and diligent appraisal of the manuscript, and for the many useful and
illuminating discussions and comments. We also thank Adi Nusser, Neil Turok, and Gert Vegter for many
encouraging discussions and the anonymous referee for helpful comments. JF acknowledges the Perimeter Institute 
for facilitating this research through the support by the Government of Canada through the Department 
of Innovation, Science and Economic Development Canada and by the Province of Ontario through the Ministry
of Research, Innovation and Science. 

%%%%%%%%%%%%%%%%%%%%%%%%%%%%%%%%%%%%%%%%%%%%%%%%%%%%%%%%%%%%%%%%%%%%%%%%%%%%%%%%%%%%%%%%%%%%%%%%%%%%%%%%%%%%

\bibliographystyle{plain}

\bibliography{mybibliography}

%%%%%%%%%%%%%%%%%%%%%%%%%%%%%%%%%%%%%%%%%%%%%%%%%%%%%%%%%%%%%%%%%%%%%%%%%%%%%%%%%%%%%%%%%%%%%%%%%%%%%%%%%%%%

\appendix
\section{Zel'dovich approximation}
\label{sec:Zel'dovich}
\noindent The Zel'dovich approximation is the first order approximation of a Lagrangian pressureless fluid evolving under self gravity, \cite{Zeldovich:1970}. The Zel'dovich approximation is the simplest example of a Lagrangian fluid with Hamiltonian dynamics and serves as a good illustration of the caustic conditions. The displacement map of the Zel'dovich approximation factors into a term depending on time and a term depending on the initial conditions
\begin{equation}
s_t(q)=-b_+(t) \nabla_q \Psi(q),
\end{equation}
with the linearized velocity potential $\Psi(q)$ and growing mode $b_+(t)$. The growing mode can be obtained from linear Eulerian perturbation theory. Up to linear order, the linearized velocity potential is proportional to the linearly extrapolated gravitational potential at the current epoch $\phi_0(q)$, i.e.
\begin{equation}
\Psi(q)=\frac{2}{3\Omega_0 H_0^2}\phi_0(q),
\end{equation}  
with current Hubble constant $H_0$ and current energy density $\Omega_0$. The linearized velocity potential $\Psi(q)$ encodes the initial conditions while the growing mode $b_+(t)$ encodes the cosmological evolution of the fluid. For the Zel'dovich approximation it is common to define the deformation tensor as
\begin{equation}
\psi_{ij} = \frac{\partial^2 \Psi(q)}{\partial q_i \partial q_j}
\end{equation}
with eigenvalues ${\lambda}_i(q)$ satisfying $\mu_i(q,t) = - b_+(t) {\lambda}_i(q)$. The density in the Zel'dovich approximation can be expressed as 
\begin{eqnarray}
\rho(x',t)&=&\sum_{q \in A(x',t)} \frac{\rho_i(q)}{|1-b_+(t)\lambda_1(q)||1-b_+ (t) \lambda_2(q)||1- b_+(t)\lambda_d(q)|},
\label{LagrangianDensityZel'dovich}
\end{eqnarray}
with $\rho_i$ the initial density field. Caustics occur at $q$ at time $t$ if and only if 
\begin{equation}
\lambda_i(q) = \frac{1}{b_+(t)} \label{eq:surface}
\end{equation}
 for at least one $i$. The eigenvalues $\lambda_i$ are functions determined by the initial gravitational field. Equation \eqref{eq:surface} can be pictured as a hyperplane at height $1/b_+(t)$. Since the Zel'dovich approximation concerns potential flow, which means that the eigenvalues are real and can be ordered such that $\lambda_1 \geq \lambda_2 \geq \lambda_3$. The intersection of this plane with the graph of the eigenvalues undergoes shell-crossing at that time. For the Zel'dovich approximation the caustic conditions in terms of the eigenvalues $\lambda_i$ are given by
\begin{eqnarray}
A_1&=&\{ q\in L| {\lambda}_i(q)\neq1/b_+(t)\mbox{ for all } t \mbox{ and } i \},\\
A_2^i(t)   &=&\{q \in L| {\lambda}_{i}(q) = 1/b_+(t)\},\\
A_3^{i}(t) &=&\{q \in L| q \in A_2^i(t) \mbox{ and } \lambda_{i,i}(q)=0\},\\
A_4^{i}(t) &=&\{q \in L| q \in A_3^i(t) \mbox{ and } \lambda_{i,ii}(q)=0\},\\
A_5^{i}(t) &=&\{q \in L| q \in A_4^i(t) \mbox{ and } \lambda_{i,iii}(q)=0\},\\
D_4^{\pm ij}(t)&=&\{q \in L| \lambda_i(q)=\lambda_j(q)=1/b_+(t) \mbox{ and sign}(S_\defm) = \pm 1\},\\
D_5^{ij}(t)&=&\{q \in L| q \in D_4^{ij}(t) \mbox{ and } (\lambda_{i}-\lambda_j)_{,i}(q) = (\lambda_{i}-\lambda_j)_{,j}(q) = 0 \},
\end{eqnarray}
and the points at which the topology of above sets changes
\begin{eqnarray}
A_3^{i+}&=&\{q \in L| q \in A_2^i \wedge \lambda_i(q) \mbox{ max-/minimum of } {\lambda}_i\},\\
A_3^{i-}&=&\{q \in L| q \in A_2^i \mbox{ saddle point of } {\lambda}_i\},\\
A_4^{i+}&=&\{q \in L| q \in A_3^i \wedge \lambda_{i,ii}(q)\mbox{ max-/minimum of }\lambda_{i,ii}|_{A_2}\},\\
A_4^{i-}&=&\{q \in L| q \in A_3^i\mbox{ saddle point of }\lambda_{i,ii}|_{A_2}\},\\
D_4^{ij \pm }&=&\{q\in L| q \in D_4^{\pm ij} \wedge\lambda_i(q)=\lambda_j(q) \mbox{ max-/minimum of } \lambda_i|_{D_4^{\pm ij}} = \lambda_j|_{D_4^{\pm ij}} \}.
\end{eqnarray}
with the direction derivatives $\lambda_{i,i}=\nabla \lambda_i \cdot v_i,\lambda_{i,ii}=\nabla \lambda_{i,i} \cdot v_i$ and $\lambda_{i,iii}=\nabla \lambda_{i,ii} \cdot v_i$. Note that the eigenvectors are defined modulo multiplication by a real number and really represent lines.

%%%%%%%%%%%%%%%%%%%%%%%%%%%%%%%%%%%%%%%%%%%%%%%%%%%%%%%%%%%%%%%%%%%%%%%%%%%%%%%%%%%%%%%%%%%%%%%%%%%%%%%%%%%%

\section{Non-diagonalizable deformation tensors}
\label{Ap:Non-diagonalizable}
In sections \ref{sec:ShellCrossing} and \ref{sec:CC} we derived the shell-crossing conditions and caustic conditions under the assumption that the deformation tensor $\frac{\partial s_t}{\partial q}=\mathcal{M}$ is diagonalizable. We here extend these conditions to non-diagonalizable deformation tensors.

The eigenvalues $\mu_i$ are the roots of the characteristic function $\chi(\lambda)=\det\left(\mathcal{M} - \lambda I\right)$ corresponding to the deformation tensor. The deformation tensor is diagonalizable if and only if the algebraic multiplicity -- the order of the root -- is equal to the geometric multiplicity -- the number of eigenvectors corresponding to the root -- for all eigenvalues. Hence, in order for the deformation tensor to be non-diagonalizable, two or more eigenvalues need to coincide while there are fewer corresponding eigenvectors. In this case we can extend the set of eigenvectors by adding the necessary generalized eigenvectors to put the deformation tensor in Jordan normal form
\begin{equation}
\label{eq:Jordan}
\mathcal{M} = \mathcal{V} \mathcal{M}_J \mathcal{V}^{-1}
\end{equation}
where $\mathcal{V}$ is the generalized modal matrix consisting of the eigenvectors and generalized eigenvectors and $\mathcal{M}_J$ is the upper triangular matrix of Jordan normal form containing the eigenvalues.

In the three-dimensional case, these non-diagonalisable deformation tensors correspond to the hyperbolic/elliptic umbilic ($D_4^\pm$) caustics. For simplicity lets restrict to the three-dimensional case where the shell-crossing occurs due to the first and the second eigenvalue fields $1+\mu_1=1+\mu_2 =0$. In this case, we extend the set of eigenvectors $\{v_1,v_3\}$ by adding the generalized eigenvector $\bar{v}_2$. The Jordan matrix now takes the form
\begin{equation}
\mathcal{M}_J = \begin{pmatrix} \mu_1 & 1 & 0 \\ 0 & \mu_2 & 0 \\ 0 & 0 & \mu_3 \end{pmatrix}\,.
\end{equation}
Substituting equation \eqref{eq:Jordan} in equation \eqref{eq:con4} we obtain the condition that there needs to exist a non-zero tangent vector of $D_4^{\pm 12}$ for which 
\begin{align}
(1+\mu_1) v_1^* \cdot T + \bar{v}_2^*\cdot T &= 0\\
(1+\mu_2) \bar{v}_2^* \cdot T &= 0\\
(1+\mu_3) v_3^* \cdot T &= 0\,,
\end{align}
replacing equations \eqref{eq:Diagonal}. We thus see that the $D_4^{12}$ variety forms a $D_5^{12}$ caustic if and only if $\bar{v}_2^*\cdot T=0$ and $v_3^* \cdot T=0$ (for a diagonalizable deformation tensor we obtain the second condition). This is equivalent to the condition that $T$ is parallel to the eigenvector $v_1$.

Finally note that the deformation tensor can only be non-diagonalizable for non-Hamiltonian dynamics for which the parabolic umbilic caustic ($D_5$) is not stable (see section \ref{sec:condnH}). This condition is thus not relevant for Hamiltonian and generic non-Hamiltonian Lagrangian fluids in three dimensions.

This analysis straightforwardly generalizes to the case in which the geometric and algebraic multiplicity of the eigenvalues differs by more than one for higher dimensional fluids.
%%%%%%%%%%%%%%%%%%%%%%%%%%%%%%%%%%%%%%%%%%%%%%%%%%%%%%%%%%%%%%%%%%%%%%%%%%%%%%%%%%%%%%%%%%%%%%%%%%%%%%%%%%%%

\section{Shell-crossing conditions: coordinate transformation}
\label{app:coordtransf}
\noindent The shell-crossing conditions are manifestly independent of coordinate choices. However, the eigenvalue and eigenvector fields generally do 
depend on the choice of coordinates. By themselves, they do therefore not provide valid descriptions of the dynamics of the fluid. Suppose the 
displacement field can be written as $s = \nabla \psi$ for some potential $\psi$. The Hessian $H_x$ of $\psi$,
\begin{equation}
H_{ij}\,=\,\frac{\partial^2 \psi}{\partial x_i \partial x_j}\,,
\end{equation}
transforms non-trivially under the local coordinate transformation $x\to X(x)$ i.e. 
\begin{equation}
H \to \tilde H\,=\, J^T H  J + J^T \nabla (J) \nabla \psi\,,
\end{equation}
with $J$ the Jacobian between the coordinate systems $X$ and $x$,
\begin{equation}
J_{ij}\,=\,\frac{\partial X_i}{\partial x_j}\,. 
\end{equation}
From this we immediately infer that the eigenvalue field and eigenvector fields are invariant if the transformation is
orthogonal and global, i.e. if $J^T=J^{-1}$ and $\nabla(J)=0$. As may be expected, these transformations include rotations and translations.

%%%%%%%%%%%%%%%%%%%%%%%%%%%%%%%%%%%%%%%%%%%%%%%%%%%%%%%%%%%%%%%%%%%%%%%%%%%%%%%%%%%%%%%%%%%%%%%%%%%%%%%%%%%%

\section{Lagrangian maps and Lagrangian equivalence}
\label{Ap:LE}
\noindent We here shortly describe the mathematical background of symplectic manifolds, Lagrangian manifolds and Lagrangian maps. For a detailed description and derivations we refer to \cite{Arnold:2012b,Arnold:2012a}.\\

\subsection{Symplectic manifolds and Lagrangian maps}
\noindent A $2n$-dimensional symplectic manifold $(M,\omega)$ is a smooth $2n$-dimensional manifold $M$, equipped with a closed nondegenerate bilinear 2-form $\omega$ called the symplectic form. Symplectic manifolds are always even dimensional for $\omega$ to be nondegenerate. In Hamiltonian dynamics the symplectic form $\omega$ can be associated to the Poisson brackets which encodes the dynamics of the theory. A Lagrangian manifold $L$ of a $2n$-dimensional symplectic manifold $(M,\omega)$ is a $n$-dimensional submanifold of $M$ on which the symplectic form $\omega$ vanishes. Let $(B,\pi)$ be a Lagrangian fibration of $(M,\omega)$, which is a $n$-dimensional manifold with a projection map $\pi:M\to B$ for which the fibers $\pi^{-1}(b)$ are Lagrangian manifolds for all $b\in B$.\\
\indent An example of a symplectic manifold is phase space consisting of position and canonical momenta $(q_1,\dots,q_n,p_1,\dots p_n)$ with the symplectic form $\omega = \sum_{i}^n \mathrm{d}q_i\wedge \mathrm{d}p_i$. An example of a Lagrangian fibration is $\{(q_1,\dots,q_n),\pi\}$ with the projection map $\pi(q_1,\dots,q_n,p_1,\dots p_n)=(q_1,\dots, q_n)$. \\
\indent Give a symplectic manifold $(M,\omega)$ with a Lagrangian fibration $(B,\pi)$ we can for every Lagrangian manifold $L$ define a Lagrangian map $(\pi\circ{} i):L\to M \to B$, with $i$ being the inclusion map sending $L$ into $M$. Two Lagrangian maps $(\pi_1\circ{} i_1):L_1\to M_1 \to B_1$ and $(\pi_2\circ{} i_2):L_2\to M_2 \to B_2$ are defined to be Lagrangian equivalent if there exist diffeomorphisms $\sigma,\tau$ and $\nu$ such that $\tau \circ{} i_1= i_2\circ{}\sigma, \nu \circ{}\pi_1=\pi_2\circ{}\tau$ and $\tau^*\omega_2=\omega_1$, or equivalently the diagram below commutes

\begin{center}
\begin{tikzpicture}
\node at (0,0) (L1) {$L_1$};
\node at (1.75,0) (M1) {$(M_1,\omega_1)$};
\node at (3.5,0) (B1) {$B_1$};
\node [below of=L1] (L2) {$L_2$};
\node [below of=M1] (M2) {$(M_2,\omega_2)$};
\node [below of=B1] (B2) {$B_2$};
\draw[->] (L1) to  (M1);
\draw[->] (M1) to  (B1);
\draw[->] (L2) to  (M2);
\draw[->] (M2) to  (B2);
\draw[->] (L1) to  (L2);
\draw[->] (M1) to  (M2);
\draw[->] (B1) to  (B2);

\draw[->](L1) to node [above] {$i_1$} (M1);
\draw[->](M1) to node [above] {$\pi_1$} (B1);
\draw[->](L2) to node [above] {$i_2$} (M2);
\draw[->](M2) to node [above] {$\pi_2$} (B2);

\draw[->](L1) to node [left] {$\sigma$} (L2);
\draw[->](M1) to node [left] {$\tau$} (M2);
\draw[->](B1) to node [left] {$\nu$} (B2);
\end{tikzpicture}
\end{center}

\subsection{Displacement as Lagrangian map}
\label{app:displlagr}
Given a Lagrangian submanifold $\mathcal{L}$ we can construct a corresponding Lagrangian map. First map the Lagrangian submanifold $\mathcal{L}$ with the inclusion map $i:\mathcal{L} \to \phasespace$ to the corresponding points in phase space $\phasespace$, i.e., $i: (q,x) \mapsto (q,x)$ for all $(q,x)\in \mathcal{L}$. Subsequently map these points to a base manifold $B$ with the projection map $\pi:\phasespace\to B$. In Lagrangian fluid dynamics it is convenient to pick the Eulerian manifold $E$ as the base manifold $B$ and define the projection map as $\pi: (q,x) \mapsto x$ for all $(q,x)\in \phasespace$. As there will always be an exact correspondence between the Lagrangian manifold $L$ and the Lagrangian submanifold $\mathcal{L}_t \subset \phasespace$ (there exists a unique point $x\in E$ such that $(q,x)\in \mathcal{L}_t$ for every $q\in L$), we can associate the Lagrangian map corresponding to $\mathcal{L}_t$ with the map $x_t$. In summary, the map $x_t$ corresponds uniquely to a Lagrangian map for fluids with Hamiltonian dynamics.\\

\indent A Lagrangian map can develop regions in which multiple points in the Lagrangian manifold are mapped to the same point in the base space. The points at which the number of pre-images of the Lagrangian map changes are known as Lagrangian singularities. Lagrangian catastrophe theory classifies the stable singularities, stable with respect to small deformations of $\mathcal{L}$, up to Lagrangian equivalence. Lagrangian equivalence is a generalization of equivalence up to coordinate transformations. For a precise definition of Lagrangian equivalence we refer to appendix \ref{Ap:LE}.\\

\subsection{Lagrangian map germs}
\noindent In catastrophe theory it is important to consider the Lagrangian map at a point. This is achieved by means of Lagrangian germs. Starting with a point $p\in M$ we can consider Lagrangian functions $F_i:U_i\to B$ for $i=1,2$ for small environments $U_i$ of $p$ which coincide on the intersection $U_1 \cap U_2$. The equivalence classes of such Lagrangian functions are Lagrangian germs. The Lagrange equivalence of Lagrangian maps straightforwardly extends to Lagrange equivalence of Lagrangian germs. These are the equivalence classes used in the classification of stable Lagrangian maps, where a Lagrangian germ is stable if and only if every sufficiently small fluctuation on the germ is Lagrange equivalent to the germ.\\

\subsection{Gradient maps}
\indent Every Lagrangian germ is Lagrange equivalent to the germ of a gradient map. That is to say, for every Lagrangian map $l=\pi \circ i:\mathcal{L} \to \phasespace \to E$ we can for a point $(q,x)\in \mathcal{L}$ locally write the map as
\begin{equation}
l(q_1,\dots, q_n,x_1,\dots,x_n)= \left(\frac{\partial S}{\partial q_1},\frac{\partial S}{\partial q_2},\dots,\frac{\partial S}{\partial q_n}\right)
\end{equation}
for some function $S:\mathbb{R}^n\to \mathbb{R}$. The corresponding map $x$ is given by
\begin{equation}
x(q_1,\dots,q_n,t)=\left(\frac{\partial S}{\partial q_1},\frac{\partial S}{\partial q_2},\dots,\frac{\partial S}{\partial q_n}\right)
\end{equation}
for some time $t$. By writing $S= \frac{1}{2} q^2+ \Psi$ for $\Psi:\mathbb{R}^3\times \mathbb{R} \to \mathbb{R}$ we obtain
\begin{equation}
x(q,t)= q+  \frac{\partial \Psi}{\partial q},
\end{equation}
with the gradient field
\begin{equation}
s= \frac{\partial \Psi}{\partial q}.
\end{equation}
The Jacobian of the displacement map
\begin{equation}
\left[\frac{\partial s}{\partial q}\right]_{ij}=\frac{\partial^2 \Psi}{\partial q_i \partial q_j}
\end{equation}
is symmetric. The set of eigenvectors $\{v_i\}$ can be taken to be orthonormal by which the dual vectors coincide with the eigenvectors, i.e., $v_i^*=v_i$ for all $i$. A Lagrangian map is locally equivalent to the Zel'dovich approximation.\\

\subsection{Arnol'd's classification of Lagrangian catastrophes}
\noindent In section~\ref{sec:CC}, we described the classification of Lagrangian singularities in up to three dimensions. However the classification extends to higher dimensional singularities. A $(n+1)$-dimensional fluid can contain stable singularities in the $A_i$, $D_i$ and $E_i$ classes with $i \leq n+2$, where the $D$-class range starts at $i=4$ and the $E$-class is only defined for $i=6,7,8$. These singularities decompose into lower-dimensional singularities as illustrated in the unfolding diagram below.

\bigskip

\begin{center}
\begin{tikzpicture}
\node (A1) {$A_1$};
\node [right of=A1] (A2) {$A_2$};
\node [right of=A2] (A3) {$A_3$};
\node [right of=A3] (A4) {$A_4$};
\node [right of=A4] (A5) {$A_5$};
\node [right of=A5] (A6) {$A_6$};
\node [right of=A6] (A7) {$A_7$};
\node [right of=A7] (A8) {$A_8$};
\node [right of=A8] (A9) {$A_9$};
\node [right of=A9] (A10) {$\dots$};
\node [below of=A4] (D4) {$D_4$};
\node [below of=A5] (D5) {$D_5$};
\node [below of=A6] (D6) {$D_6$};
\node [below of=A7] (D7) {$D_7$};
\node [below of=A8] (D8) {$D_8$};
\node [below of=A9] (D9) {$A_9$};
\node [below of=A10] (D10) {$\dots$};
\node [below of=D6] (E6) {$E_6$};
\node [below of=D7] (E7) {$E_7$};
\node [below of=D8] (E8) {$E_8$};
  \draw[<-] (A1) to  (A2);
  \draw[<-] (A2) to  (A3);
  \draw[<-] (A3) to  (A4);
  \draw[<-] (A4) to  (A5);
  \draw[<-] (A5) to  (A6);
  \draw[<-] (A6) to  (A7);
  \draw[<-] (A7) to  (A8);
  \draw[<-] (A8) to  (A9);
  \draw[<-] (A9) to  (A10);
  \draw[<-] (D4) to  (D5);
  \draw[<-] (D5) to  (D6);
  \draw[<-] (D6) to  (D7);
  \draw[<-] (D7) to  (D8);
  \draw[<-] (D8) to  (D9);
  \draw[<-] (D9) to  (D10);
  \draw[<-] (E6) to  (E7);
  \draw[<-] (E7) to  (E8);
  \draw[<-] (A3) to  (D4);
  \draw[<-] (A4) to  (D5);
  \draw[<-] (A5) to  (D6);
  \draw[<-] (A6) to  (D7);
  \draw[<-] (A7) to  (D8);
  \draw[<-] (A4) to [out=302,in=152] (E6);
  \draw[<-] (A5) to  (E6);
  \draw[<-] (A5) to [out=302,in=152] (E7);
  \draw[<-] (A6) to  (E7);
  \draw[<-] (A6) to [out=302,in=152] (E8);
  \draw[<-] (A7) to  (E8);
\end{tikzpicture}
\end{center}

%%%%%%%%%%%%%%%%%%%%%%%%%%%%%%%%%%%%%%%%%%%%%%%%%%%%%%%%%%%%%%%%%%%%%%%%%%%%%%%%%%%%%%%%%%%%%%%%%%%%%%%%%%%%

\begin{table}
\begin{center}
  \begin{tabular}{ llll }
    \hline
    \\
$A_1:\ x(q,1)=(q_1,q_2,q_3)$ 							&	$1+\mu_1 =1$ 			& $1+\mu_2 = 1$ 		& $1+\mu_3 = 1$ \\ 
$A_2:\ x(q,1)=(q_1,q_2,q_3^2)$ 						&	$1+\mu_1 = 1$ 			& $1+\mu_2 = 1$ 		& $1+\mu_3 = 2q_3$ \\ 
									&	&	& 	$\mu_{3,3} = 2$ \\  
$A_3:\ x(q,1)=(q_1,q_2,q_1 q_3 +q_3^3)$					&	$1+\mu_1 =1$ 			& $1+\mu_2 = 1$ 		& $1+\mu_3 = 3 q_3^2 +q_1$ \\ 
									&	&  	& 	$\mu_{3,3} = 6 q_3$ \\ 
									&	&  	& 	$\mu_{3,333} = 6$ \\  
$A_4:\ x(q,1)=(q_1,q_2,q_1 q_3 + q_3^4)$					&	$1+\mu_1 = 1 $ 			& $1+\mu_2 = 1$ 		& $1+\mu_3 = q_1 + 4 q_3^3$ \\ 
									&	& 	& 	$\mu_{3,3} = 12 q_3^2$ \\ 
									&	&  	& 	$\mu_{3,33} = 24 q_3$ \\ 
									&	&  	& 	$\mu_{3,333} = 24 $ \\
$A_5:\ x(q,1)=(q_1,q_2,q_1 q_3 + q_2 q_3^2 + q_3^5)$			&	$1+\mu_1 = 1 $ 			& $1+\mu_2 = 1$ 		& $1+\mu_3 = q_1 + 2 q_2 q_3 + 5 q_3^4 $ \\ 
									&	&	& 	$\mu_{3,3} = 2 q_2+20 q_3^3$ \\ 
									&	&	& 	$\mu_{3,33} = 60 q_3^2$ \\ 
									&	&	& 	$\mu_{3,333} = 120 q_3 $ \\
									&	&	& 	$\mu_{3,3333} = 120$\\
$A_3^\pm:\ x(q,1)=(q_1,q_2,(q_1^2 \pm q_2^2) q_3 + q_3^3 )$	&	$1+\mu_1 = 1 $ 			& $1+\mu_2 = 1$ 		& $1+\mu_3 =  q_1^2 \pm q_2^2 + 3q_3^2$ \\ 
									&	&	& 	$\mu_{3,3} = 6 q_3$ \\ 
									&	&  	& 	$\mu_{3,33} = 6$ \\ 
$A_4^\pm:\ x(q,1)=(q_1,q_2,q_1 q_3 \pm q_2^2 q_3^2 + q_3^4)$	&	$1+\mu_1 = 1 $ 			& $1+\mu_2 = 1$ 		& $1+\mu_3 = q_1 \pm 2 q_2^2 q_3 + 4 q_3^3$ \\ 
									&	 & 	& 	$\mu_{3,3} = \pm 2 q_2^2 + 12 q_3^2$ \\ 
									&	 &  	& 	$\mu_{3,33} = 24 q_3$ \\ 
									&	 &  	& 	$\mu_{3,333} = 24 $ \\
  \ \\
  \hline
\end{tabular}
\end{center}
\caption{The caustic conditions of the normal forms of the $A$ 
singularity classes}
\label{table:eigenvalues}
\end{table}

%%%%%%%%%%%%%%%%%%%%%%%%%%%%%%%%%%%%%%%%%%%%%%%%%%%%%%%%%%%%%%%%%%%%%%%%%%%%%%%%%%%%%%%%%%%%%%%%%%%%%%%%%%%%
\section{Caustic conditions of the normal forms}
\label{ap:normalform}
\vskip 0.5truecm

\noindent We here verify the caustic conditions for the normal forms in the generic classification of singularities given in section \ref{sec:condnH}. The normal forms of the the Lagrangian singularities given in section \ref{sec:LagrangianSingularities} follow analogously.

\medskip
The eigenvalue fields and corresponding derivatives in the direction of the eigenvector fields are given in table \ref{table:eigenvalues}. The eigenvalues of the normal form for the trivial ($A_1$) case equal $1$ and thus satisfy the condition $1+\mu_i \neq 0$ for all $i$. The third eigenvalue of the normal form of the fold ($A_2$) singularity equals $-1$ in the origin. The derivative of the eigenvalue field in the direction of the corresponding eigenvector field does not vanish in the origin. The normal form thus satisfies the caustic conditions of the fold singularity. The normal forms of the remaining singularities follow analogously.

\end{document}